\documentclass[fleqn,usenatbib]{mnras}

\usepackage[T1]{fontenc}
\usepackage{ae,aecompl}
\usepackage{hyperref}
\hypersetup{draft}

\usepackage{graphicx}	% Including figure files
\usepackage{amsmath}	% Advanced maths commands
\usepackage{amssymb}	% Extra maths symbols
\usepackage{rotating}
\usepackage{appendix}
\title[{\sc HR-pyPopStar} evolutionary synthesis model]{{\sc HR-pyPopStar}: high wavelength-resolution stellar populations evolutionary synthesis model}
\author[Mill\'an-Irigoyen et al.]{
I. Mill\'an-Irigoyen$^{1}$\thanks{E-mail: iker.millan@ciemat.es},
M. Moll\'a$^{1}$,
M. Cervi\~{n}o$^{2}$,
Y. Ascasibar$^{3}$,
M.L. Garc\'{i}a-Vargas$^{4}$ 
\newauthor
\& P.R.T. Coelho$^{5}$ 
\\
$^{1}$ Departamento de Investigaci\'on B\'asica, CIEMAT, Av. Complutense 40, E-28040, Madrid, Spain\\
$^{2}$ Centro de Astrobiolog\'{i´}a (CSIC/INTA), ESAC Campus, Camino Bajo del Castillo s/n, E-28692 Villanueva de la Ca\~{n}ada, Spain\\
$^{3}$ Departamento de F\'isica Te\'orica, Universidad Aut\'{o}noma de Madrid, E-28049, Cantoblanco (Madrid), Spain\\
$^{4}$ FRACTAL S.L.N.E., C/ Tulip\'{a}n 2, p13, 1A, E-28231, Las Rozas de Madrid, Spain \\
$^{5}$ Universidade de S\~{a}o Paulo, Instituto de Astronomia, Geof\'{i}sica e Ci$\rm \hat{e}$ncias Atmosf\'{e}ricas, Rua do Mat\~{a}o 1226, 05508-090, S\~{a}o Paulo, Brazil
}

\date{Accepted XXX. Received YYY; in original form ZZZ}

\pubyear{2020}

\begin{document}
\label{firstpage}
\pagerange{\pageref{firstpage}--\pageref{lastpage}}

\maketitle

\begin{abstract}
We present the {\sc HR-pyPopStar} model, which provides a complete set (in ages) of high resolution (HR) Spectral Energy Distributions of Single Stellar Populations. The model uses the most recent high wavelength-resolution theoretical atmosphere libraries for main sequence, post-AGB/planetary nebulae and Wolf-Rayet stars. The Spectral Energy Distributions are given for more than a hundred ages ranging from 0.1\,Myr to 13.8\,Gyr, at four different values of the metallicity ($Z=0.004$, 0.008, 0.019 and 0.05), considering four different IMFs. The wavelength range goes from 91 to $24\,000$\,\AA\ in linear steps $\delta\lambda= 0.1$\,\AA, giving a theoretical resolving power $R_{th,5000} \sim 50\,000$ at $5000$\,\AA. This is the main novelty of these spectra, unique for their age and wavelength ranges. The models include the ionising stellar populations that are relevant both at young (massive hot stars) as well as old (planetary nebulae) ages. We have tested the results with some examples of HR spectra recently observed with MEGARA at GTC. We highlight the importance of wavelength-resolution in reproducing and interpreting the observational data from the last and forthcoming generations of astronomical instruments operating at 8-10m class telescopes, with higher spectral resolution than their predecessors.
\end{abstract}

%%%%%%%%%%%%%%%%%%%%%%%%%%%%%%%%%%%%%%%%%%%%%%%%%%

%%%%%%%%%%%%%%%%% BODY OF PAPER %%%%%%%%%%%%%%%%%%

\section{Introduction}
\label{Sec:Intro}

Observations of composite stellar populations (like clusters, star forming regions or galaxies) need to be analysed and interpreted in terms of their basic components or {\sl building blocks} to infer their star formation histories.
In classical synthesis models \citep[see e.g.][]{santos02}, these building blocks are observed spectra of stars or stellar clusters with different ages and abundances.
Alternatively, one may search for the best combination of theoretical models of Single Stellar Populations (SSP) obtained with evolutionary synthesis models, which is currently the most usual method.
In an evolutionary synthesis model, a SSP is assumed to be composed of stars that formed simultaneously, according to an assumed Initial Mass Function (IMF) and metallicity.
The stars then evolve with time according to the predictions of stellar tracks up to a certain set of ages, creating the isochrones. For each isochrone, individual stellar spectra are added to produce a synthetic Spectral Energy Distribution (SED) associated to the SSP.

There are two different approaches for assigning a combination of SSPs to an observed SED of a galaxy or galactic region.
On the one hand, an inverse technique is used in codes such as {\sc starlight} \citep{cid05}, {\sc vespa} \citep{tojeiro07} or {\sc fado} \citep{gomes18}.
These algorithms decompose the whole observed spectrum as a sum of contributions from individual SSPs selected from a base intended to cover a wide range in age and metallicity.
They find the optimum combination that fits the observed spectrum, minimising the residuals. On the other hand, chemical evolution models track the formation of different generations of stars and metal enrichment of the gas according to the infall and star formation history of the galaxy or region \citep{Mancone_Gonzalez2012, Molla14}. These models follow the evolution of the stars and the gas up to a certain age, and combine the results corresponding to these stellar generations to compute the final theoretical SED.

In both techniques, the SEDs computed for SSPs by an evolutionary synthesis model are an essential ingredient.
For a meaningful comparison between models and observations, it is required that both SEDs, synthetic and observed, share a common spectral resolution.
In the case of synthetic SEDs with lower resolution than observations, the latter would have to be degraded.
Ideally, a high resolution synthetic SED should be convolved with all the contributions responsible for the broadening of the spectral lines (intrinsic physical phenomena in the observed galaxy, atmosphere and instrumental profile) before being compared with an observed spectra.
It is more adequate, in any case, to degrade the synthetic SED than observed spectra.

{\sc PopStar} models \citep[][hereinafter MOL09, MAN10 and GV13]{molla09,mman10,GarciaVargas_Molla_MartinManjon2013} based on previous works by \citet{garcia-vargas94,garcia-vargas95,garcia-vargas98}, were specifically devoted to the study of young stellar clusters, with a similar scope to the {\sc STB99} models \citep{lei99}.
MOL09 models included the most updated isochrones and atmosphere models at that time, and a very wide age range to cover both very young and very old stellar populations.
The inclusion of WR stars spectra, present at the early stages, and planetary nebulae (PNe), for evolved clusters, changed the ionising photons budget and made a difference in both computed SED and magnitudes, with respect to previously published SSP models \citep{CER1994,Fioc_Rocca_Volmerange1997,Kodama_Arimoto1997,lei99,bc03,Gonzalez+2005,Maraston2005,fritze06,Coelho+2007,elst09, Conroy_Gunn_White2009, Maraston+2009}, and even with respect to later models \citep{Conroy_VanDokkum2012, Maraston_Stromback2011, lei14, vazdekis2015, vaz16, Fioc_Rocca_Volmerange2019, Maraston+2020, Coelho+2020}. However, {\sc PopStar} SEDs had low wavelength-resolution, $R_{th,5000}\sim 800$.

The term {\sl resolution} in the context of spectral library models (either for stars or for SSPs) \citep[hereinafter C14]{Coelho2014} and in the observational spectroscopy may have different meanings.
The theoretical model resolution is often characterised the wavelength sampling of the output spectrum, or wavelength step $\delta\lambda$.
Here we will use the term {\sl wavelength-resolution}, defined as $R_{\mathrm{th}} = \lambda/\delta\lambda$.
In observations, {\sl spectral resolution} refers to the ability to resolve spectral lines, broadened due to different factors, such as instrument design, seeing in ground-based observations and intrinsic velocity dispersion.
Thus, the resolving power is defined as $R=\lambda/\Delta\lambda$, where $\Delta\lambda$ is the resolution element calculated as the full-width half maximum (FWHM) of the line-spread function.
Although the terms $R$ and $R_{\mathrm{th}}$ cannot be compared, the wavelength step, and the reciprocal linear dispersion of an instrument, both named $\delta\lambda$, can be.

The resolution $R_{th,5000}\sim 800$ of old PopStar models was enough for many purposes, like the use of these ionising spectra to predict the gas emission lines (MAN10) and their contribution to broadband magnitudes and colours (GV13).  In fact, over the years, a large number of galaxy spectra obtained by low-resolution instrumentation were analysed with similarly low-resolution theoretical synthesis models.
Age, metallicity and dust content were estimated through continuum fitting, continuum slope determination and/or colours.
Nevertheless, {\sc PopStar} cannot be used to fit observed spectra with resolving power $R$ larger than 1000.  
The need for higher wavelength-resolution theoretical SSP spectra appeared when spectrographs with $R \sim ~1500 - 3000$ in the visible range became the workhorse instruments in 4m class telescopes for some decades.
The field of extragalactic astronomy was then revolutionised thanks to a more detailed view provided by the study of absorption lines, whose fluxes and equivalent widths contain information on the metal abundance $\rm [M/H]$ and age $t$ of the composite stellar populations.
Thus, the old method of using {\sl fitting functions} of the stellar spectral absorption indices as a function of the star physical parameters, and then combine them in the composite spectra to compute the spectral indices in SSP evolutionary models \citep{worthey94,garcia-vargas98,mgv00, cen09,Thomas+2011} was not necessary any more.
With the increase in spectral resolution, the simultaneous measurement of continuum and absorption lines started to be the way to estimate age and metallicity \citep{vaz03,Maraston_Stromback2011}.

The increase in spectral resolution in the last generations of instruments for the largest ground-based telescopes has made possible to measure both flux and equivalent width of only stellar de-blended lines, therefore detected with "intermediate" or moderately high resolution models ($R \sim 5\,000 - 10\,000$). The interpretation of these observations demanded new models with higher $R_{\mathrm{th}}$ that were computed by some groups, providing new SEDs, flux and equivalent width of stellar lines, mostly in the optical range, for SSP and combined populations. Some of these evolutionary models are \citet{LeBorgne+2004,Gonzalez+2005,Coelho+2007,Percival+2009,Maraston_Stromback2011, vaz16, conroy18, Coelho+2020} 

Over the last years, high-resolution ($R$ in the range $10\,000 - 20\,000$) was the opportunity niche of stellar astronomy, since stars were bright enough to allow very fine observations in the 4m class and the 8-10m class telescopes, while at the same time very high resolution echelle-type instruments ($R$ in the range $50\,000 - 100\,000$) were mostly devoted to the study of individual stars in the search of exoplanets. The high resolution ($R \approx  10\,000 - 20\,000$) was abandoned in the context of composite stellar systems. 

However, some instruments recently in operation in 8-10m class telescopes took the baton with powerful high-resolution spectroscopy modes.
This new generation of instruments, in very large telescopes with high photon-collection capability, produce high Signal-to-Noise (SNR) HR spectra ($R \approx 20\,000$) of star clusters and galaxies with magnitude in the range 20 - 22 in few hours of exposure time.
This is the case of MEGARA, in operation at the GTC 10.4m telescope since 2018, with LR, MR and HR spectroscopy modes at $R \approx$ 6\,000, 12\,000, and 18\,000, respectively.
MEGARA LR and MR modes have VPH gratings covering the whole wavelength range between 3650 and 9750 \,\AA, while HR modes are centred at 6606.5\,\AA\ (HR-R) and 8633.0\,\AA\ (HR-I), with $\Delta\lambda\,(\rm FWHM)$ = 0.392\, and 0.520\,\AA, corresponding to reciprocal linear dispersion $\delta\lambda$ = 0.098 and 0.130\,\AA/pix for HR-R and HR-I, respectively.
VLT/MOONS will operate soon at the VLT 8.2m telescope, with two HR modes: HR-I (7650 - 8980\,\AA) at $R\approx 9\,200$, $\Delta\lambda = 1.11$\,\AA\ at 8315.0\,\AA\ and HR-H (1.521 - 1.641 \,$\mu$m) at $R\approx 18\,300$, $\Delta\lambda = 1.15$\,\AA\ at 1.58 $\mu$m.
These new observations demand synthetic high wavelength-resolution SEDs of SSPs from young stellar populations formed during a recent burst of star formation to old populations created at the beginning of the universe, over the widest possible range in metallicity.

\begin{table}
  \begin{center}
	\caption{Summary of some intermediate and high resolution SSP models of the literature, compared to the present work.}
\label{Table:1}
\begin{tabular}{clccr}
\hline
Reference & Code &  Range (\AA) & $\delta\lambda$\,(\AA) & $ R_{\rm th}$ \\
\hline
 LEB04 & {\sc Pegase-HR} & $4000 \-- 6800$ & 0.55 & 9091 \\
 GON05 & Sed@      & $3000 \-- 7000$ & 0.30 & 16667 \\
 COE07 & {\sc galaxev}   &$3000 \-- 13400$ & 0.20 & 25000 \\
 PER09 & {\sc BASTI} iso.     &$2500 \-- 10500$ & 1.00 & 5000 \\
 M\&S11 & Maraston & $1000 \-- 25000$ & 0.25 & 20000 \\
 VAZ16 & {\sc miles}    & $1680 \-- 500000$ & 0.90 & 5556 \\
 CON18 & {\sc fsps}     & $3700 \-- 24000$ & 1.67 & 3000\\
 CBC20 & {\sc galaxev}  & $3540 \-- 7410 $ & 0.90 & 3000 \\
 This work & {\sc pyPopStar} & 91 \-- 24000 & 0.10 & 50000 \\
\hline
\end{tabular}
\end{center}
\footnotesize{References. LEB04: \citet{LeBorgne+2004};  GON05: \citet{Gonzalez+2005}; COE07: \citet{Coelho+2007}; PER09: \citet{Percival+2009}; M\&S11: \citet{Maraston_Stromback2011};
VAZ16: \citet{vaz16}; CON18: \citet{conroy18}; CBC20: \citet{Coelho+2020}. }
\end{table}

Table~\ref{Table:1} summarises the evolutionary synthesis models
from the literature with medium or high spectral resolution (considering only those with $R_{th} \ge 3000$). The columns indicate: (1) the reference of each model, (2) the synthesis code used to compute the SED,
(3) the wavelength range, (4) the wavelength step $\delta\lambda$ at $\lambda = 5000$\,\AA\, and (5) the value $R_{\mathrm{th}}=\lambda/\delta\lambda$, as defined for theoretical models, at that wavelength.
Most of the current models lack the wavelength-resolution to manage data with $R_{\mathrm{th}} \ge 15\,000$, except GON05, COE07, and M\&S11, which are limited to the optical range between 3000 and 10\,000\,\AA. 
Thus, the main aim of this work is to present our {\sc HR-pyPopStar} models, a new grid of {\sc PopStar} models with high wavelength-resolution, $R_{\textrm{th,5000}}=50\,000$, based on available HR theoretical stellar libraries, as a tool to support the interpretation of observations from the state-of-the-art HR instruments that are now starting to operate in large telescopes. 
We describe the model in Section ~\ref{Sec:Model}, mainly introducing the stellar libraries used in this work and how we have assigned these atmosphere models to the stars of each isochrone.
The resulting SEDs are described in Section~\ref{Section:Results} together with some derived products, such as broadband magnitudes and number of ionising photons of different atomic elements.
Section \ref{Sec:Discussion} discusses the results and shows a comparison between our SEDs and other theoretical models in the literature.
We also illustrate the power of our models by showing an analysis of the absorption spectral lines and a comparison with published HR observed spectra.
The conclusions are summarized in Section~\ref{Sec:Conclusions}.
Two appendices in electronic format describe some details of the model ingredients (Appendix A) and a comparison of results with the previous {\sc PopStar} models (Appendix B).

%-------------------------------------------------------------------
\section{Model Description}
\label{Sec:Model}
%-------------------------------------------------------------------

The main ingredients in any population synthesis code are the isochrones, which contain the stellar parameters of each star present in the system at a given age; and the spectral library, the collection of the spectra - observed or modelled - for each type of star. Other assumptions include the IMF, the way to compute the SEDs, and the prescription to compute the contribution of the nebular emission continuum.

As summarized in Appendix A, many features of these new {\sc HR-pyPopStar} models are inherited from {\sc PopStar} (MOL09). The main differences are the use of different theoretical stellar atmosphere libraries, and a new code to manage HR spectra.
We have computed the number of stars for each mass interval using four different IMF: \citet{Salpeter1955}, \citet{Ferrini_Penco_Palla_1990}, \citet[][with a slope $\alpha=-2.7$ for massive stars]{Kroupa2002}, and \citet{Chabrier2003}, hereinafter referred to as SAL, FER, KRO and CHA, respectively.
The stellar isochrones are from the Padova group \citep{Bressan_Bertelli_Chiosi1993, Fagotto+1994a, Fagotto+1994b,Girardi+1996}, for metallicities $Z =0.004$, 0.008, 0.02 and 0.05, and ages $t$ between 0.1 and 13.8\,Gyr.

%%%%%%%%%%%%%%%%%%%%%%%%%%%%%%%%%%%%%%%%%%%%%%%%%%

\subsection{High wavelength-resolution stellar libraries}
\label{Subsec:Spectra}

The stellar spectral libraries are the key piece in this SSP code building. The spectral coverage and resolution of the stellar libraries available in the literature are crucial to produce the desired results. The spectral libraries, or collections of stellar spectra, would ideally have the same wavelength range and resolution, to construct easily the SSP spectra as the addition of all of them in the adequate contribution, as defined by the IMF,
for a certain age (isochrone) in the evolutionary synthesis models.
The stars comprising these libraries are usually classified according to the main stellar atmospheric parameters that govern their SED: effective temperature ($T_\mathrm{eff}$), surface gravity ($\log g$) and metallicity, usually given in terms of iron [Fe/H] or global abundance [M/H], although WR stars are characterized by other quantities, as the mass loss or the surface abundance.
In order to produce accurate synthetic spectra, the stellar library should cover a range in these stellar parameters that encompasses the values predicted by the isochrones.

Stellar libraries may be classified as empirical (based on observed data) and theoretical (i.e. computed from stellar models including radiative-transfer processes as a function of physical parameters).
%Each type of library has its advantages and disadvantages.
Thus, empirical libraries have the advantage of being composed by real observed stars.
Most of them, however, have relatively low resolution, reaching $R \sim$  2500, 2000 and 1800 in MILES \citep{sb06}, E-IRTF \citep{villaume2017} and MaStar \citep[][see this last work for a review about empirical libraries]{Yan+19}, respectively.
The other important issue is the limited coverage in terms of the parameter space, being constrained to the ranges of $T_{\rm eff}$, $\log{g}$ and $\rm [M/H]$ spanned by the stars in the Milky Way galaxy and its satellites.
These libraries are often biased towards the brightest stars to save observing time and/or the most frequent types, associated to the length of each evolutionary stage.
For all these causes, empirical libraries are scarce in low metallicity stars, young massive stars, and cool dwarfs, which can make a big difference in the final synthetic spectrum.
Thus, in the MANGA library, MaStar \citep{Yan+19}, even after improving the catalogue with low metallicity stars, there are still no stars hotter than 18\,000\,K, missing the early O-type and WR populations.
Part of our group is building a HR empirical stellar library for MEGARA at the GTC \citep[{\sc megastar};][]{gv20} that will accurately sample these hot stars; details of the first release can be found in \citet[][]{carrasco21}.
A similar caveat applies to post-AGB stars, which are not always well represented in empirical libraries, although they may even dominate the spectrum at certain wavelength intervals due to their high luminosity.

On the other hand, theoretical stellar libraries are based on atmosphere models, which allow a wide coverage of the stellar parameter space: gravity, effective temperature and metallicity (although some types of stars are still missing).
Examples of theoretical libraries used for population synthesis models are \citet{Lejeune+1997, Coelho+2005, mar05, mun05, rm05, Fremaux06, Gustafsson+08, lei10, k11, dl12, Coelho2014} and  \citet{Bohlin+2017} among others.
These libraries require a comprehensive and reliable database of both atomic and molecular line opacities, not always complete or available, and suffer from potential uncertainties coming from the atmosphere model limitations (convection properties, line-blanketing, expansion, non-radiative heating, non-Local Thermodynamic Equilibrium -- non-LTE -- effects, etc.).
Since a lot of progress in some of the research issues mentioned above has been made in recent years, we will use theoretical stellar spectra in this work, considering them good enough to be included in our evolutionary synthesis model.

From a careful comparison between empirical and theoretical stellar libraries, \citet[][hereinafter CBC20]{Coelho+2020} conclude that magnitudes and ages are more affected by the limited coverage of stellar parameters in an empirical library, while the metallicity, being robust against the limited coverage, is underestimated when using synthetic libraries.
That is, there is not an optimum solution when selecting the stellar library to estimate and predict all quantities simultaneously, although we expect that the metallicity determination will improve with the use of these new models due to the higher number of metallic lines resolved at HR.

As mentioned above, {\it hot} stars (early O, B and WR, as well as PNe) are important in order to compute the ionising spectra for the youngest and intermediate/old ages, respectively. Therefore, we classify the stellar spectral models we have used in four types: 1) Normal stars, NS, with spectral types A to M (effective temperature lower than 25\,000\,K); 2) Hot stars with O and B spectral types, 3) WR stars, and 4) post-AGB/Planetary Nebulae (PNe). 

\begin{table}
	\begin{center}
	\caption{Summary of the characteristics of the used atmosphere models.}
\label{Table:2}
\begin{tabular}{|c|c|}
\hline
Properties & Range \\
\hline
\multicolumn{2}{c}{NS models}\\
\multicolumn{2}{c}{C14}\\
\hline
$ T_{\rm eff}$ &  $3000 \-- 24\,000$\,K  \\
$ \Delta T_{\rm eff}$ & 250 \-- 1000\,K\\
$\log{g} $ & $0.0 \-- 5.0$  \\
$\Delta \log g$ & 0.5\\
$Z$ & $0.0017, 0.005, 0.017, 0.03$  \\
Wavelength range & $900\-- 24\,000$ \AA \\
$\delta\lambda$ & 0.02\,\AA\ \\
\hline
\multicolumn{2}{c}{O and B stars models}\\ 
\multicolumn{2}{c}{\citep{Hainich+2019}}\\
\hline
$ T_{\rm eff}$ &  $25\,000 \-- 50\,000$\,K  \\
$ \Delta T_{\rm eff}$ & 1000\,K \\
$\log g$ &$2.2 \-- 4.2$  \\
$\Delta \log{g}$ & 0.2\\
$Z$ & $0.0024, 0.008, 0.017$  \\
Wavelength range & $10$ \-- $24\,000$ \AA\ \\
$\delta\lambda $ & 0.1 \AA\ \\
\hline
\multicolumn{2}{c}{ WR stars models}\\ 
\multicolumn{2}{c}{\citep{Hainich+2019}}\\
\hline
$ T_{\rm R20}$ &  $39\,810 \-- 199\,526$ \\ % $0.37 \-- 9.36$
$ \Delta T_{\rm R20}$ & $4858$ {\--} $21\,700$  \\ % $0.1$ to $1.92$ 
$\dot{M}$ & $-4.34 \-- -6.14$  \\
$\Delta \dot{M}$ & $ 0.15$  \\
$Z$ & $0.0024, 0.008, 0.017$  \\
Wavelength range & $10$ \-- $24\,000$ \AA\ \\
$\delta\lambda $ & 0.1 \AA\ \\
\hline
\multicolumn{2}{c}{Post-AGB/PN models} \\
\multicolumn{2}{c}{\citep{Rauch2003}}\\
\hline
$ T_{\rm eff}$ &  $50\,000 \-- 190\,000$ K \\
$ \Delta T_{\rm eff}$ & 10\,000 \\
$\log g$ &$5.0 \-- 8.0$ \\
$\Delta \log g$ & 1.0\\
$Z$ & $0.0017,  0.017$ \\
Wavelength range & $10$ \-- $160\,000$ \AA\ \\
$\delta \lambda $ & 0.1 \AA\ \\
\hline
			\end{tabular}
	\end{center}
\end{table}

For NS stars, we use the stellar library from C14.
This is a theoretical stellar library of spectral types A to M based on previous models from \citet{Coelho+2005}.
C14 presents plane-parallel atmosphere models and synthetic spectra corrected for line blanketing.
The models provide very high spectral dispersion $R_{th}\approx\,300\,000$ in the visible range, and have been extended to a wavelength range from 900 to 24\,000\,\AA\ for this particular work (see Table~\ref{Table:2}).
We have further expanded the C14 spectra for NS, using the low resolution stellar library of \citet[][BaSel library hereafter]{Lejeune+1997} to cover the wavelength range from 91 to 900\,\AA.
This expansion allows to us to compute the numbers of ionising photons, and to have a homogeneous wavelength range for all stellar libraries used in {\sc HR-pyPopStar} code.

\begin{figure}
\centering
\includegraphics[width=0.47\textwidth]{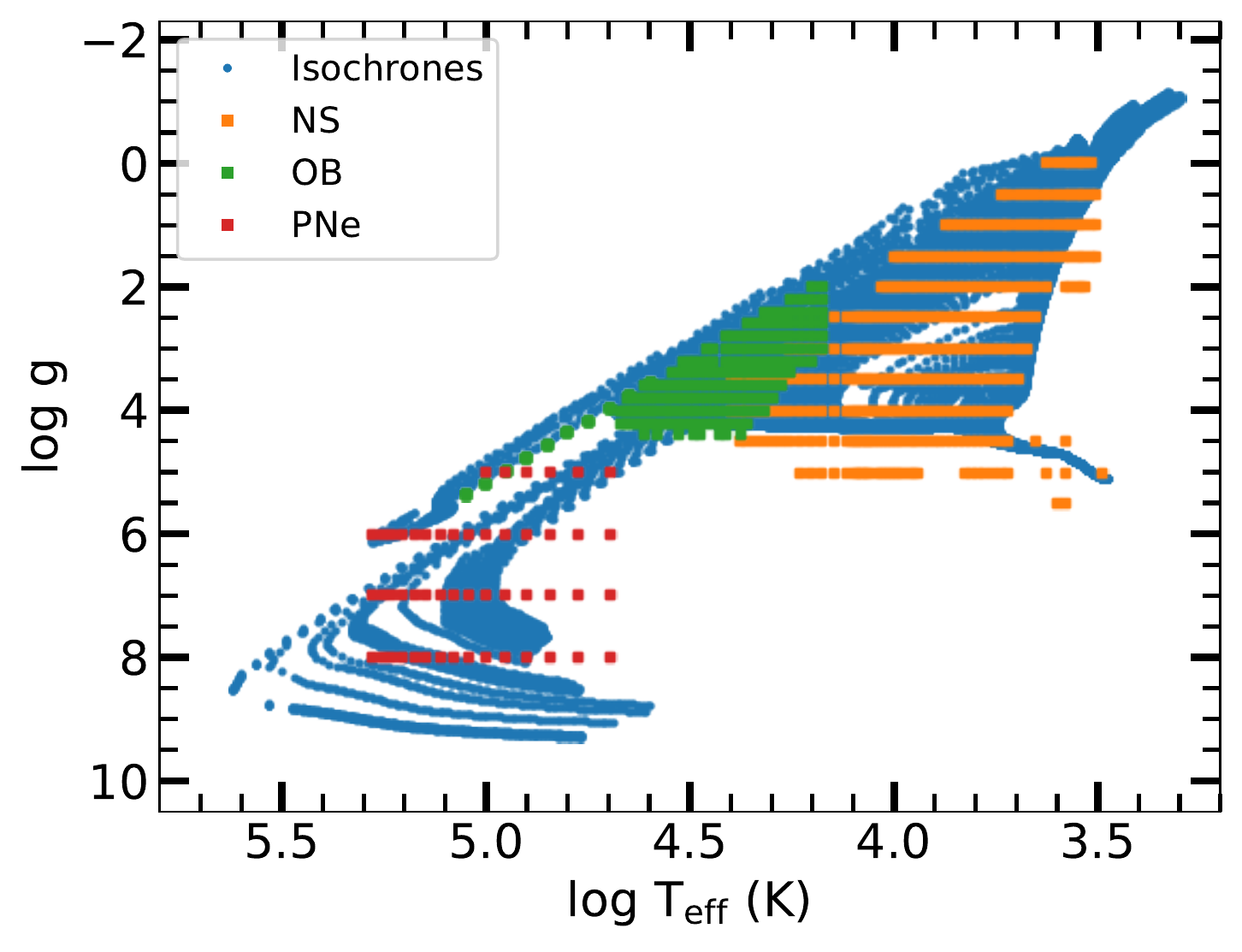}
\caption{The $T_\mathrm{eff} - \log{g}$ diagram showing the points from the isochrones (in blue) at all ages and solar metallicity, and the values from stellar spectral libraries plotted in  orange, green and red for NS, OB and post-AGB/PNe stars, respectively (WR stars are not included).}
\label{Fig:2}
\end{figure}

This library has a range and step resolution in $T_{\rm eff}$ and $\log{g}$ appropriate to match the values of isochrones for most normal stars (see Figure~\ref{Fig:2} and Table~\ref{Table:2}), albeit the lowest effective temperature in C14 models, $T_\mathrm{eff,min} \sim 3000$~K, is somewhat higher than the minimum reached by the isochrones, $T_\mathrm{eff,min} \sim 2000$~{K}.
This limitation might affect the oldest populations ($t > 10$\,Gyr), where the low-mass cool stars make a significant contribution to the final spectra.
In order to quantify this effect, we have compared in Appendix B (see supplementary information) the photometry obtained from our new {\sc HR-pyPopStar} models with the one from the old {\sc PopStar} SEDs, which contained stars of the BASEL library as cool as 2000\,K.
We find no significant differences in the resulting magnitudes at the ages that could be affected. Nevertheless, some spectral lines dominated by the contribution of late-K and M stars might be underestimated; in particular those of Vanadium Oxide (VO), not included in C14 atmosphere models.
Therefore, some caution must be exercised when using the models of the oldest stellar populations.

O and B spectral types correspond to massive stars, $m \ge 25\,\mathrm{M}_{\sun}$, with a significant mass loss.
This demands a more detailed treatment, as a spherical expansion in NLTE including line blanketing.
We have chosen the model atmospheres from the PoWR team \citep{Hainich+2019}, computed using their own code \citep{Grafener+2002,HamannGrafener2003,Sander+2015}, for O and B stars of 3 different metallicities ($\mathrm{Z}_{\sun}$, $0.5\,\mathrm{Z}_{\sun}$ and $0.14\,\mathrm{Z}_{\sun}$) simulating the Milky Way Galaxy, Large Magellanic Cloud and Small Magellanic Cloud conditions, respectively. The coverage in the $T_\mathrm{eff}-\log g$ plane can be seen in Figure~\ref{Fig:2} (green squares) and Table~\ref{Table:2}. \citet{Hainich+2019} compute the lowest metallicity set of models (SMC) for 3 different wind strength parameters. We have used the models they call {\it moderate wind strength}, calculated with the same mass loss rate, $\log \dot{M}= -7.0$, as for the other two metallicities. 

Post-AGB and planetary nebulae are low and intermediate mass stars in the late stage of their lives with high surface gravity and effective temperature, but low luminosity.
We have used the theoretical stellar spectral library from \citet{Rauch2003} extending from 10 to 160\,000\,\AA.
These authors assume NLTE a plane-parallel atmosphere in their models, computed for ranges of $T_{\rm eff}$ between 50\,000\ and 190\,000\,K; and $\log{g}$ from 5.0 to 8.0.
Since the highest value of $T_{\rm eff}$ is lower than the hottest PNe found in the isochrones, $T_{\rm eff}\sim 300\,000$\,K, we have used blackbody spectra for stars with $T_{\rm eff} \ge 225\,000\,\mathrm{K}$. 
In general, the coverage in $T_\mathrm{eff}$ (with steps as small as $\Delta\,T_\mathrm{eff} = 250$\,K for the cool NS, and $\sim 1000$\,K for the hottest stars is proportionally much better than the gravity steps $\Delta \log g$ for the whole stellar library, but this is particularly relevant for the \citet{Rauch2003} models, where $\Delta \log g = 1$.

Figure~\ref{Fig:2} compares the $T_\mathrm{eff} - \log{g}$ diagram from both isochrones and stellar libraries. Blue points solar metallicity isochrones at all ages, while the other points correspond to the different stellar libraries we are using for NS. WR stars are not shown in the diagram because of the different physical parameters which define them, as discussed below.
The parameter coverage is good enough for NS, with $3000\,\mathrm{K} < T_\mathrm{eff} < 24\,000\, \mathrm{K}$ and $0.0 < \log{g} < 5.0$, OB stars and PNe.

For WR stars, with strong winds and consequently high mass loss rates, we have also used the models by the PoWR group. Given the extreme conditions of WR stars, the surface abundances of key chemical elements, such as C, N and O, change quickly. For this reason, in addition to the classical classifications of WR by metallicity \citep{Smith_Norris_Crowther2002},  
they provide models with varying hydrogen surface abundance like the ones from \citet{HamannGrafener2004,SanderHamannTodt2012, Todt+2015}. We have used their models with H abundance  $X_{H}=0$ for WR-WC and those with $X_{H}=0.2$ for WR-WN, since the presently used isochrones do not provide the information of surface abundances of each star, only giving the classification as WR-WC or WR-WN. WR atmosphere models are not characterised by $T_{\rm eff}$ and $\log{g}$ as in the other stars, but by $T_{\rm R20}$, the effective temperature in $R_{20}$, the radius where the atmosphere of the star has an opacity $\tau=20$, and the mass loss $\dot{M}$.  

As a final comment, let us note that we have homogenised the wavelengths for all stellar libraries, rebinning the spectra to a same wavelength range 91 - 24\,000 \AA\ with $\delta\lambda = 0.1$\,\AA.

%%%%%%%%%%%%%%%%%%%%%%%%%%%%%%%%%%%%%%%%%%%%%%%%%%

\subsection{Description of the code: Stellar models assignment}
\label{Subsec:code}

\begin{figure}
\centering
\includegraphics[width=\linewidth]{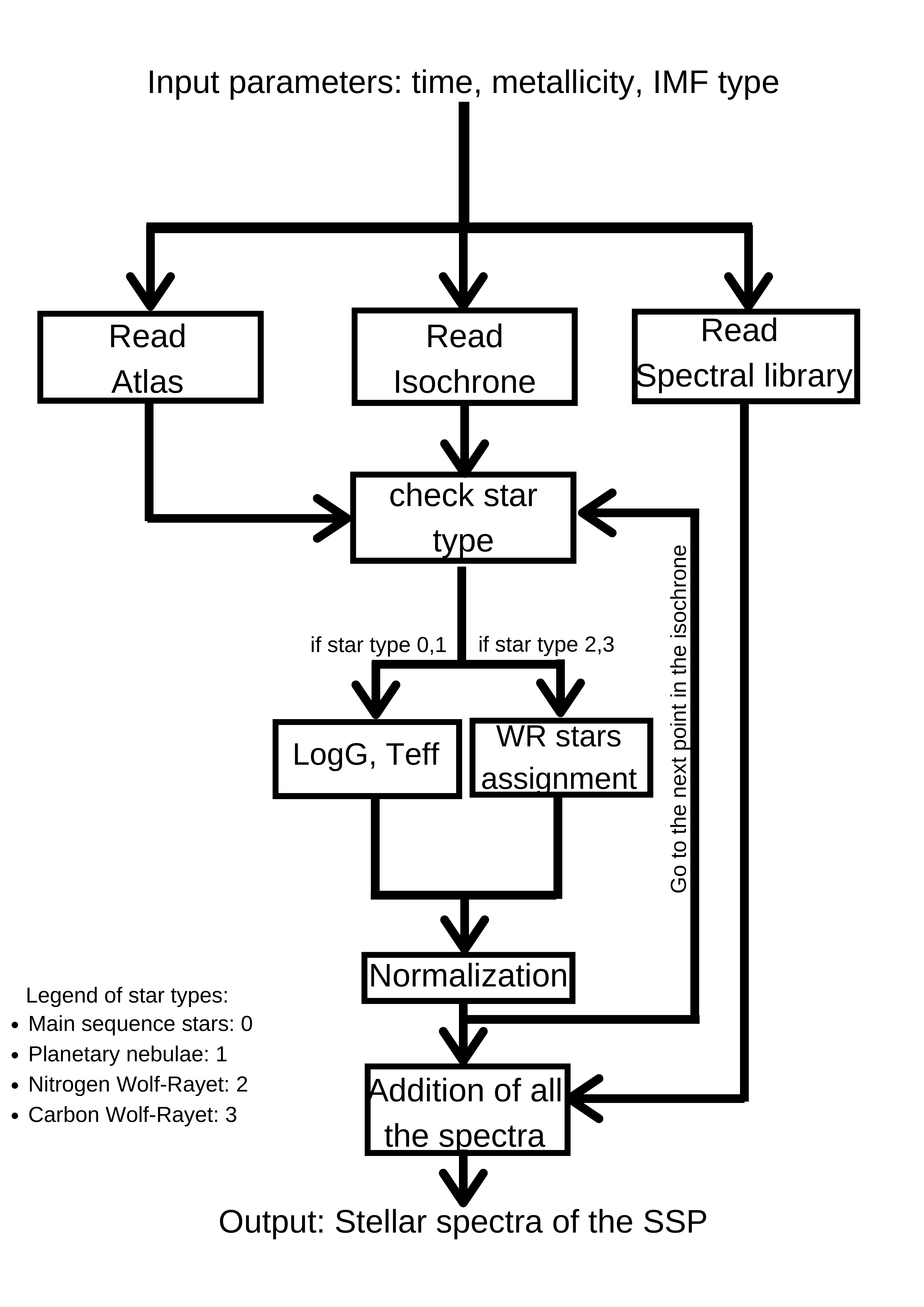}
\caption{Flow chart of the pyPopStar code.}
    \label{Fig:1}
\end{figure}

{\sc HR-pyPopStar} uses very high wavelength-resolution spectra, with $\sim$240\,000 wavelength values.
This requires the use of an appropriate technique to manage them without using extremely long arrays.
We have developed a new, highly modular code in Python language (hence the name {\sc HR-pyPopStar}) that makes possible to change its individual ingredients in a very flexible way.
In the current version, we can use any isochrone set (as long as initial mass, current mass, mass intervals, $T_{\rm eff}$, $L_{\rm bol}$, $\log{g}$ and $Z$ are given) or spectral libraries, at any resolution, other than those used here.
Code improvements have also resulted in a strong optimisation, both in terms of memory resources and execution time.
The current version computes a SSP model at HR in $\sim$10~s and stores 7.3~MB for 239\,900 wavelengths. 

A scheme that explains how {\sc HR-pyPopStar} runs is shown in Figure~\ref{Fig:1}.
As a first step, the code needs the selected age, metallicity and IMF of the models to be computed.
Once the required information is specified, the code reads the isochrone for that age and metallicity, the stellar library of the specified metallicity and an atlas associated to this stellar library containing all its complementary information. Then, it iterates over each point of the isochrone to assign the nearest neighbour in the stellar library.

For non-WR stars (i.e. main sequence, O, B, post-AGB and planetary nebulae), the code looks, for each mass interval in the selected isochrone, for the closest model in $T_\mathrm{eff}$, and then for the closest model in $\log{g}$ at that temperature.
For O and B stars in the range covered by both PoWR and C14 libraries, we establish, for each metallicity, a boundary temperature (for each gravity) between the last value of the C14 library and the first one of the PoWR O and B models, defined as the average value between these two temperatures.
Since there is no overlapping between the post-AGB and PNe stars and O stars, this question is avoided in that part of the HR diagram.

For NS stars and OB stars, the error $\Delta T_\mathrm{eff}$ is smaller than a 10\%, except for the coolest ones, due to the already mentioned lack of models of M-type stars. For PNe, it is also lower than a $10 \%$ except for the lowest and the highest temperature cases, due to the lack of low $T_\mathrm{eff}$ models, and to the use of a black-body spectrum when $ T_\mathrm{eff}$ > 225\,000,K, respectively.
The errors in $\log g$ are simply those associated to sampling $\Delta \log{g}$ of the corresponding stellar library, quoted on Table~\ref{Table:2}.

For WR stars, the assignment is more complex, because the isochrones give the effective temperature at the radius $R_\mathrm{hydro}$, where the optical depth is $\tau=2/3$, while WR atmosphere models are computed at an optical depth $\tau=20$.
We have followed a similar method as the one presented in MOL09, using the radius of the WR atmospheres at $\tau=20$, $R_{\tau=20}$ (instead of $R_{\tau=10}$ as before) and the mass loss $\dot{M}$ to assign the WR spectral templates (see details in Appendix A).

Now, we must add the luminosities of all the assigned spectra, taking into account the IMF, or number of stars in each point defined by the isochrone of given age $t$ and metallicity $Z$, to compute the final SED of each SSP:
\begin{equation}\label{Eq:sum_of_spectra}
    F_{\lambda, \mathrm{SSP}}(Z,t) = \int_\mathrm{m_{low}}^{m_\mathrm{lim}(t)}\, \phi(m)\, F_{*,\lambda}(m, Z)\,\mathrm{d}m,
\end{equation}
where $m_\mathrm{low}$ is the lower limit of the IMF, $\phi(m)$, and $m_\mathrm{lim}(t)$ is the most massive star still alive in the isochrone of age $t$.
Thus, the integration is performed for the total number of points along each isochrone.

In this case, however, as quoted before, the size of a single metallicity stellar library is 1.8\,Gb
that is $\sim$ 300 times larger than a typical low Resolution set. Therefore, the problem is to read and operate with these new spectra, which is very time consuming due to the high number of different  wavelengths. In order to manage them and to optimize the memory and time performance of the code, we have following here the same strategy as in the Sed@ code \citep[][]{Gonzalez+2005, cer06}, that is to work in the atmosphere models domain, instead of the usual isochrone domain, (where the final flux was obtained by adding all stellar fluxes assigned for each point of the isochrone as explained).

In the present case, the final SSP flux of a certain isochrone age $t_i$ and metallicity $Z_i$ is therefore computed as:
\begin{equation}
    F_{\lambda, \mathrm{SSP}}(Z_i,t_i) = \sum_{j=1}^{\mathrm{N_{tot\,lib}}(Z_i)} w_j(Z_i,t_i) \times  {F_{\lambda,\,j}^\mathrm{lib}}(Z_i), 
\end{equation}
where $\mathrm{N_{tot\,lib}}(Z_i)$ is the number of elements in the library set with metallicity $Z_i$, $F_{\lambda,\,j}^\mathrm{lib}(Z_i)$ is the monochromatic flux of the element $j$ of the library with metallicity $Z_i$, and $w_{j}(Z_i,t_i)$ is the weight of each library element $j$ in the isochrone associated to index $i$.

The strategy is as simple as computing the $w_{j}(Z_i,t_i)$ values.
Each isochrone point has a contribution given by the IMF:
\begin{equation}
w_\mathrm{IMF}(m_j, Z_i, t_i) =
\int_{m_{\mathrm{low},j}(Z_i, t_i)}^{m_{\mathrm{up},j}(Z_i,t_i)}\ \phi(m)dm,
\end{equation}
\noindent where $m_{\mathrm{low},j}(Z_i,t_i)$ and $m_{\mathrm{up},j}(Z_i,t_i)$ define the mass intervals along the isochrone that share the same element $j$ in the atmosphere library.
We correct the normalisation based on the bolometric luminosities $L$ obtained for the isochrone and the stellar library model:
\begin{equation}
    w_j(Z_i,t_i) = w_\mathrm{IMF}(m_j, Z_i, t_i)\ 
     \frac{L_\mathrm{iso}(m_i,Z_i,t_i)}
    {L_{j,\mathrm{lib}}(Z_i)},
\end{equation}
where $L_{j}^{\mathrm{lib}}(Z_i)$ is the integral over wavelength of ${F_{\lambda,\,j}^\mathrm{lib}}(Z_i)$ in the stellar library, which can be pre-computed or obtained form the library parameters.
In this case, for plane-parallel non-expanding atmosphere models as C14 or \cite{Rauch2003}, ${F_{\lambda,\,j}^\mathrm{lib}}(Z_i)$ is the flux at the surface per $\mathrm{cm}^2$, so we compute the luminosity for a given $T_\mathrm{eff}$ as $L_\mathrm{bol}/R^2 = \sigma_\mathrm{SB} T_\mathrm{eff}^4$, being $\sigma_\mathrm{SB}$ the Stefan-Boltzmann constant. In the case of expanding models \cite[][models]{Hainich+2019} such integral has been previously calculated for each stellar model by the own authors, being an information included in their libraries. 

We note that to compute the weights $w_j(Z_i,t_i)$ we do not require the particular ${F_{\lambda,\,j}^\mathrm{lib}}(Z_i)$ values, but just the parametric structure of the atmosphere library.
Hence computing all the required weights first, the whole library is never stored in memory.
The final spectra are calculated on a wavelength by wavelength basis, which substantially decreases the memory requirements and computation time when the number of wavelength points is much larger than the number of isochrone points.

\begin{table}
	\begin{center}
\caption{Metallicities of isochones and stellar models}
\label{Table:3}
			\begin{tabular}{c|c|c|c|}
				\hline
				Isochrones & C14  & Rauch & PoWR \\
				\hline
				0.004 & 0.0017 & 0.002 & 0.0024  \\
			    0.008 & 0.008 & 0.002 & 0.008  \\
				0.020 & 0.017 & 0.02 & 0.017  \\
                0.050 & 0.03 & 0.02 & 0.017  \\
				\hline
			\end{tabular}
	\end{center}
\end{table}

Finally, we note that in this piece of work we use isochrones with $Z = 0.004$, 0.008, 0.02 and 0.05. Then, we use the C14 set of stellar models with $Z = 0.0017$, 0.008, 0.017 and 0.03, respectively; the PoWR models with $Z=0.0024$, and 0.008 are used for isochrones of $Z=0.004 $ and 0.008, and the ones of  $Z=0.017$ for the last two sets of isochrones: $Z = 0.02$ and 0.05.
The PN models were computed only for two values of the metallicity, and therefore we use the sub-solar set for $Z = 0.004$ and 0.008 isochrones, and the solar one for $Z = 0.02$ and 0.05.
We name our model sets with the isochrones metallicities: $Z = 0.004$, 0.008, 0.02 and 0.05, as summarised in Table~\ref{Table:3}.

The {\sc HR-pyPopStar} code has been tested against the original {\sc PopStar} for all ages, metallicities and IMFs, finding, as expected, similar SEDs, Johnson-Cousins and SDSS magnitudes, and number of ionising photons of different species, whenever using the same input ingredients: IMF, isochrones and atmosphere stellar libraries. This comparison is given in Appendix~B in electronic format. We will devote the rest of the paper to outline and discuss the results from {\sc HR-pyPopStar} genuinely obtained by the use of the high-resolution stellar libraries.

%-------------------------------------------------------------------
\section{Results}
\label{Section:Results}
%-------------------------------------------------------------------

Based on the ingredients mentioned in Section \ref{Sec:Model}, we have computed, using the {\sc HR-pyPopStar} code, the stellar and the total (nebular $+$ stellar) SEDs for four different IMFs, four values of the metallicity, and a total of 106 ages, from $0.1$\,Myr to $\sim 15$\,Gyr, except for the SAL IMF, which has $m_\mathrm{low}=1\,\mathrm{M}_{\sun}$, and therefore the oldest computed SSP is younger than 10\,Gyr.

\begin{table}
  \begin{center}
	\caption{Summary of the characteristics of models in this work.
}
\label{Table:4}
			\begin{tabular}{lll}
%\hline
%Characteristic & & Range \\
\hline
Wavelength range &$\lambda_{i}$--$\lambda_{f}$ & $91 \-- 24\,000$\,\AA\ \\
Wavelength step & $\delta \lambda $ & $0.1$\,\AA\ \\
Age (yr) & $\log{t}$ & $5.0 \-- 10.18$ \\
Abundance & $Z$ &  $0.004,\ 0.008,\ 0.02,\ 0.05$\\
Initial Mass Function & IMF & SAL, FER, KRO, CHA \\
\hline
			\end{tabular}
	\end{center}
\end{table}
Table~\ref{Table:4} summarizes the characteristics of these new {\sc HR-pyPopStar} models. As said, they cover a wide range in age and metallicity to reproduce the spectra from both young and old populations with different metal content, and extend over a wide wavelength range, with a wavelength step $\delta \lambda =0.1$\,\AA. This combination of characteristics (high wavelength-resolution, long wavelength range, computed for young and old stellar populations, ionising populations included), is not produced by any evolutionary synthesis model up to now, and therefore represent a significant improvement with respect to the state of the art. The final size of each set of models for a single metallicity and IMF is around 0.8\,GB, the complete set of  SSPs for one IMF is around 6\,GB.

\tiny
\begin{table*}
\caption{Magnitudes in broad band filters for different IMFs, abundances $Z$'s and ages. Columns 2 and 3 are the HST UV magnitudes, columns 4 to 12 are Johnson-Cousins-Glass system magnitudes in the Vega system, while columns 13 to 17 refer to the SDSS filters AB system. The whole table is available in electronic format.}
\label{Table:5}
\begin{tabular}{lcccccccccccccccc}
\hline
$\log t$ & UV$_{1}$ & UV$_{2}$ & U & B$_{1}$ & B$_{2}$ & V & I & R & J & H & K &  u & g & r & i & z   \\
(yr) & \multicolumn{14}{c}{(mag)} \\
\hline
5.00& -3.393& -0.887&  0.809&  1.764&  1.753&  1.969&  2.078&  2.229&  2.455&  2.501&  2.668&  1.930&  2.199&  2.614&  2.996&  3.255\\
5.48& -3.416& -0.921&  0.778&  1.736&  1.724&  1.942&  2.052&  2.206&  2.435&  2.483&  2.652&  1.898&  2.172&  2.587&  2.972&  3.231\\
5.70& -3.452& -0.954&  0.746&  1.707&  1.695&  1.914&  2.025&  2.181&  2.413&  2.463&  2.632&  1.866&  2.143&  2.560&  2.946&  3.207\\
5.85& -3.496& -0.986&  0.715&  1.679&  1.667&  1.888&  1.999&  2.155&  2.388&  2.438&  2.608&  1.834&  2.116&  2.534&  2.920&  3.182\\
6.00& -3.518& -1.035&  0.664&  1.630&  1.618&  1.838&  1.950&  2.107&  2.342&  2.395&  2.567&  1.783&  2.066&  2.484&  2.871&  3.133\\
6.10& -3.583& -1.109&  0.589&  1.561&  1.549&  1.771&  1.883&  2.041&  2.280&  2.335&  2.508&  1.707&  1.998&  2.417&  2.805&  3.069\\
\hline 
\end{tabular}
\end{table*}
\normalsize

For each parameter combination we give: (a) The SEDs of both stellar and stellar $+$ nebular continuum emission; (b) the magnitudes in the Johnson-Cousins and SDSS systems and (c) the number of ionising photons ($Q$) of \ion{H}{I}, \ion{He}{I}, \ion{O}{ii} and \ion{He}{II}. The last calculations follow the same prescriptions as in MOL09 (details in Appendix~A).

All this information, the computed SEDs, as well as the tables with the results, will be available on our web page:
\url{http://www.pypopstar.com}. The user will be able to download the complete set of models or a required subset (with all ages) just selecting by IMF and $Z$. Moreover, since the code is totally flexible, any other model, with different stellar isochrones or stellar libraries, IMFs or any input, might be computed upon request to the authors.

\subsection{Magnitudes}
\label{mag}
We have computed with {\sc HR-pyPopStar} the Johnson-Cousins-Glass and SDSS magnitudes for the whole set of models. For the Johnson-Cousins-Glass filter system, we have used the prescription from \citet{Girardi+2002}. In the case of SDDS filters in the AB system, we have also used the formulae from \citet{Girardi+2004}. 

The complete table of SSP magnitudes is given in electronic format for all the combinations of IMF ($\times$4), $Z$ ($\times$4) and age ($\times$106). Table~\ref{Table:5} shows, as an example, the corresponding magnitudes of Johnson-Cousins-Glass in Vega system, the SDSS filters in the AB system and two ultraviolet magnitudes computed using the HST filters for KRO IMF and solar metallicity, stellar spectra, and ages younger than $\log{t} < 6.1$. We give in column 1 the age, in logarithmic scale as $\log{t}$, the magnitudes in the 11 filters: UV1, UV2 (computed with the two HST UV filters), U, B1 (used for U-B), B2 (used for B-V), V, R, I, J, H, and K, in columns 2 to 12. Columns 13 to 17 list the corresponding SDSS broad band filter magnitudes, u, g, r, i, and z. 

\begin{figure}
    \includegraphics[width=0.45\textwidth]{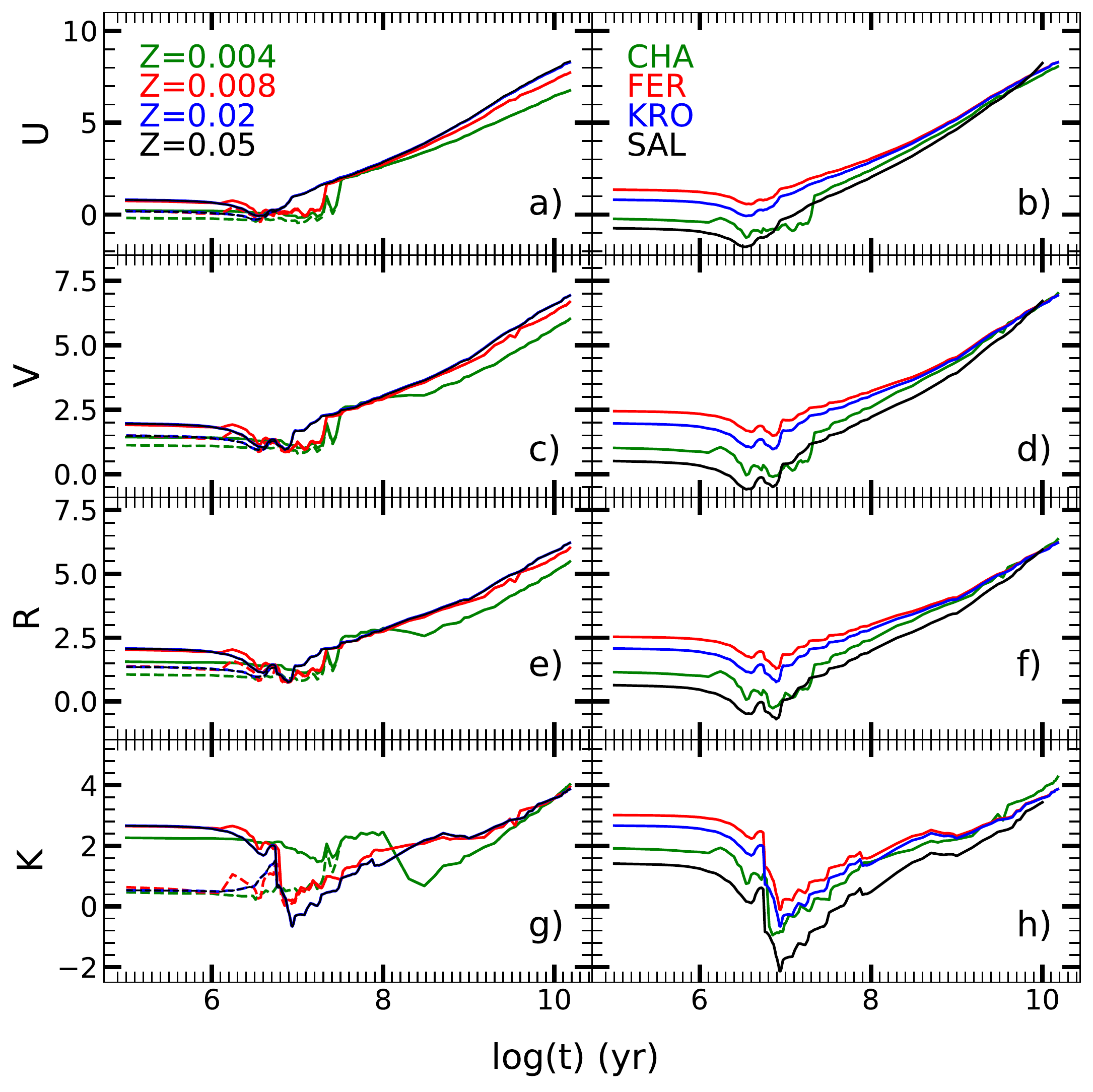}
    \caption{Evolution of Johnson-Cousins magnitudes as a function of age. Left panels a), c), e) and g) gives \textbf{U, V, R and K} magnitudes, computed with KRO IMF and the four metallicities labelled in plot a). Right panels b), d), f) and h) displays the same magnitudes for solar metallicity and different IMFs (CHA, FER, KRO, SAL) as labelled in plot b). Solid lines represent the magnitudes computed using the total spectra, stellar $+$ nebular, while dotted lines represent the magnitudes computed using the stellar SED.}
    \label{Fig:3}
\end{figure}

Figure~\ref{Fig:3} shows the evolution of magnitudes, U, B, V and K for the 4 values of metallicity $Z$ using the KRO IMF in left panels); and for the solar abundance and the four different IMF in right panels. The SDSS magnitudes, also computed, have a similar behaviour. Moreover, as we have verified in Appendix B that the broad band filter magnitudes are coherent with the ones obtained with {\sc PopStar}. 
\begin{figure}
\centering
\includegraphics[width=0.45\textwidth]{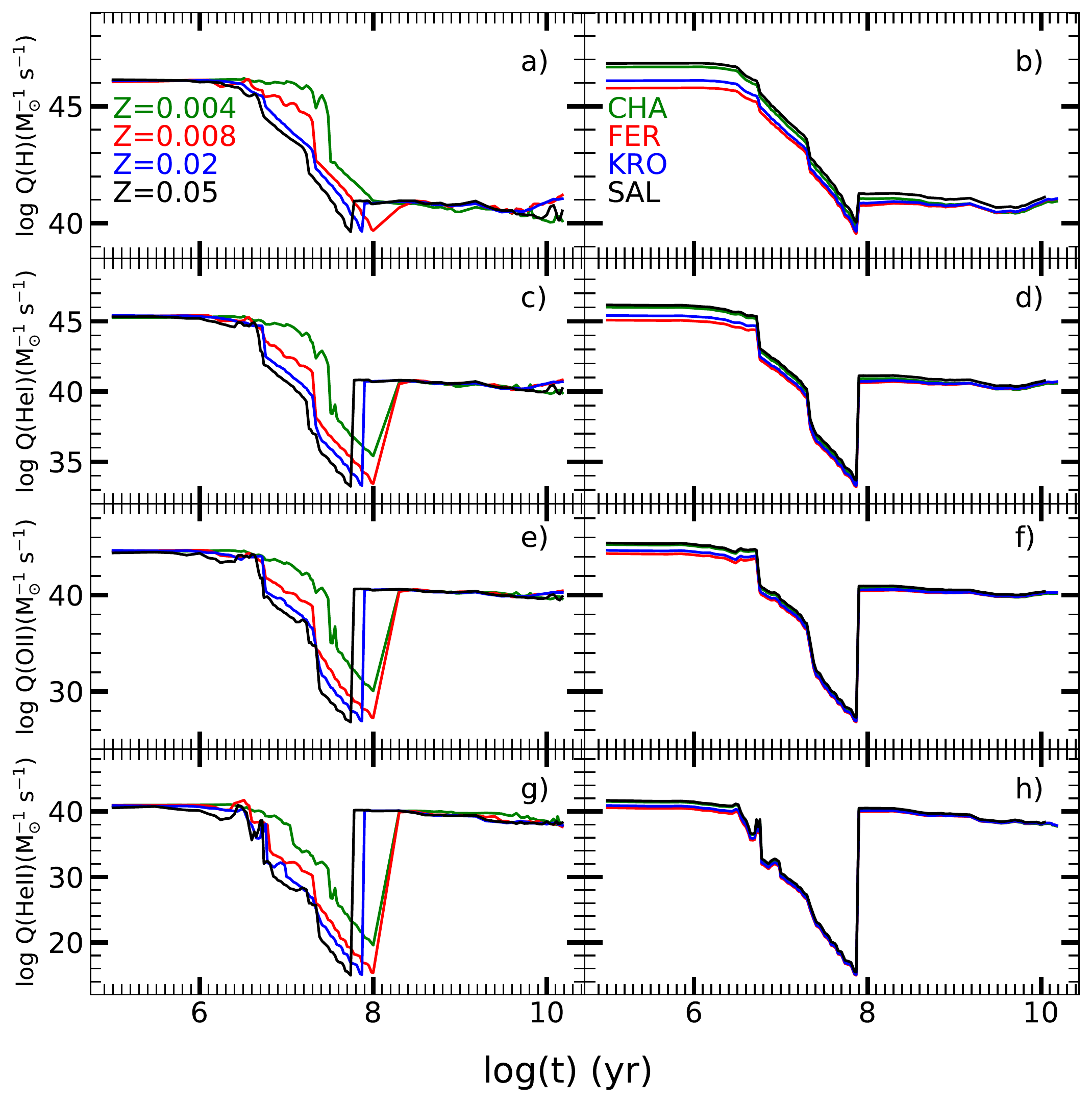}
\caption{Evolution of Q(\ion{H}{I}), Q(\ion{He}{I}), Q(\ion{O}{ii}) and Q(\ion{He}{II}) as a function of age: left panels show the results for the KRO IMF and different metallicities Zs as labelled in plot a); right panels plot the results for Z$_{\sun}$ and the four IMFs used in this work, as labelled in the plot b).}	
\label{Fig:4}
\end{figure}

\begin{figure}
\includegraphics[width=0.47\textwidth]{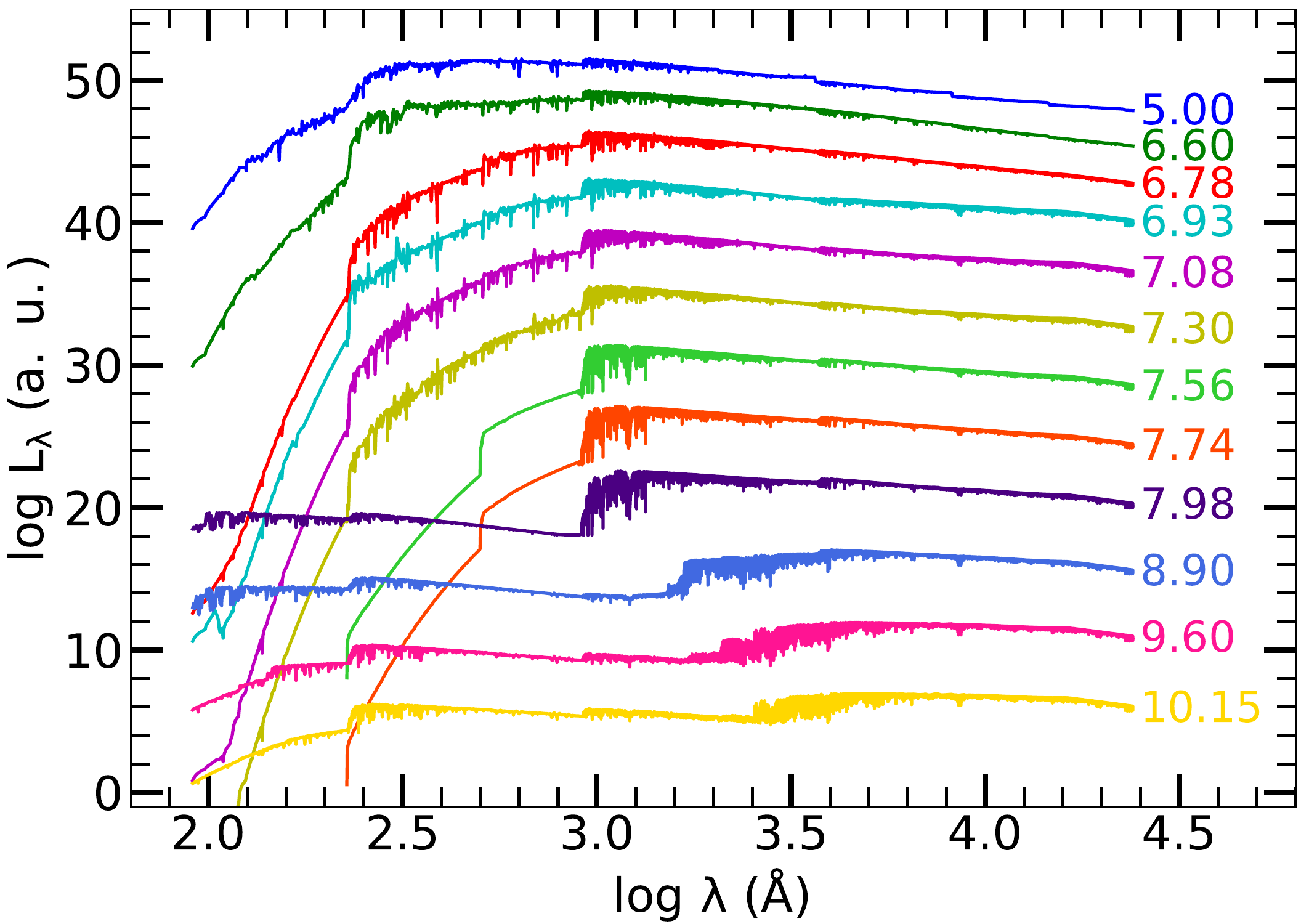}
\caption{Total, stellar $+$ nebular, SEDs of abundance $\rm Z=0.02$ and a Kroupa IMF for SSP of different ages, the units of the age labels are log(yr). The luminosity of each SED, represented in logarithmic scale, is shifted $-$2.0\,dex with respect to the previous one for the sake of clarity.}
\label{Fig:5}
\end{figure}

In the left panels of Figure~\ref{Fig:3}, we have over-plotted with dotted lines the evolution of magnitudes when computed with the stellar spectra only, without including the nebular component. Magnitudes computed with the total spectra are redder than the ones computed using only the stellar spectra. This is specially important in K-band, where there are almost 2 magnitudes of difference, since this band is usually assumed as the best one to compute the stellar masses through the ratio $M/L$. Although this is a well known effect, we insist in the need of taking into account the possible error produced in the estimates of the stellar masses obtained from this band when the nebular contribution is not included in the models. 

\subsection{The evolution of the ionising population}
\label{Subsec:qs}

\begin{table}
\begin{center}
\caption{Number of Ionising photons by solar mass of formed stars for \ion{H}{I}, \ion{He}{I}, \ion{O}{ii} and \ion{He}{II}, for different IMFs, abundances $Z$'s and ages. The whole table is available in electronic format.}
\label{Table:6}
\begin{tabular}{ccccccc}
\hline
 IMF & $Z$ & $\log{t}$ & $Q$(\ion{H}{I}) & $Q$(\ion{He}{I})  & $Q$(\ion{O}{ii}) & $Q$(\ion{He}{II})	  \\
 \hline
CHA &  0.004  &     5.00 & 39.976 &  39.800   & 39.559  &  38.403  \\
CHA &  0.004  &     5.48 & 39.908 &  39.718   & 39.470  &  38.177  \\
CHA &  0.004  &     5.70 & 39.929 &  39.769   & 39.549  &  38.745  \\
CHA &  0.004  &     5.85 & 40.209 &  39.987   & 39.691  &  37.834  \\
CHA &  0.004  &     6.00 & 40.132 &  39.983   & 39.775  &  39.076  \\
\hline
\end{tabular}
\end{center}
\end{table}

We have computed the number of ionising photons, $Q$, 
for the selected species as explained in Appendix A, giving the resulting $Q$s in Table~\ref{Table:6}, which is fully available in electronic format. We have included here a sample: for each IMF (column 1), abundance $Z$ (column 2) and age (column 3), in logarithmic scale, the logarithmic of the number of ionising photons, normalised to 1 solar mass of the SSP, of \ion{H}{I}, \ion{He}{I}, \ion{O}{ii} and \ion{He}{II}, labelled as $Q$(\ion{H}{I}), $Q$(\ion{He}{I}), $Q$(\ion{O}{ii}) and $Q$(\ion{He}{II}) in columns 4, 5, 6 and 7 respectively.

Figure~\ref{Fig:4} shows the time-evolution of $Q$(\ion{H}{I}), $Q$(\ion{He}{I}), $Q$(\ion{O}{ii}) and $Q$(\ion{He}{II}) for the 4 $Z$'s of KRO IMF in left panels. Although there exist differences in the behaviour of each $Q_{i}$ when comparing results with different $Z$'s, they all share a similar overall shape: the number of ionising photon decreases after the first 10-20\,Myr, and then increases at the time of the appearance of the first PNe (at $\sim$100\,Myr, this age depends on metallicity).

\begin{figure*}
\includegraphics[width=0.495\textwidth]{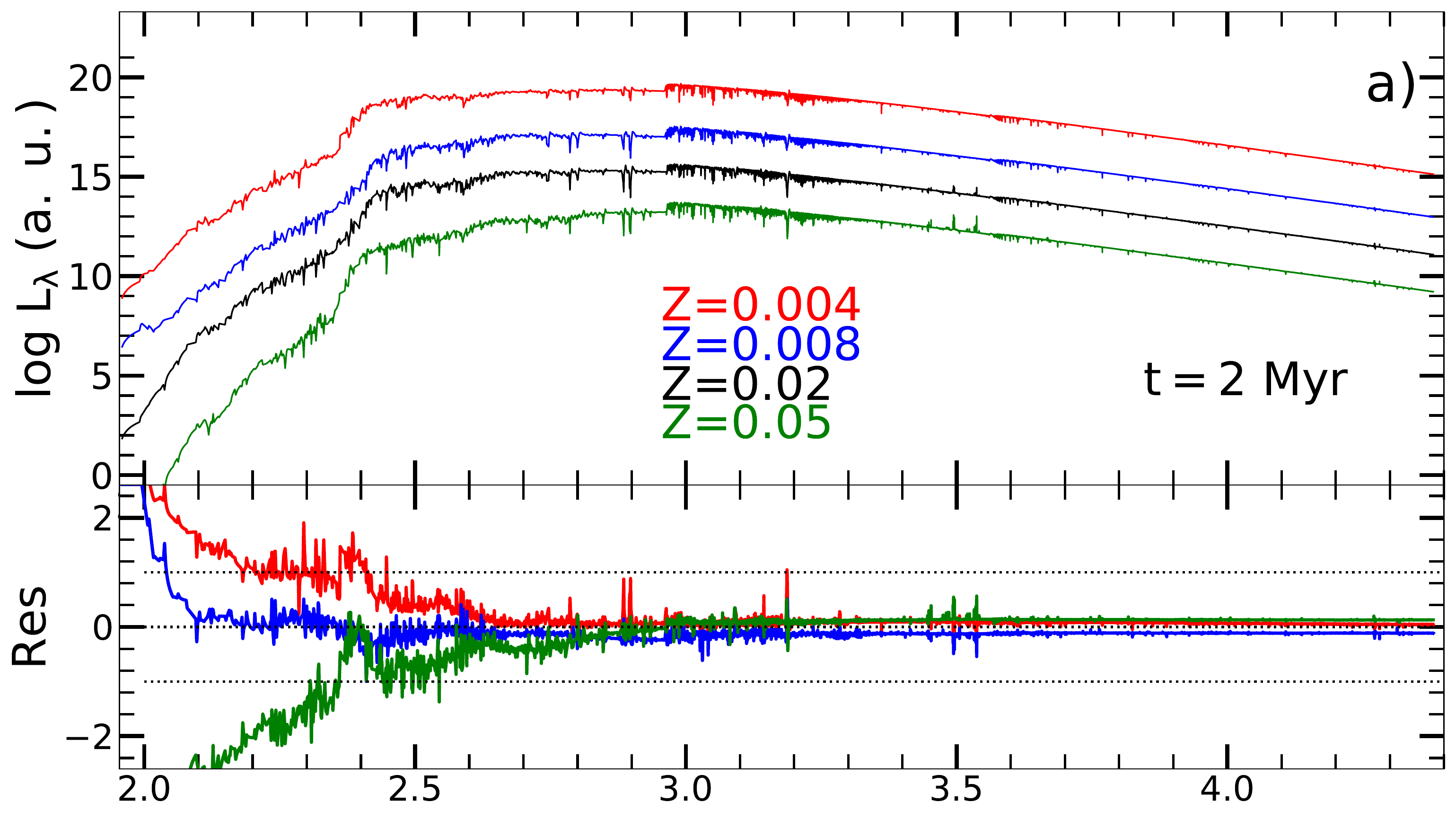}
\includegraphics[width=0.495\textwidth]{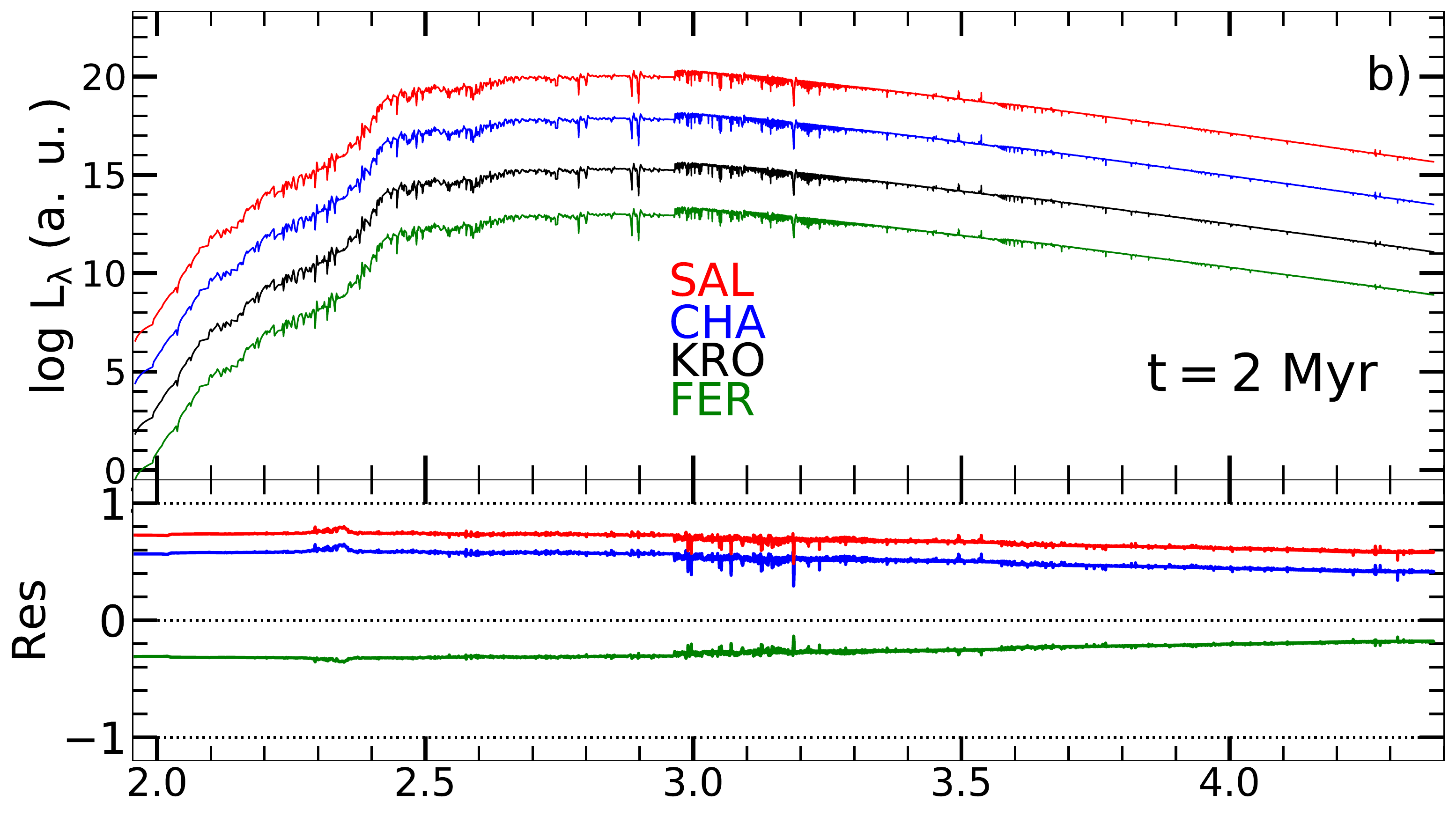}
\includegraphics[width=0.495\textwidth]{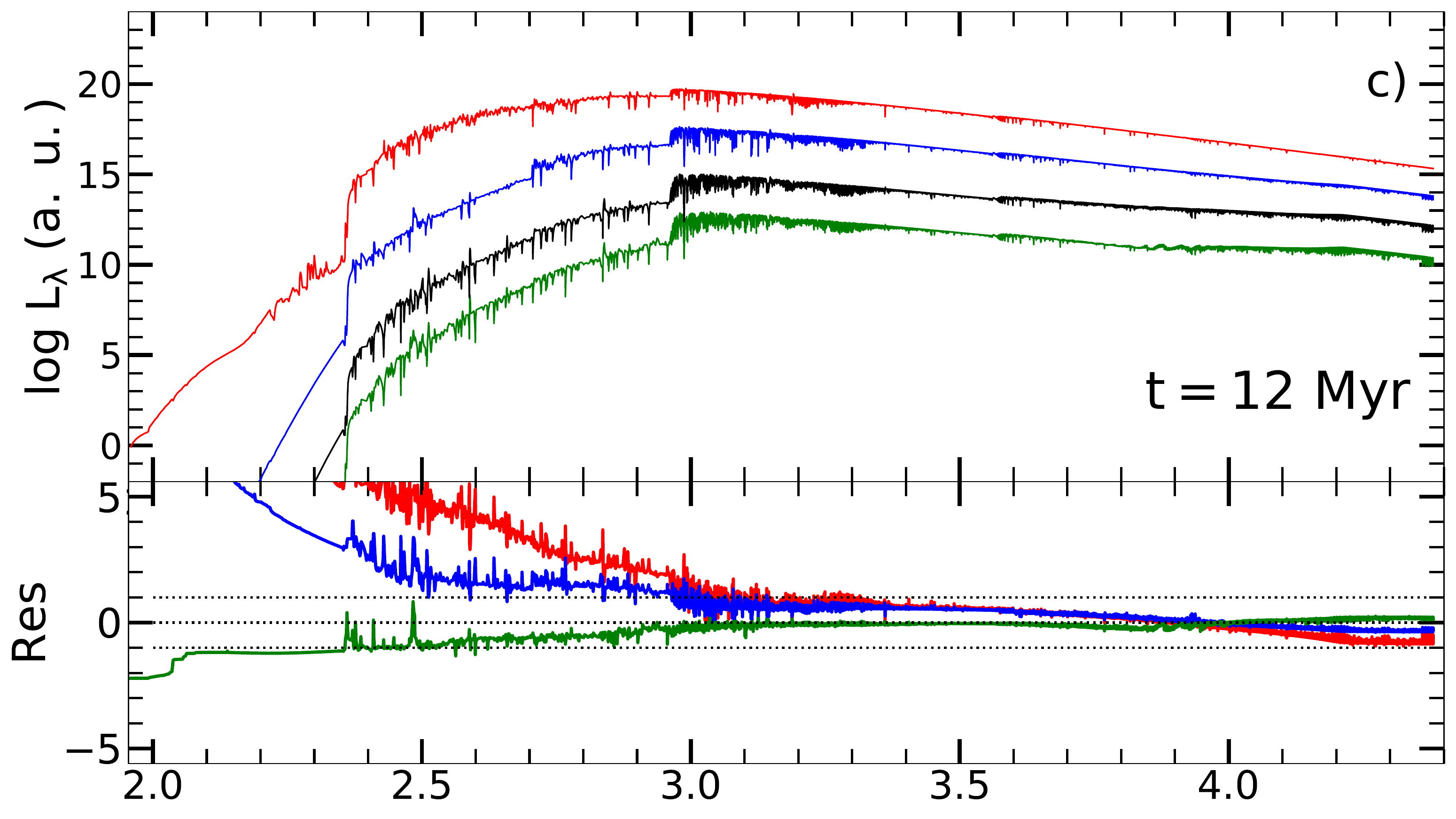}
\includegraphics[width=0.495\textwidth]{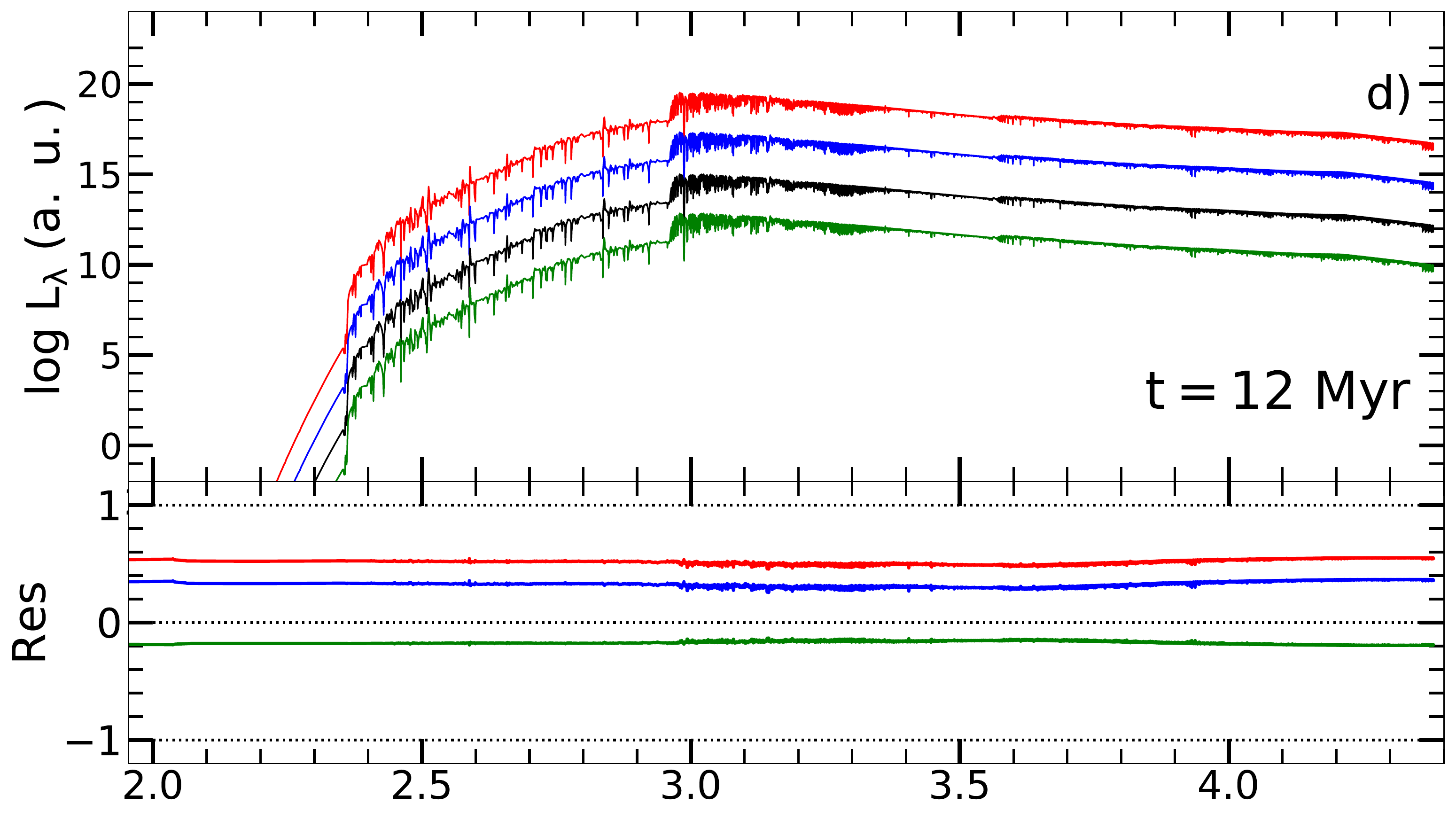}
\includegraphics[width=0.495\textwidth]{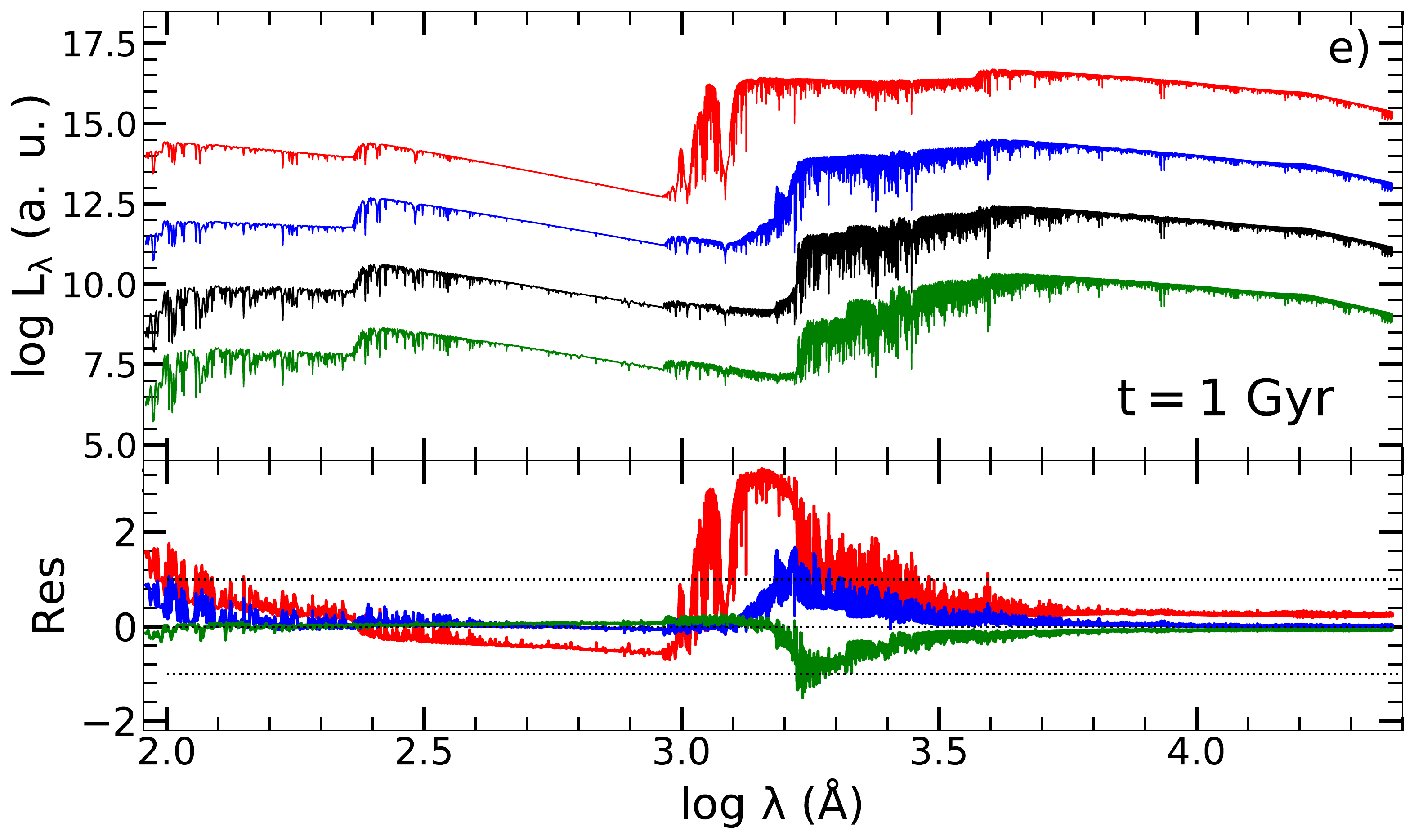}
\includegraphics[width=0.495\textwidth]{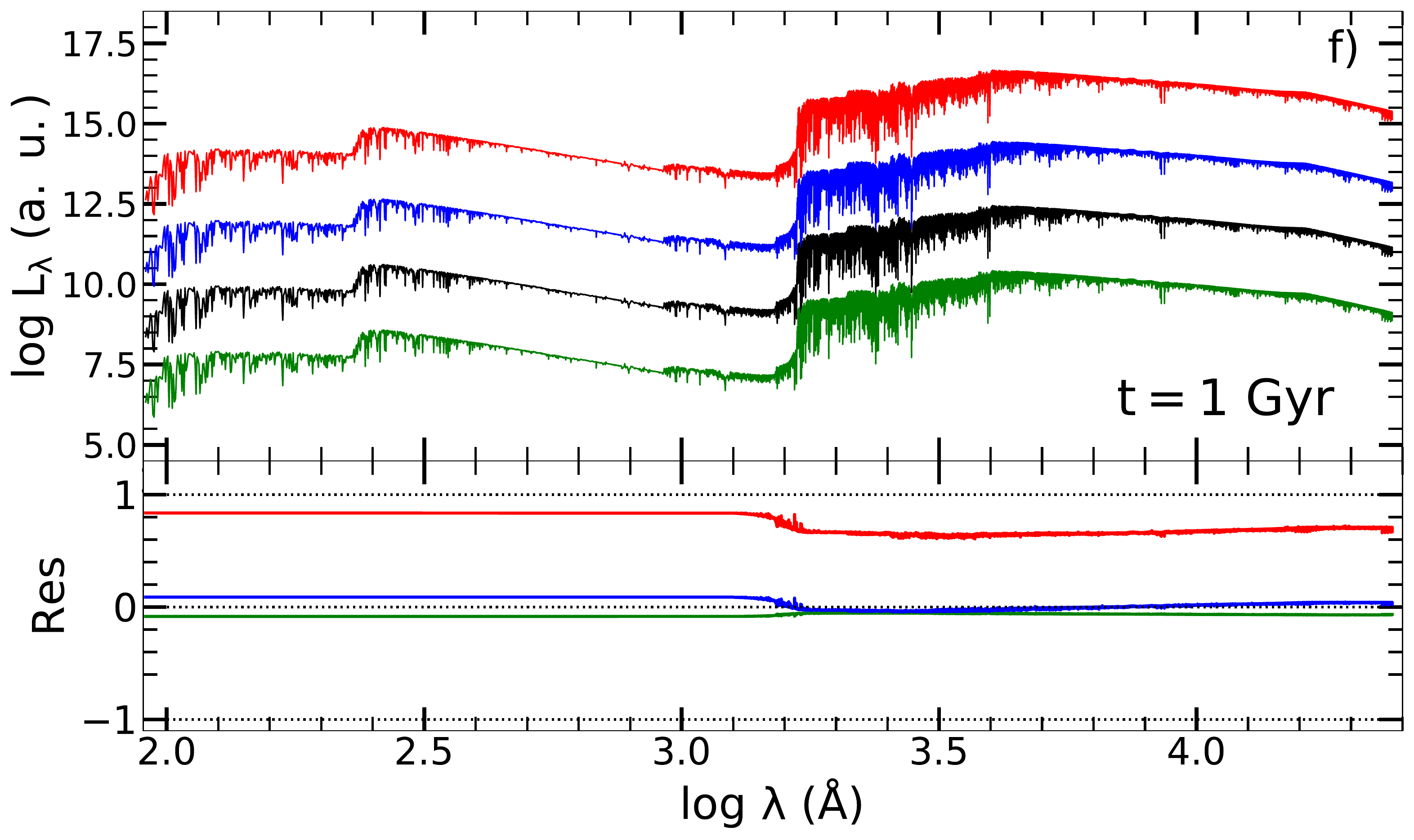}
\caption{Total (stellar + nebular) SEDs of one age, a) and b) $t=2.00$\,Myr, c) and d) $t=12$\,Myr and e) and f) $t=1$\,Gyr. Left panels show models for a KRO IMF and different metallicities as labelled in a); Right panels show models at solar metallicity and different IMFs, as labelled in b). In each panel the residuals, as $\textrm{Res} = \log{L} - \log{L_\mathrm{ref}}$ are given, $L_\mathrm{ref}$ being the solar KRO model. The value 0 is marked with dashed lines, while the two dotted lines indicate the $\pm 1$ values of residuals.}
\label{Fig:6}
\end{figure*}

As the vast majority of the UV photons are produced by the most massive stars, the ionisation rate depends, obviously, on the slope of the IMF in the upper range of stellar masses. This effect is shown on the right panels, where the ionising photons time-evolution is plotted for the 4 IMFs and at the same solar metallicity. The differences due to the IMF choice appear only for the youngest ages, $\log{t} < 6.6$,  thus modifying the ionising characteristics of stellar clusters. After this age, differences are insignificant, with only slight variations for $Q$(\ion{H}{I}) during the PNe phases

The ionisation level reached in this second phase of ionisation by PNe is not the same for all elements. $Q$(\ion{H}{I}) in the PNe stages is 5 to 6 orders of magnitude below the values produced by the youngest clusters of the same mass, whereas the rates of $Q$(\ion{He}{I}) and $Q$(\ion{O}{ii}) are only reduced by 4 - 5 orders of magnitude. In the case of  $Q$(\ion{He}{II}), the values produced by young and by PNe populations are very similar, without barely any dependence on IMF, being around $10^{40}$\,$\mathrm{s}^{-1}$ at both ages. This opens an exciting interpretation of observations with no WR features in the optical spectrum, but where the number of \ion{He}{ii} photons is still quite large, as derived from \ion{He}{ii} lines $\lambda\lambda$4686\,\AA\ in the optical range or $\lambda\lambda$1640\,\AA\ in the UV one. Could these He$^+$ photons come from PNe and not from a young cluster? This would imply that we would be detecting simultaneously emission lines from the two populations (if the continuum levels were similar) or we would be detecting only the older population while thinking that is the young one? We will explore the viability of this interesting scenario in a further work. 

\subsection{Resulting SSPs SED}
\label{subsec:SED}

Figure~\ref{Fig:5} illustrates the total (stellar $+$ nebular) SEDs for a selected number of ages, representing different stages of the SSP across the whole age range, for metallicity $\mathrm{Z}=0.02$. The luminosity of each SED, represented in logarithmic scale, is shifted $-$2.0\,dex with respect to the previous one for the sake of clarity. In next plots the default IMF used is KRO, unless we explicitly indicate otherwise. 

We notice that the ionising part of the spectra gets smooth for $\log{t}$ between 7.5 and 7.98 ($t \sim 40 - 100$\,Myr). At these ages, the stars contributing to the ionising spectrum are NS stars, whose spectra are modelled using the old low-resolution BaSeL library because models from C14 do not reach such low wavelength range. As explained above, the low spectral resolution BaSeL library have been used to extend C14 spectra in the UV from 91 to 900\,\AA, for stars with $T_\mathrm{eff}$ between 10\,000\,K and 24\,000\,K.

The nebular continuum increases the flux at long wavelengths, producing a reddening in the magnitudes for the youngest ages at all bands. To this nebular component, it would be necessary to add, to calculate the magnitudes in the broad band filters, the contribution of the emission lines originated in the surrounding gas. Although the use of high wavelength-resolution libraries should not change these results, we will investigate in the near future the contribution to the broad band filters magnitudes from the new OB and WR spectra - in particular the WR bumps, included in our SSPs SEDs.

Figure~\ref{Fig:6} shows the SEDs for ages 2\,Myr, 12\,Myr, and 1\,Gyr on the top, middle and bottom panels, respectively. The KRO IMF is used on the left panels, with the four metallicities, $Z= 0.004$, 0.008, 0.02 and 0.05 represented in different colours, as labelled in plot (a), whereas the SEDs plotted on the right panels correspond to $Z$= 0.02 with different colours representing different IMFs as labelled in the plot (b).
The total flux is represented in arbitrary units and in logarithmic scale, with a shift of 2\,dex between consecutive SEDs for the sake of clarity 
The bottom part of each panel displays the residuals, computed as $\mathrm{Res} = \log{L} - \log{L_\mathrm{ref}}$, $L_\mathrm{ref}$ being the KRO solar spectrum for each age in both panels.
In the case of metallicity variations, the stronger differences are in the UV and ionising flux ($\log{\lambda} < 3.5$; $\lambda < 3000$\,\AA), which can reach relative variations larger than 1 order of magnitude. In the case of IMF, most of the variation is in the continuum level except at young ages when there is also variation in the continuum shape. We note that our KRO models have a high mass exponent of $-2.7$, whereas the exponent for SAL and CHA is $-2.35$ and \textbf{$-2.3$} respectively. At ages near 1\,Gyr the differences among the SSPs using KRO and CHA IMF almost disappear since for low mass ranges both IMF are almost equivalent. Finally, we note that in both cases (metallicity and IMF variations) there are variations in different spectral lines, hence the line profiles of high resolution models could be used for metallicity and IMF studies as shown by other authors \cite[e.g.][among others]{MartinNavarro2015}.

\section{Discussion: the power of the high wavelength-resolution}
\label{Sec:Discussion}

\subsection{Comparison with other theoretical models}
\label{subsec:Other_models}

\begin{figure}
\hspace{-0.35cm}
\includegraphics[width=0.47\textwidth]{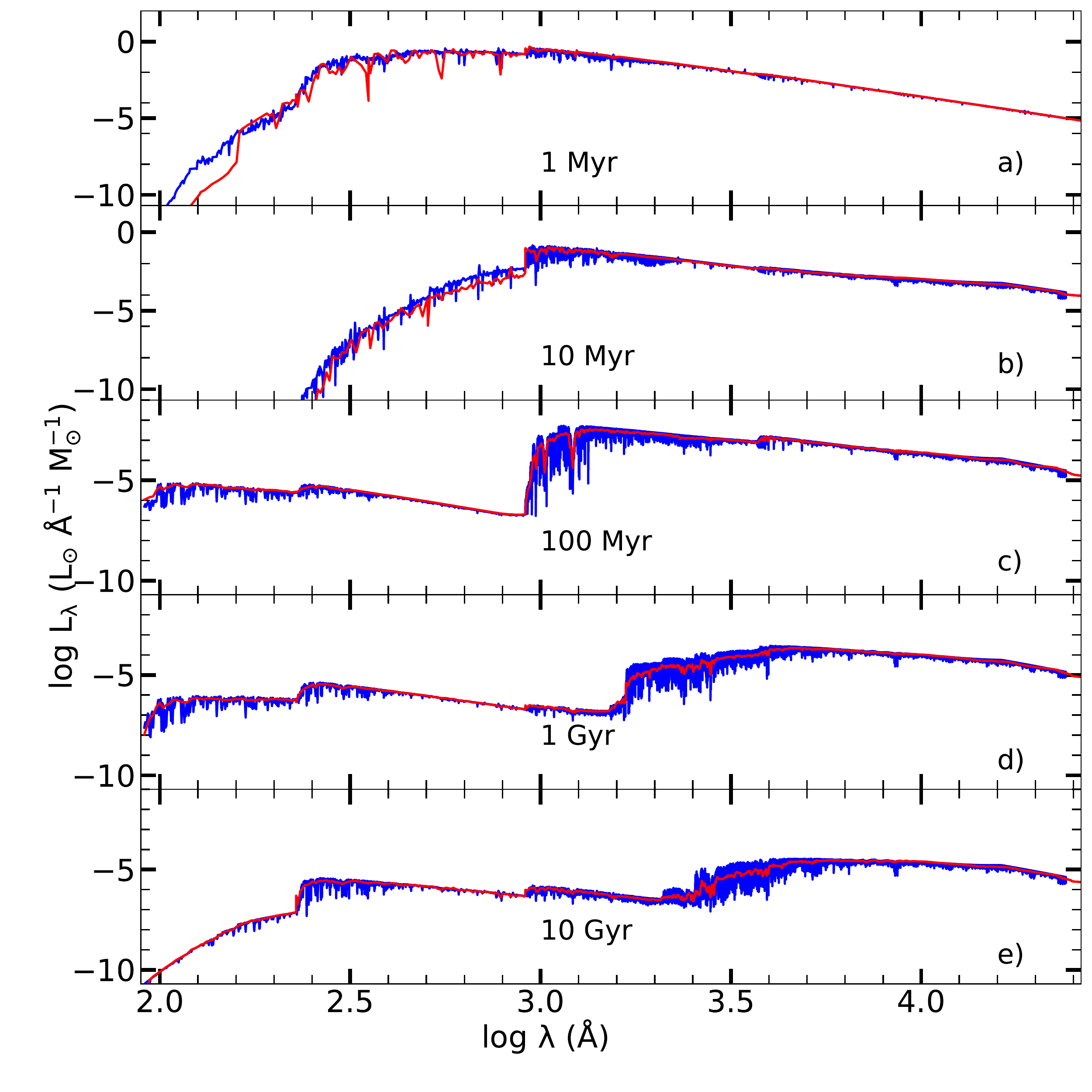}
\caption{Comparison of {\sc PopStar} and {\sc HR-pyPopStar} models, as labelled, for solar metallicity and five ages: a) 1\,Myr; b) 10\,Myr; c) 100\,Myr; d) 1\,Gyr and e) 10\,Gyr.}
\label{Fig:7}
\end{figure}

Since we present here the spectra for high wavelength-resolution, we show the differences between the {\sc  PopStar} and {\sc HR-pyPopStar} in Figure~\ref{Fig:7}. There, we have the SEDs for SSPs of five ages: 1, 10, and  100\,Myr, and 1 and 10\,Gyr.
As can be observed in this figure, there are no great differences in the shape of the both models SEDs, as expected. The new models have a slightly higher luminosity in the ionising region of the spectra, due to a small increase in hardness of the new O and B stars models used in this work. However, it is very clear that the HR in the spectra of {\sc pyPopStar}, can allow us to see well defined absorption lines that do not appear, or appear smoothed, in the {\sc PopStar} SEDs. We have compared with more details these new models with the old ones in Appendix B (see Supplementary Information). 

In the following, we present the comparison of the new HR models with existing models with high or intermediate wavelength-resolution and we refer to MOL09 for a comparison with other low resolution models.

\begin{figure*}
\hspace{-0.5cm}
\includegraphics[width=0.95\textwidth]{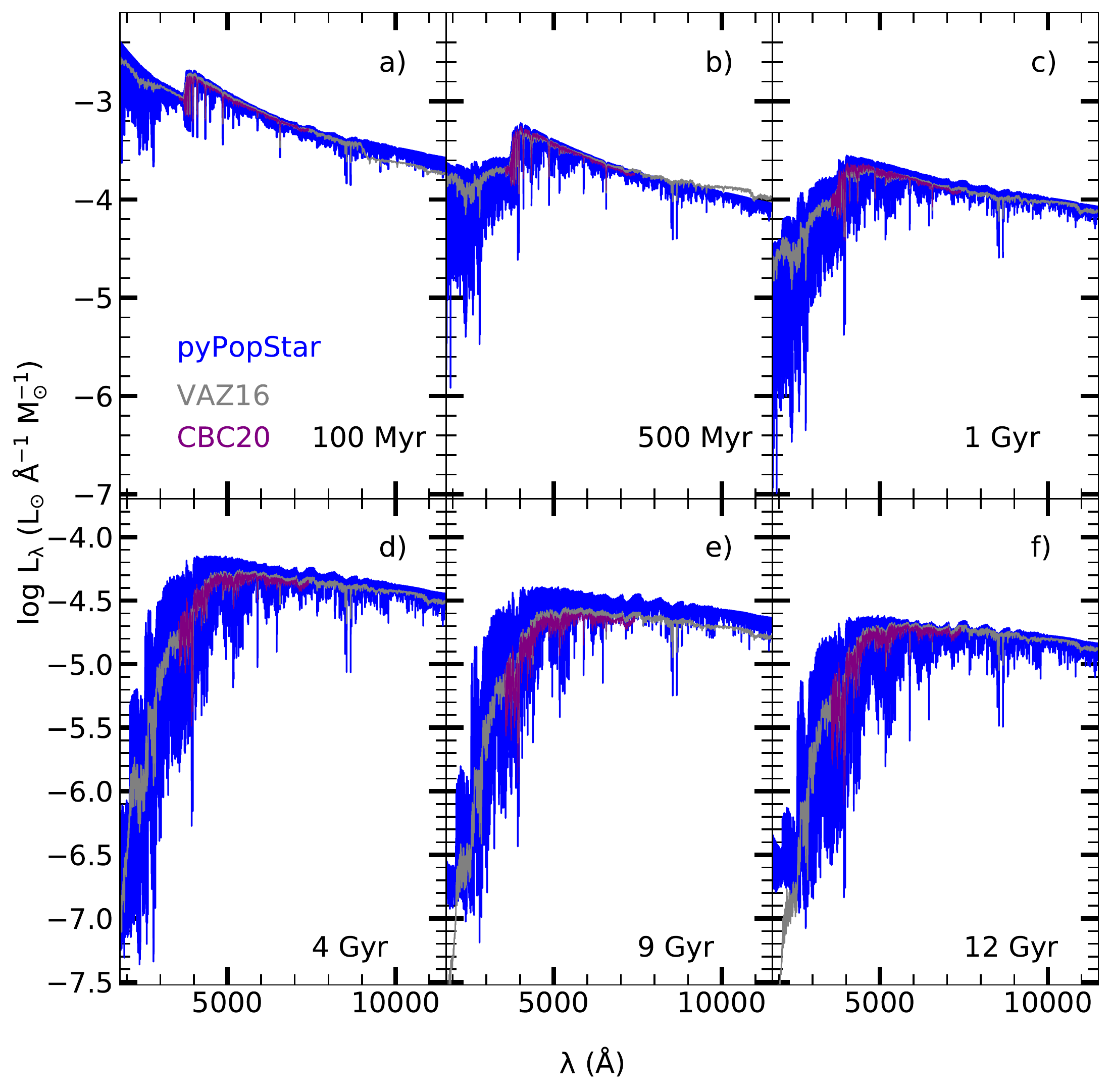}
\caption{Comparison of our SEDs using the CHA IMF with VAZ16 and CBC20 intermediate wavelength-resolution models, as labelled.}
\label{Fig:8}
\end{figure*}

\begin{figure*}
\hspace{-0.5cm}
\includegraphics[width=0.95\textwidth]{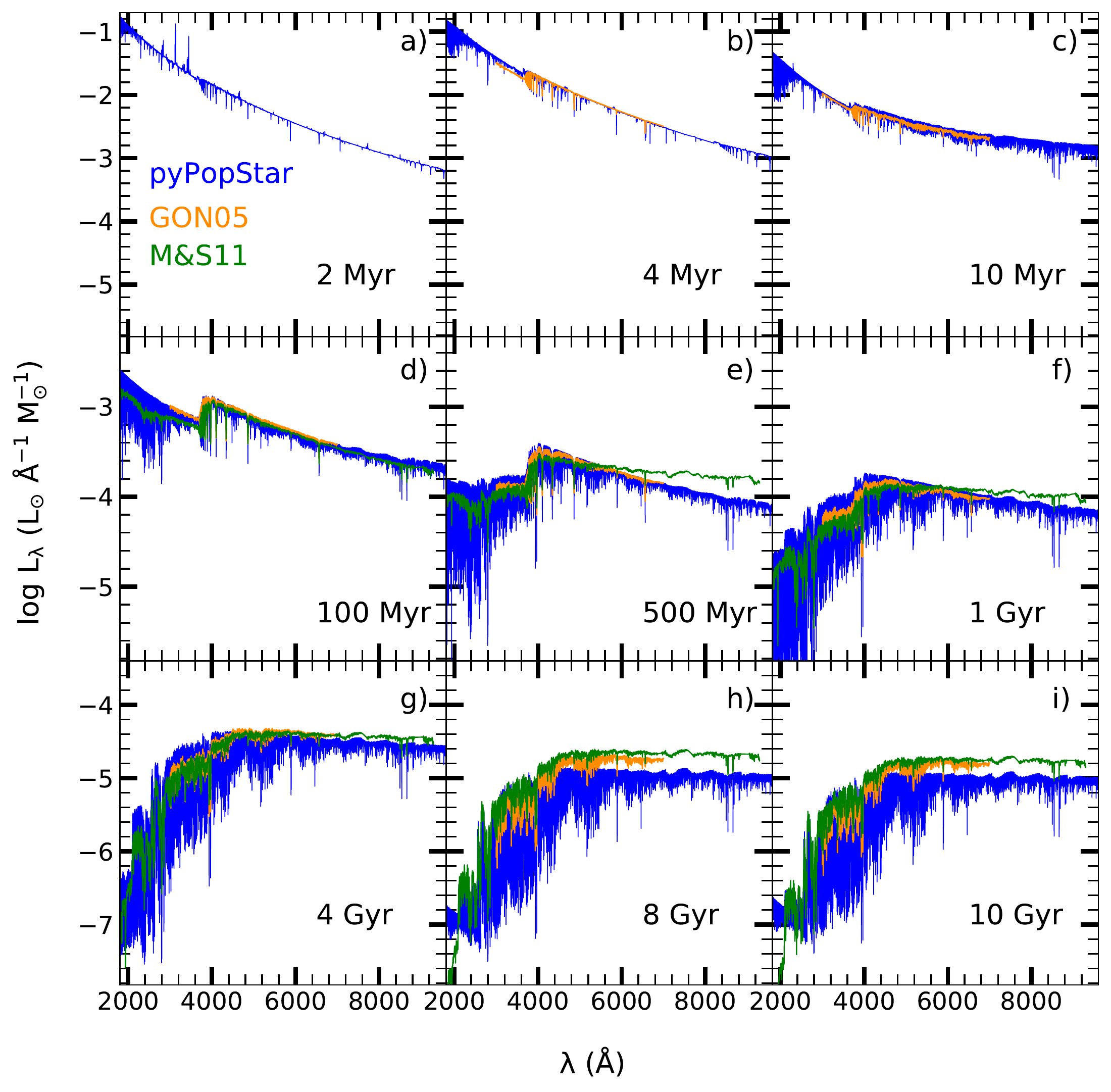}
\caption{Comparison of our SEDs using the SAL IMF with GON05 and M\&S11 high wavelength-resolution models, as labelled.}
\label{Fig:9}
\end{figure*}

In Figure~\ref{Fig:8} and Figure~\ref{Fig:9}, we show the comparison among {\sc HR-pyPopStar} SEDs and other intermediate and high resolution models. We summarize the models used in this comparison in Table~\ref{Table:7}, where we give for each one, the name in column 1, the wavelength step in 5000\,\AA, $\delta\lambda_{5000}$, in column 2, the isochrones and stellar libraries used in each model in columns 3 and 4, the colour we use in next figures in column 5, and the reference of each one in column 6.

In Figure~\ref{Fig:8}, we compare our models with the intermediate resolution spectra from VAZ16 and CBC20 --\textbf{grey and purple} lines, respectively, both using the CHA IMF for 6 ages: 100 and 500\,Myr and 1, 4, 9 and 12\,Gyr. Although all of them are very similar in the optical and NIR ranges (at the ages in which all of them exist), while at old ages VAZ16 models are systematically below {\sc HR-pyPopstar} for wavelengths shorter than 2300\,\AA (not shown in the plot), because these models do not include post-AGB/PNe spectra. Regarding CBC20 models, also very similar to the VAZ16 ones and computed to simulate the same wavelength-resolution, they also show smaller luminosities than ours at the ionising region, caused by the lack of spectra of ionising spectra. This is the reason for which neither VAZ16 nor CBC20 have models for younger ages $t < 70$\,Myr, and in the case of VAZ16, even the own authors claim that their models are sure only for $t > 500$\,Myr.

Then, in Figure~\ref{Fig:9}, we compare our results with the high wavelength-resolution models from GON05 and M\&S11, computed with a SAL IMF, drawn as orange and green lines, respectively, for 9 ages as labelled. It is necessary to say that the low and upper mass limits of the IMF are different for the three sets:  $m_\mathrm{low}=0.1\,\mathrm{M}_{\sun}$ for GON05 and M\&S11, while is 1\,$\mathrm{M}_{\sun}$ in our models; and 
$m_\mathrm{upper}=120\,\mathrm{M}_{\sun}$ for GON05 and ours, but it is 100\,$\mathrm{M}_{\sun}$ for M\&S11. Therefore, we have divided our spectra by a factor 2.56 to take into account the different normalization compared with the two others. In this case, we have plotted the SEDs only for the visible wavelength, between 3000\,\AA\ and 9000\,\AA\ for which all models were computed, and in order to see  more clearly the differences among them. Our  models are very similar to the others at young and intermediate ages up to 100\,Myr, as shown. In the range between 500\,Myr to 1\,Gyr, our models are still quite similar to GON05, but they differs from M\&S11, which shows flatter spectra, specially higher at the near-IR band; it is due probably due to a different treatment of the AGB stars. The flux level of our models is lower than the ones from GON05 and M\&S11 for ages larger than 1\,Gyr as expected. This is an effect due to our SAL IMF, with a lower mass limit of 1\,$\mathrm{M}_{\sun}$, implying the no existence of very low --cool-- mass stars.

\begin{table*}
\caption{Characteristics of models to compare with {\sc HR-pyPopStar}.
}
\label{Table:7}
    \begin{tabular}{lccccr}
    \hline
  Model & $\delta\lambda_{5000}$ & Isochrones & Stellar library & Colour & Ref.\\
             & [\AA]                 &                    &                       &             & \\
  \hline
GON05 & 0.30 & \citet{Bertelli+1994,Girardi+2000,Girardi+2002} & \citet{mar05} & yellow & 1 \\
M\&S11 & 0.25 & \citet{Cassisi_Castellani_Castellani1997} & \citet{Gustafsson+08} & green & 2\\
     & & \citet{Girardi+2000} & & & \\
VAZ16 & 0.90 & \citet{Girardi+2000} & \citet{Gregg+2006} & black & 3 \\
       &     &                      & \citet{vazdekis2012} (MILES) &  &  \\
CBC20 &  0.90 & \citet{Bressan+2012,Chen+2015} & $\rm SynCoMiL^{a}$ & cyan & 4 \\
%M20 & \citet{Cassisi_Castellani_Castellani1997} & \citet{Yan+19} & brown & 5 \\
%     & \citet{Girardi+2000} & & & \\
%& & & {Rauch2003} & & \\
& & & \citet{Smith_Norris_Crowther2002,Rauch2003} & \\ 
This work & 0.10 &  \citet{Fagotto+1994a,Fagotto+1994b,Bertelli+1994} & \citet{Coelho2014,Rauch2003}  & blue & \\
 &  &   & \citet{Hainich+2019} &  & \\
\hline
\end{tabular}
\footnotesize{References. 1: \citet{Gonzalez+2005}; 2: \citet{Maraston_Stromback2011}; 3: \citet{vaz16}; 4: \citet{Coelho+2020}.

(a) SynCoMiL is a stellar library that mimics MILES library using synthetic spectra computed with the same ingredients as in C14. }
\end{table*}

\subsection{Spectral Absorption Lines}
\label{Subsec:Lines} 

\begin{table}
\begin{center}
\caption{Definition of some absorption spectral lines indices.}
\label{Table:8}
\begin{tabular}{cccccc}
\hline
Line name & $\rm \lambda_c$ & line bandpass & Reference\\
    & (\AA) &   (\AA)   \\%& (\AA) & (\AA)  \\
 BL1302 & 1302.0 & 1292.0 -- 1312.0 & F92\\% & 1270.0 -- 1290.0 & 1345.0 -- 1365.0 & F92 \\ 
 \ion{Si}{iv} 1397 & 1397.00 & 1387.0 -- 1407.0& VID-GAR17\\% & 1345.0 -- 1365.0 & 1475.0 -- 1495.0 & VID-GAR17 \\ 
  BL1425 & 1425.0 & 1415.0 -- 1435.0 & F92\\%1345.0 -- 1365.0 & 1475.0 -- 1495.0 & F92 \\ 
 \ion{Fe}{i} 1453  & 1453.00 & 1440.0 -- 1466.0 & VID-GAR17\\%1345.0 -- 1365.0 & 1475.0 -- 1495.0 & VID-GAR17 \\
 \ion{C}{iv} 1540  & 1540.00 & 1530.0 -- 1550.0  & VID-GAR17\\%1500.0 -- 1520.0 & 1577.0 -- 1597.0  & VID-GAR17 \\ %CIV in absorption\\ 
 \ion{C}{iv} 1548  & 1548.0 & 1540.0 -- 1560.0 & VID-GAR17\\% 1500.0 -- 1520.0  & 1577.0 -- 1597.0  & VID-GAR17 \\
 \ion{C}{iv} 1560  & 1560.00 & 1550.0 -- 1570.0  & VID-GAR17\\% 1500.0 -- 1520.0  & 1577.0 -- 1597.0  & VID-GAR17 \\ % CIV in emision\\
BL1617 & 1617.0 & 1604.0 -- 1630.0 & VID-GAR17\\%1577.0 -- 1597.0 & 1685.0 -- 1705.0 & F92 \\ 
BL1664 & 1664.0 & 1651.0 -- 1677.0 & VID-GAR17\\%1577.0 -- 1597.0 & 1685.0 -- 1705.0 & F92 \\ 
 \ion{Fe}{ii} 2402 & 2402.00 & 2382.0 -- 2422.0 & VID-GAR17\\%2285.0 -- 2325.0 & 2432.0 -- 2458.0 & VID-GAR17 \\
 \ion{Fe}{ii} 2609 & 2609.00 & 2596.0 -- 2622.0 & VID-GAR17\\%2562.0 -- 2588.0 & 2647.0 -- 2673.0 & VID-GAR17 \\ 
 \ion{Mg}{ii} 2800 & 2800.00 & 2784.0 -- 2814.0 & VID-GAR17\\%2762.0 -- 2782.0  & 2818.0 -- 2838.0 & VID-GAR17 \\
 \ion{Mg}{i} 2852  & 2852.00 & 2839.0 -- 2865.0 & VID-GAR17\\%2818.0 -- 2838.0  & 2906.0 -- 2936.0 & VID-GAR17 \\ 
 \ion{Fe}{i} 3000  & 3000.00 & 2965.0 -- 3025.0 & VID-GAR17\\%2906.0 -- 2936.0 & 3031.0 -- 3051.0 & VID-GAR17 \\ 
 \ion{Ca}{i} 3934  & 3933.66 & 3926.6 -- 3940.6 & ROD-MER20\\%3906.0 -- 3912.0  & 3948.0 -- 3954.0 & ROD-MER20 \\ % Ca I H line
 \ion{He}{i} 4026  & 4026.00 & 4020.0 -- 4031.0 & GON05\\%4012.0 -- 4020.0 & 4158.0 -- 4169.0 & GON05 \\
 \ion{Fe}{i} 4046  & 4045.81 & 4044.8 -- 4046.8  & ROD-MER20\\%4036.0 -- 4041.0  & 4052.0 -- 4057.0 & ROD-MER20 \\ 
  $H_\delta$       & 4101.73 & 4092.0 -- 4112.0 & GON05\\%4012.0 -- 4020.0 & 4158.0 -- 4169.0 & GON05 \\ 
  $H_\gamma$       & 4340.47 & 4330.0 -- 4350.0 & GON05\\%4262.0 -- 4270.0 & 4445.0 -- 4453.0 & GON05 \\ 
 %\ion{He}{i} 4471  & 4471.00 & 4464 --4478.0 &  &  \\
 \ion{Mg}{i} 4480  & 4481.13 & 4480.8 -- 4481.9 & ROD-MER20\\%4473.0 -- 4478.0 & 4484.0 -- 4489.0 & ROD-MER20 \\
  $H_\beta$        & 4861.35 & 4852.0 -- 4872.0 & GON05\\%4770.0 -- 4782.0 & 4942.0 -- 4954.0 & GON05 \\ 
 \ion{He}{i} 5876  & 5876.00 & 5871.0 -- 5880.0 & GON05\\%5835.0 -- 5845.0 & 5904.0 -- 5912.0 & GON05 \\
  $H_\alpha$       & 6562.79 & 6553.0 -- 6573.0 & GON05\\%6506.0 -- 6514.0 & 6612.0 -- 6620.0 & GON05 \\
% Pa1               & 8467.27 & 8461.0 -- 8474.0 &  & 8474.0 -- 8484.0  & GV20 \\
 Pa14               & 8598.40 & 8577.0 -- 8619.0 & GV20\\%8563.0 -- 8577.0 & 8619.0 -- 8642.0 & GV20 \\
 Pa13               & 8750.47 & 8730.0 -- 8772.0 & GV20\\%8700.0 -- 8725.0 & 8776.0 -- 8792.0  & GV20 \\
 CaT1              & 8498.03 & 8482.0 -- 8512.0 & GV20\\%8450.0 -- 8460.0 & 8565.5 -- 8575.0 & GV20 \\
 CaT2              & 8542.09 & 8531.0 -- 8554.0 & GV20\\%8450.0 -- 8460.0 &  8565.5 -- 8575.0 & GV20 \\
 CaT3              & 8662.14 & 8650.0 -- 8673.0 & GV20\\%8619.5 -- 8642.5 & 8700.5 -- 8710.0 & GV20 \\
 \ion{Mg}{i} 8807  & 8807.00 & 8802.5 -- 8811.0 & CEN09\\%8781.0 -- 8789.0 & 8831.0 -- 8835.5 & CEN09 \\
 \hline
\end{tabular}
\end{center}
\footnotesize{References. F92: \citet{Fanelli+1992}, CEN01: \citet{cen01}, GON05: \citet{Gonzalez+2005}, CEN09: \citet{cen09}, VID-GAR17: \citet{vidal-garcia2017}, ROD-MER20: \citet{rodriguez-merino20}, and GV20: \citet{gv20}.}
\end{table}

\begin{figure*}
\includegraphics[width=0.3\textwidth,angle=0]{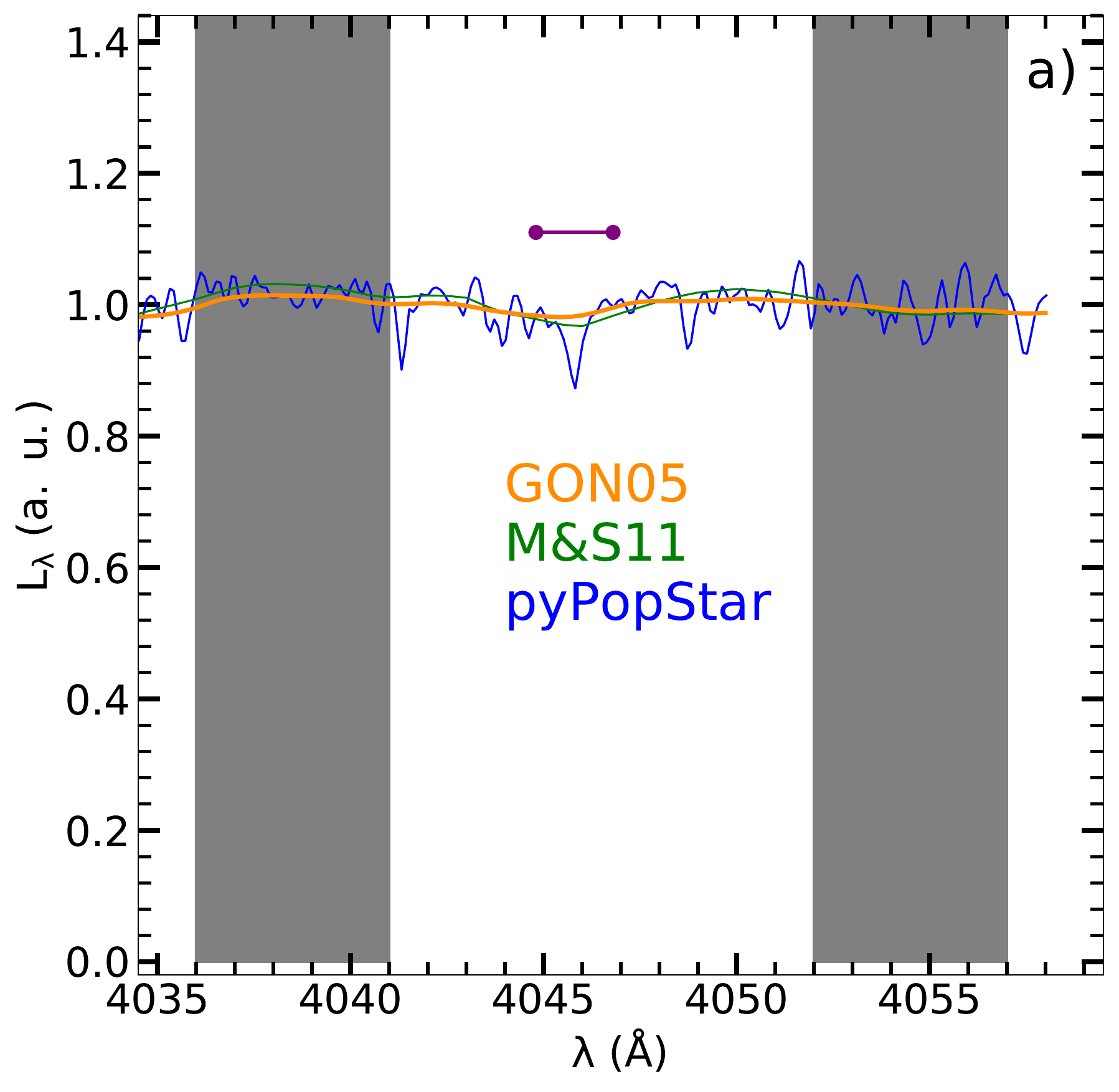}
\includegraphics[width=0.3\textwidth,angle=0]{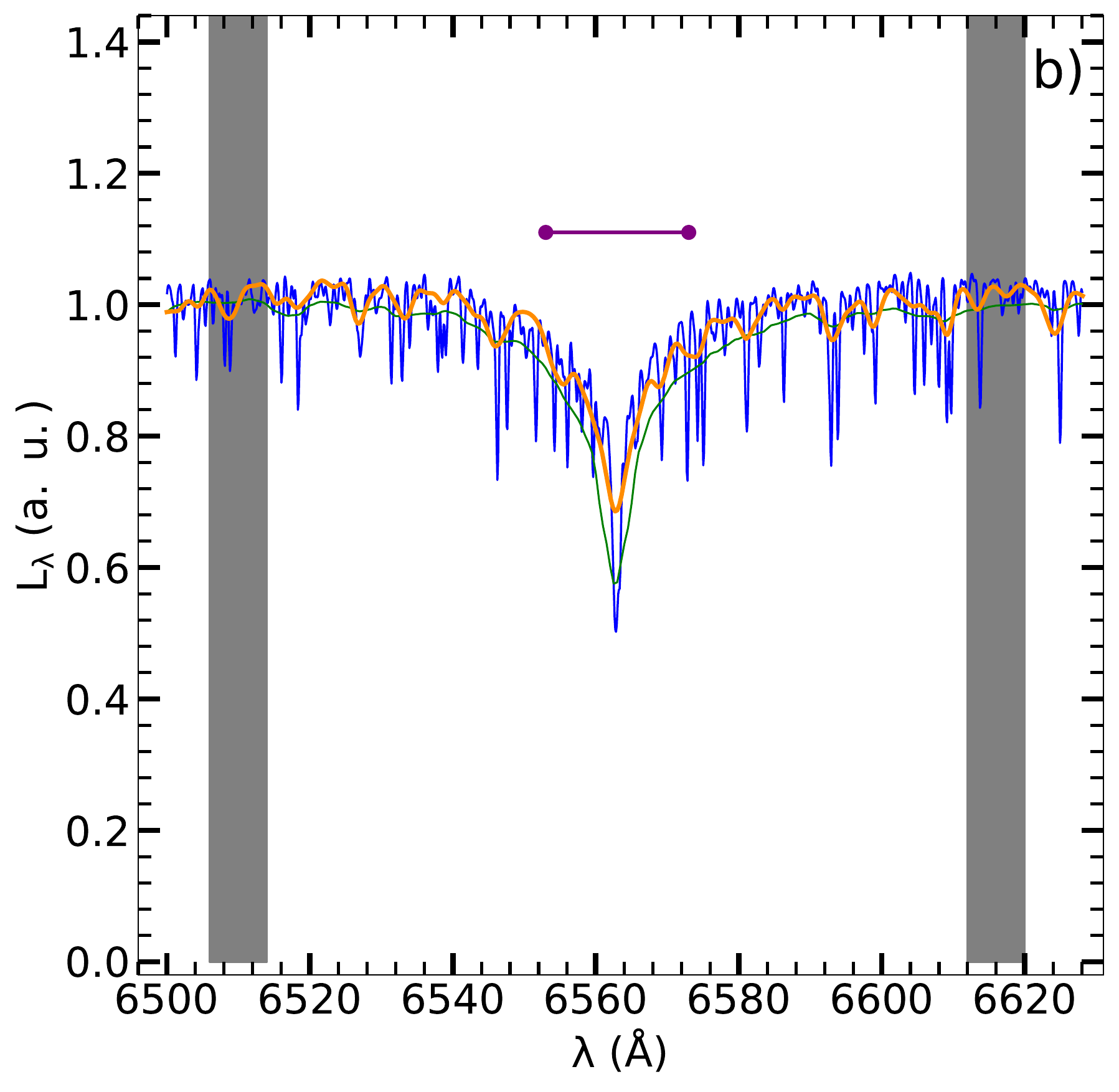}
\includegraphics[width=0.3\textwidth,angle=0]{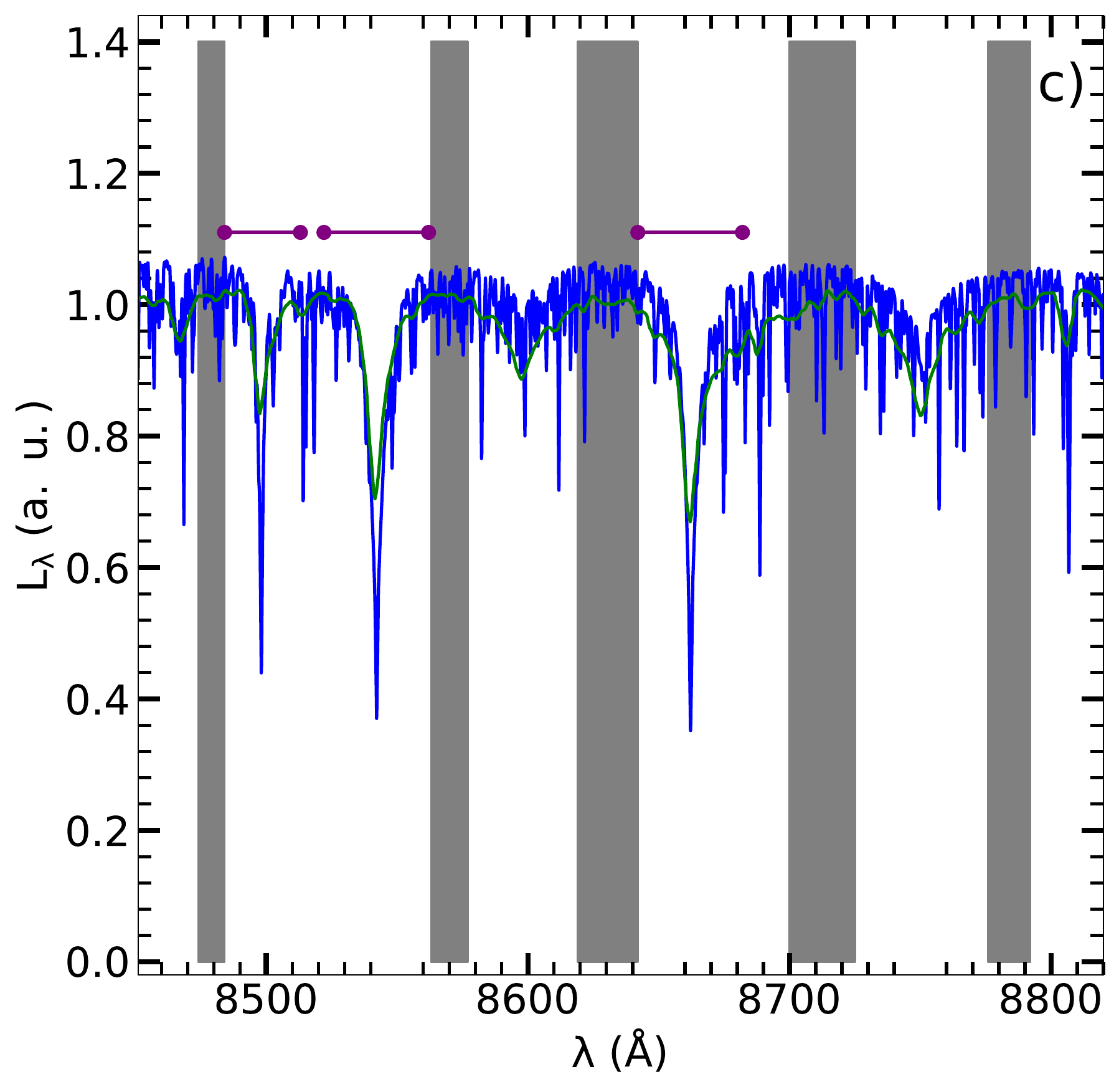}
\includegraphics[width=0.3\textwidth,angle=0]{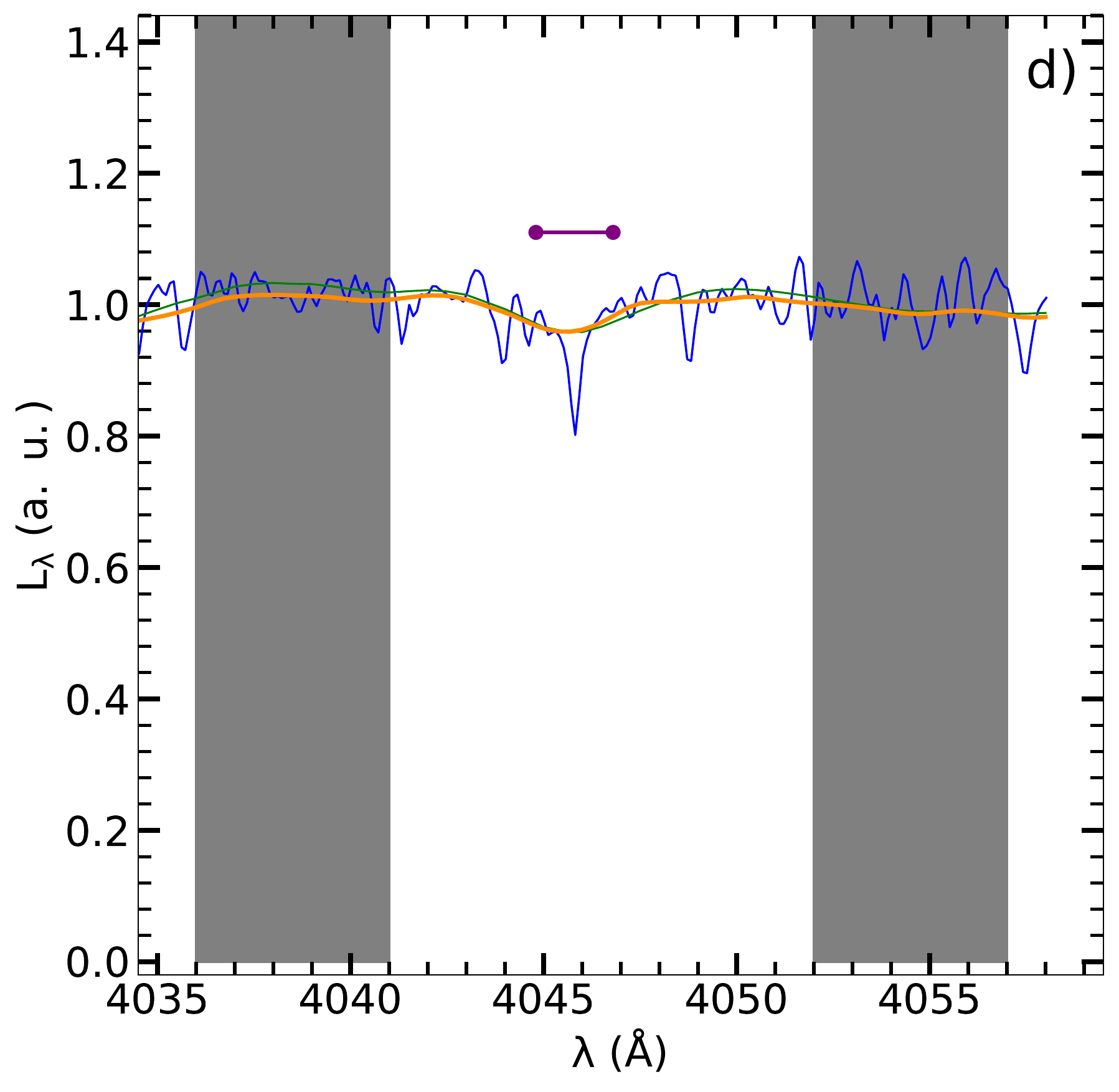}
\includegraphics[width=0.3\textwidth,angle=0]{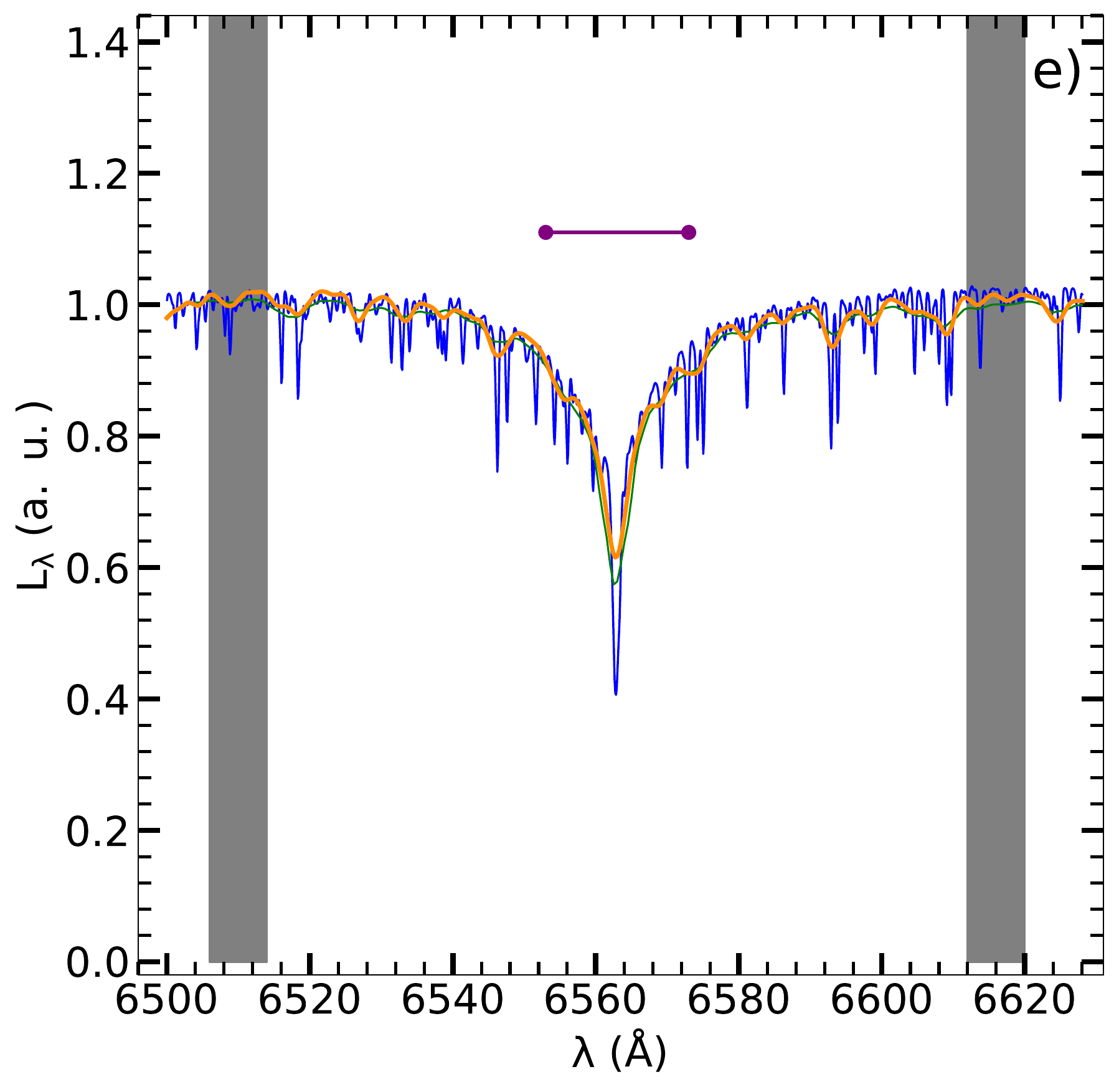}
\includegraphics[width=0.3\textwidth,angle=0]{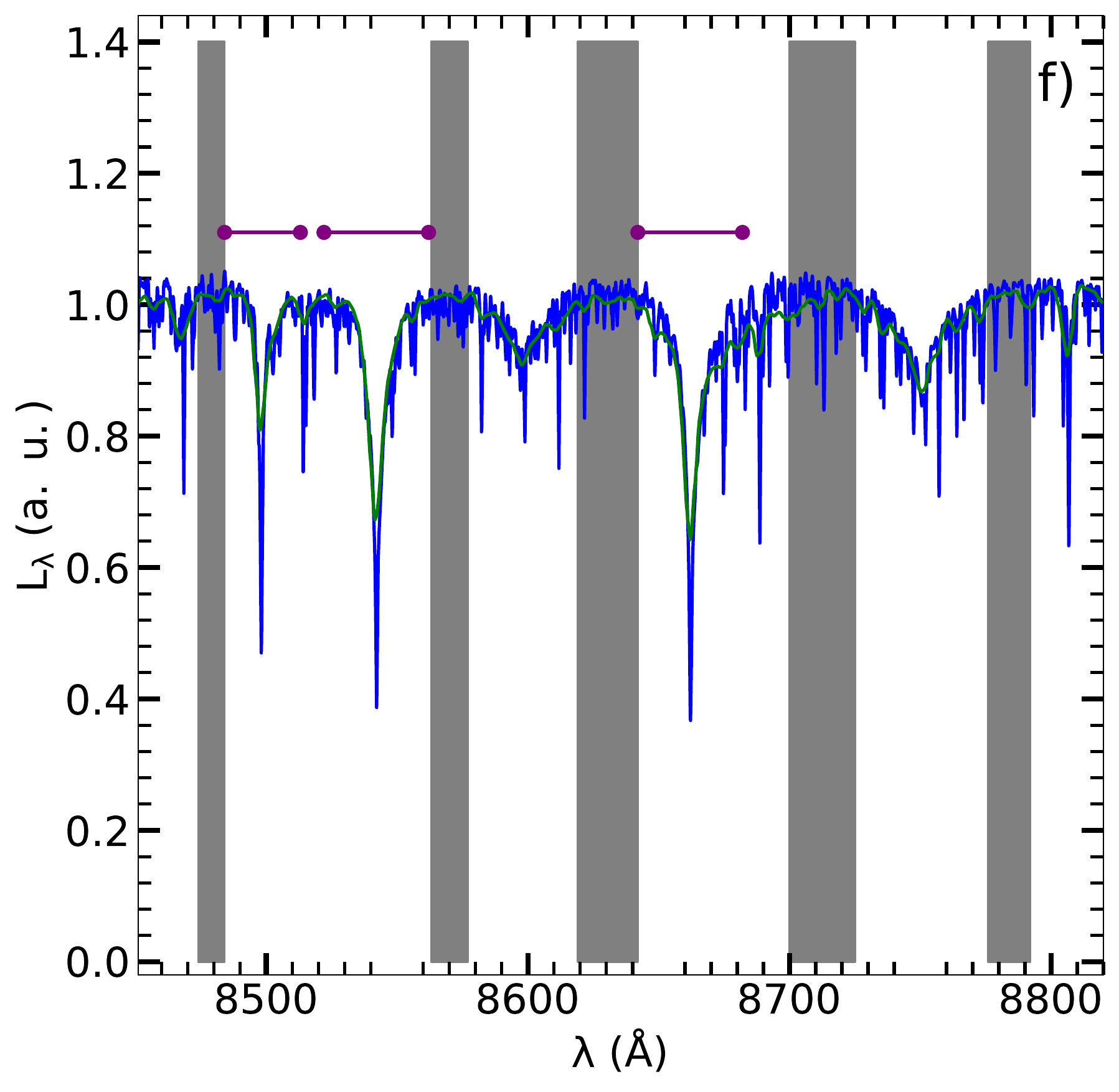}
\includegraphics[width=0.3\textwidth,angle=0]{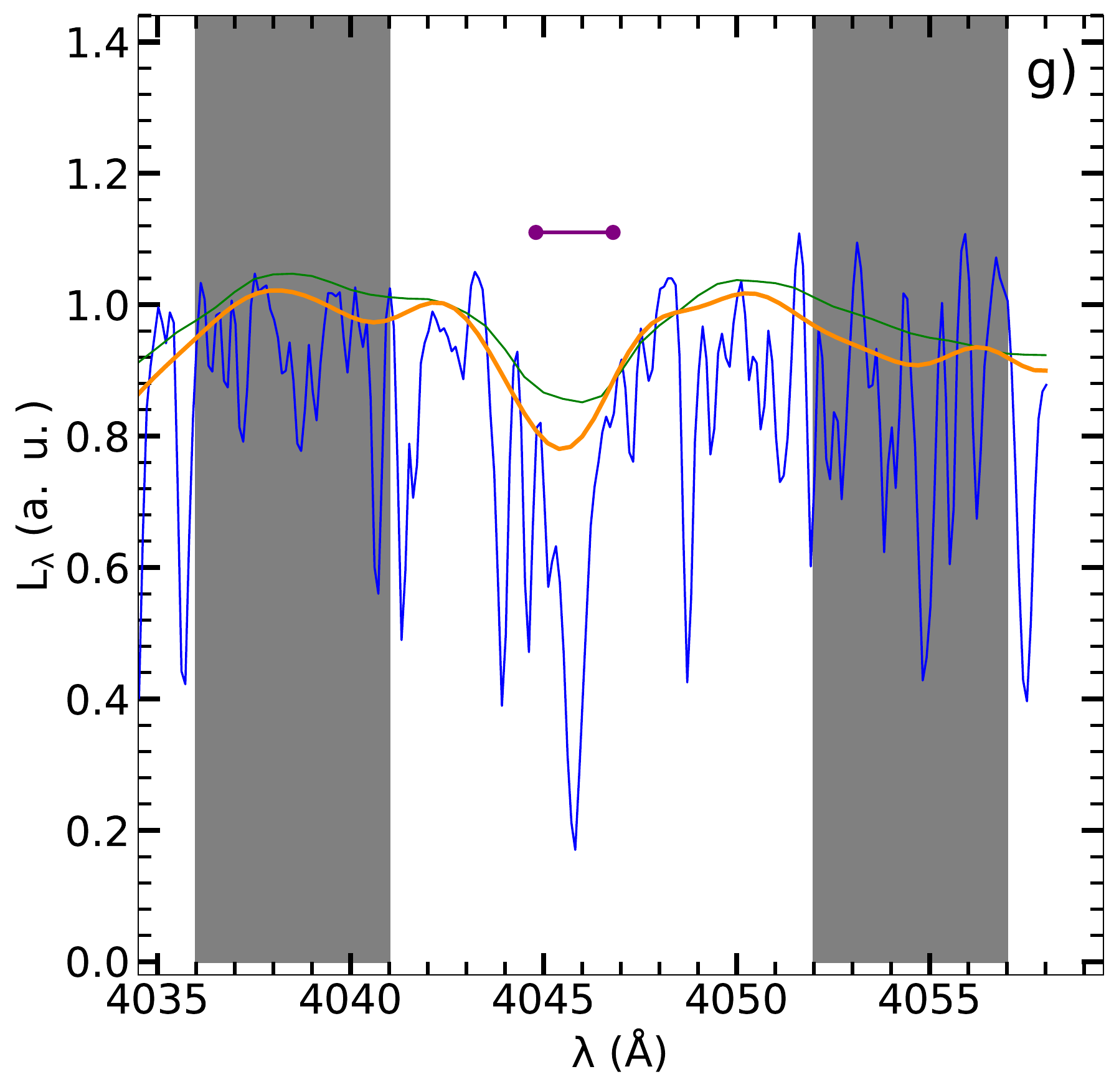}
\includegraphics[width=0.3\textwidth,angle=0]{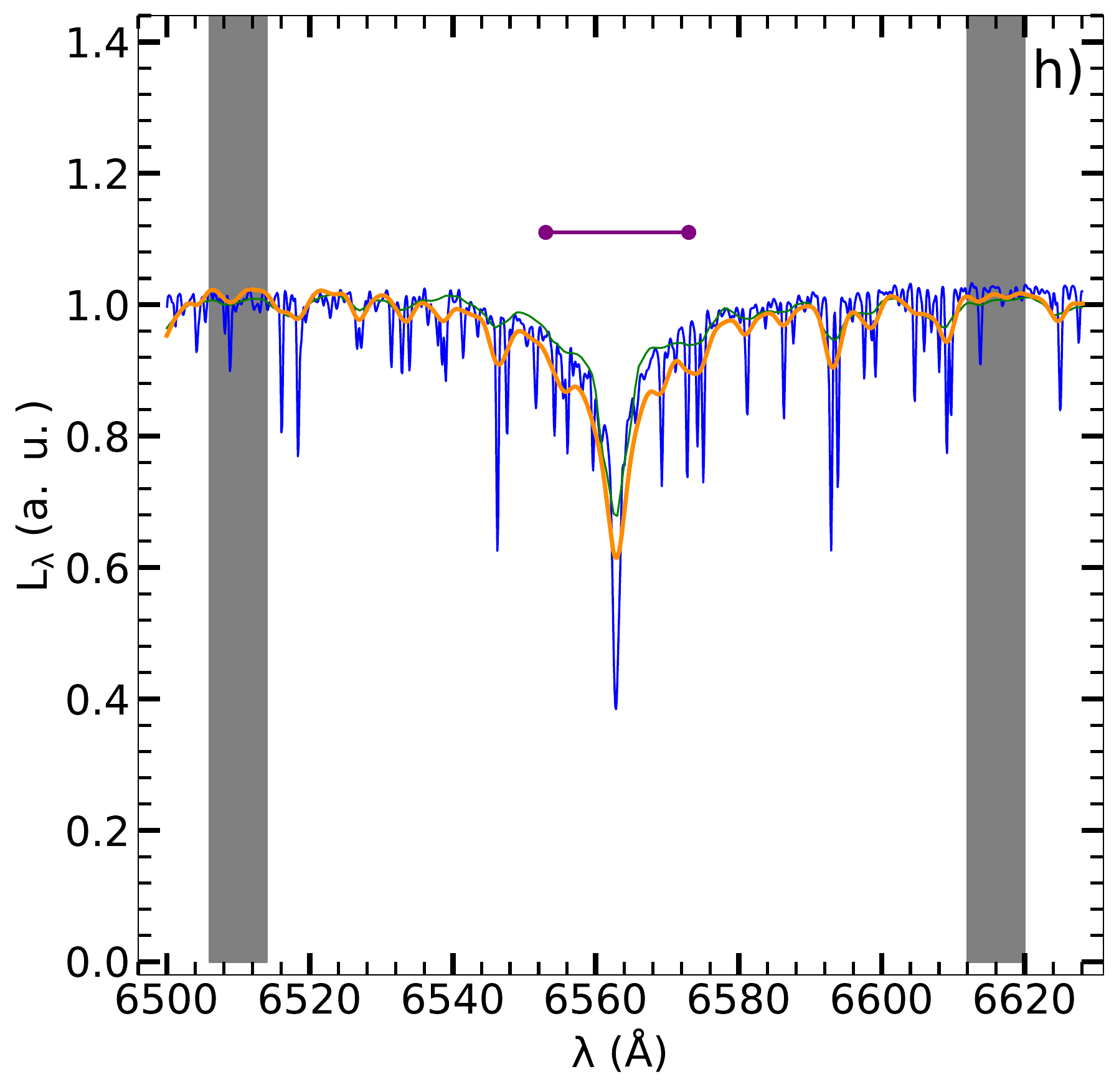}
\includegraphics[width=0.3\textwidth,angle=0]{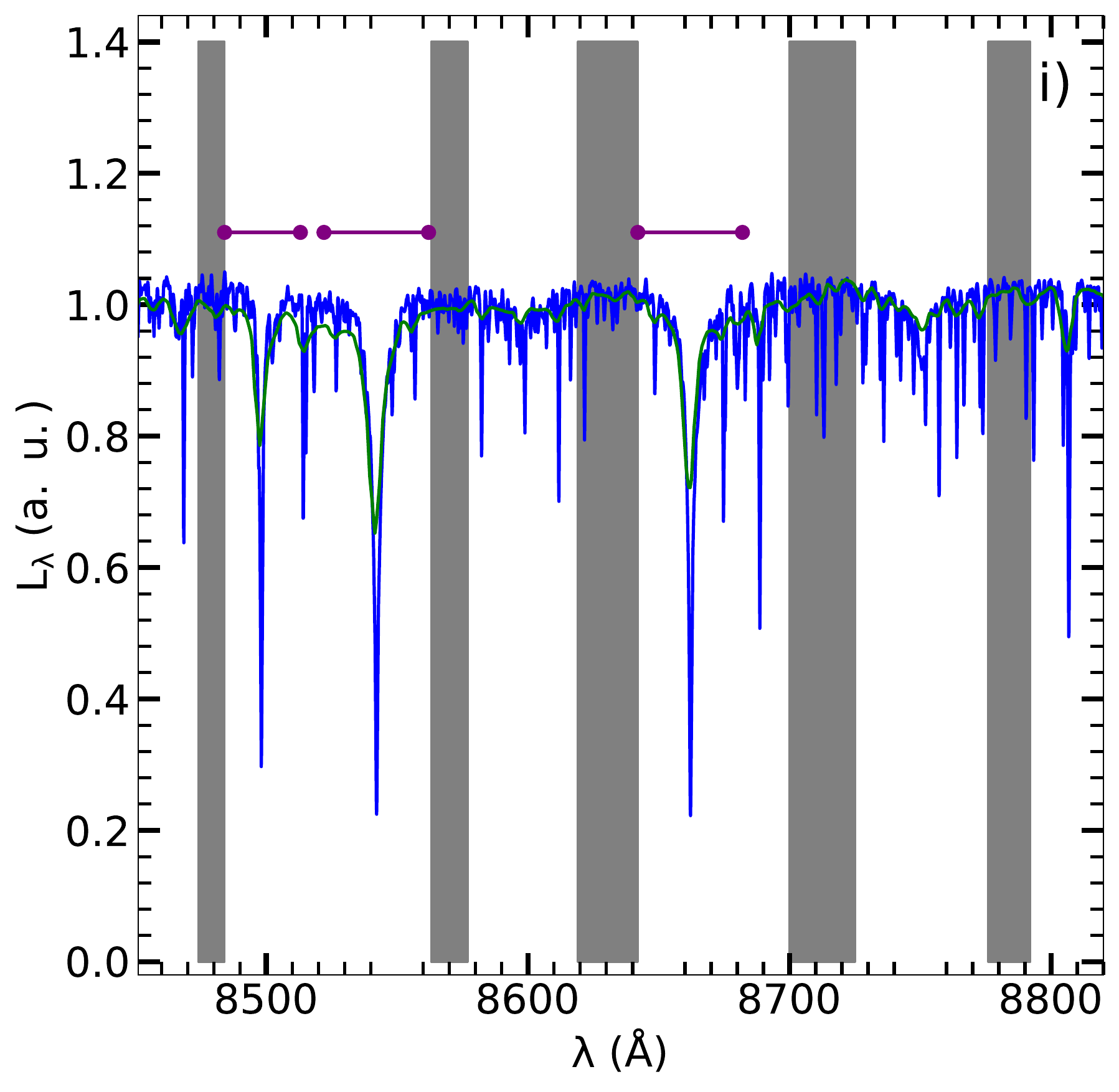}
\includegraphics[width=0.3\textwidth,angle=0]{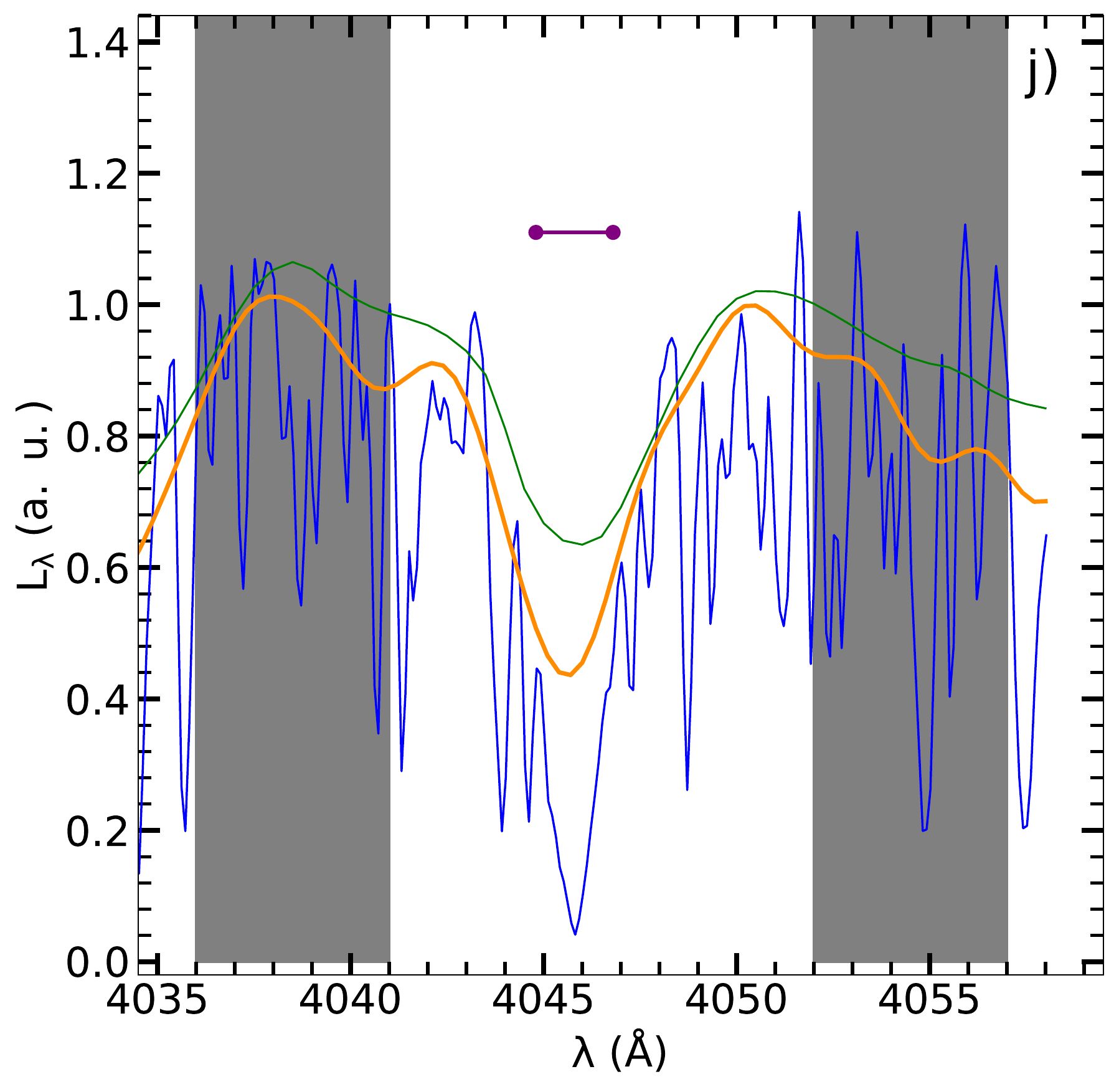}
\includegraphics[width=0.3\textwidth,angle=0]{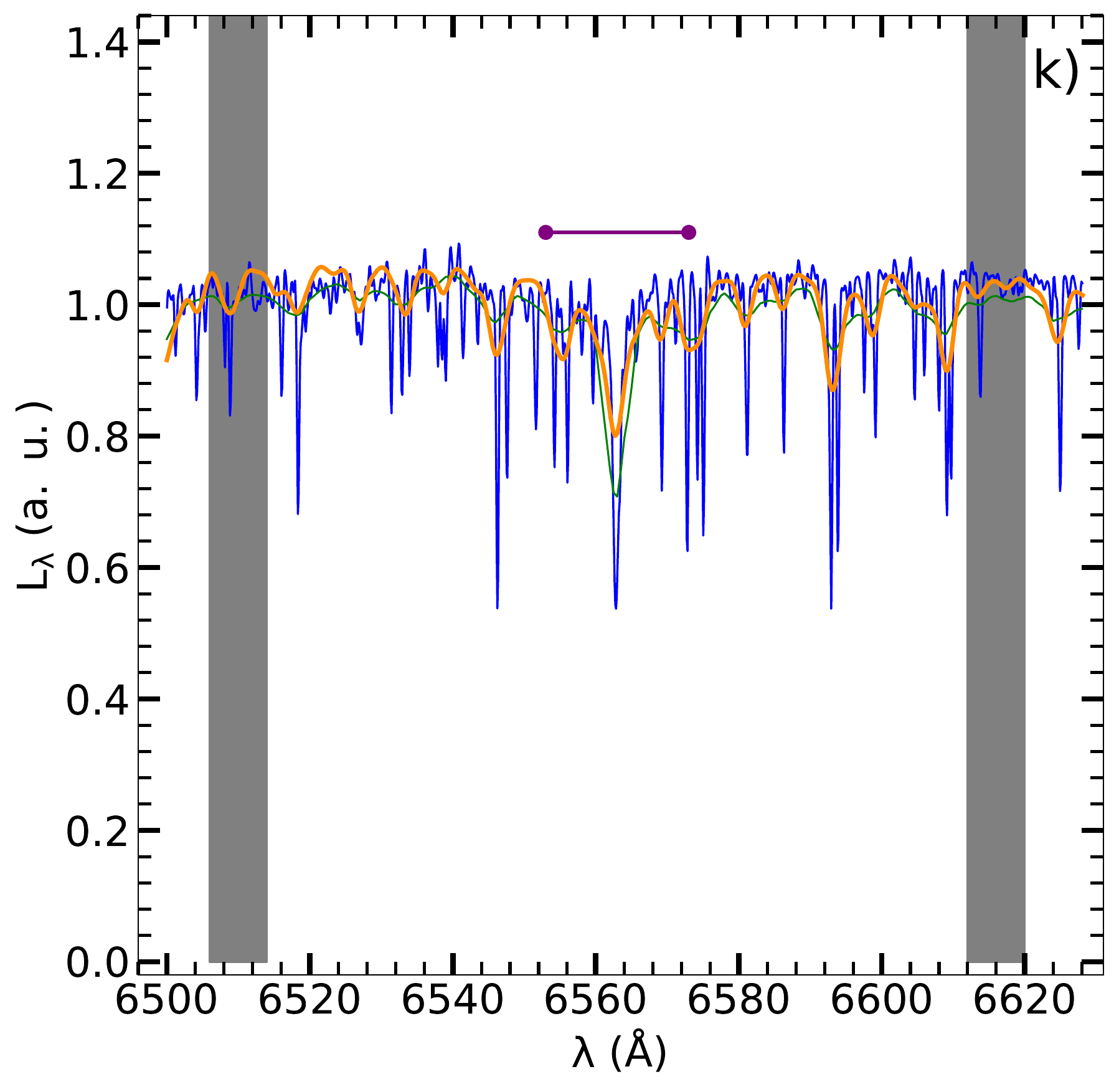}
\includegraphics[width=0.3\textwidth,angle=0]{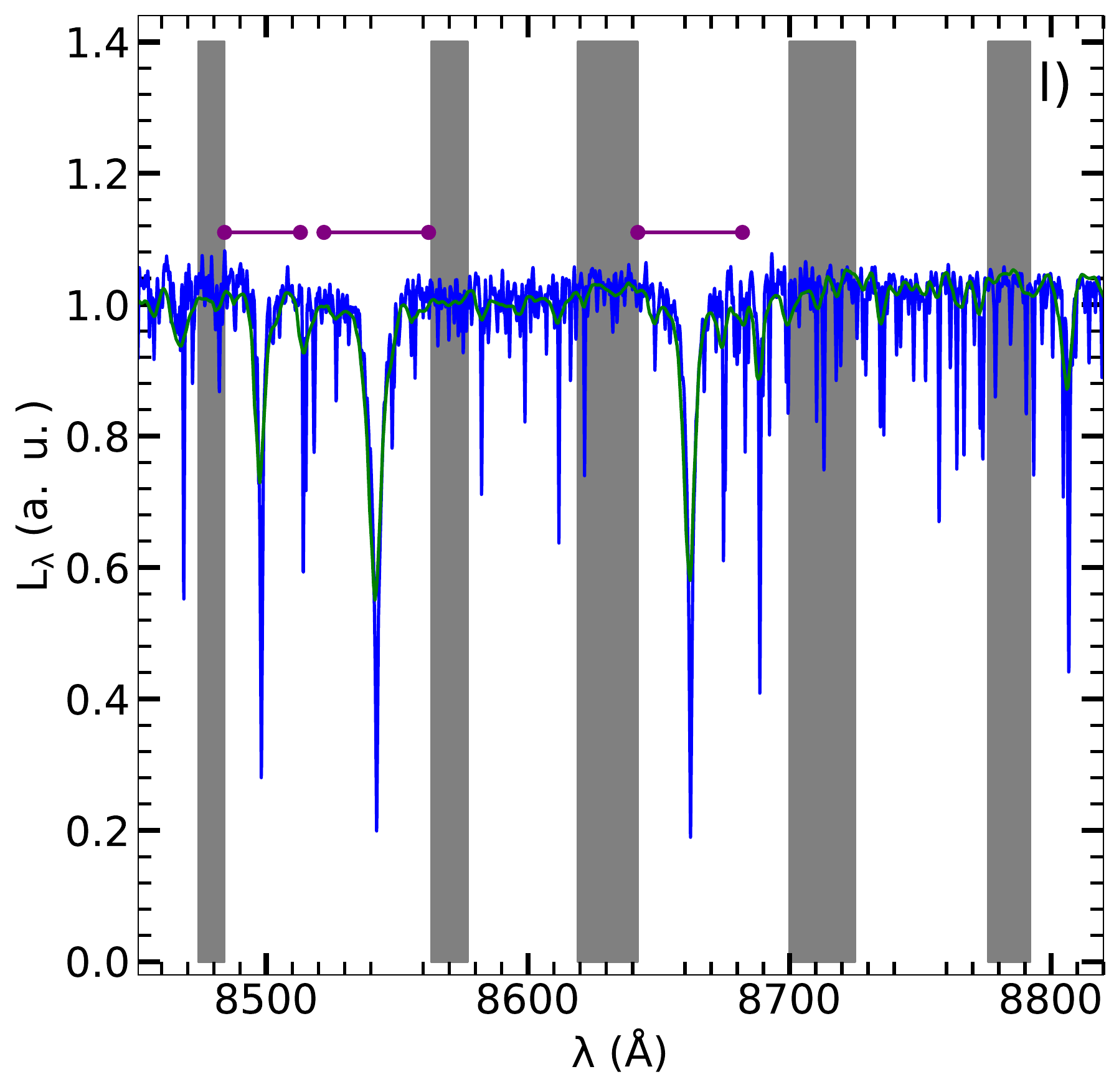}
\caption{The profiles for Fe4046, H$_{\alpha}$ and the triplet of Calcium CaT lines in columns left, middle and right, respectively, for ages $t=30$ and 100\,Myr and 1 and 12\,Gyr, from top to bottom rows, for $Z=\mathrm{Z}_{\sun}$ and KRO IMF. The continuum definitions in each panel are marked with grey zones, and the windows for the lines are shown with red lines over the spectra. In each panel, blue lines are these {\sc HR-pyPopStar} models, while green and orange lines represent M\&S11 and GON05  HR spectra.}
\label{Fig:11}
\end{figure*}

The most important point in these new SEDs resides in the HR, which allows to see clearly a large number of stellar absorption lines.
\begin{figure*}
\includegraphics[width=0.32\textwidth]{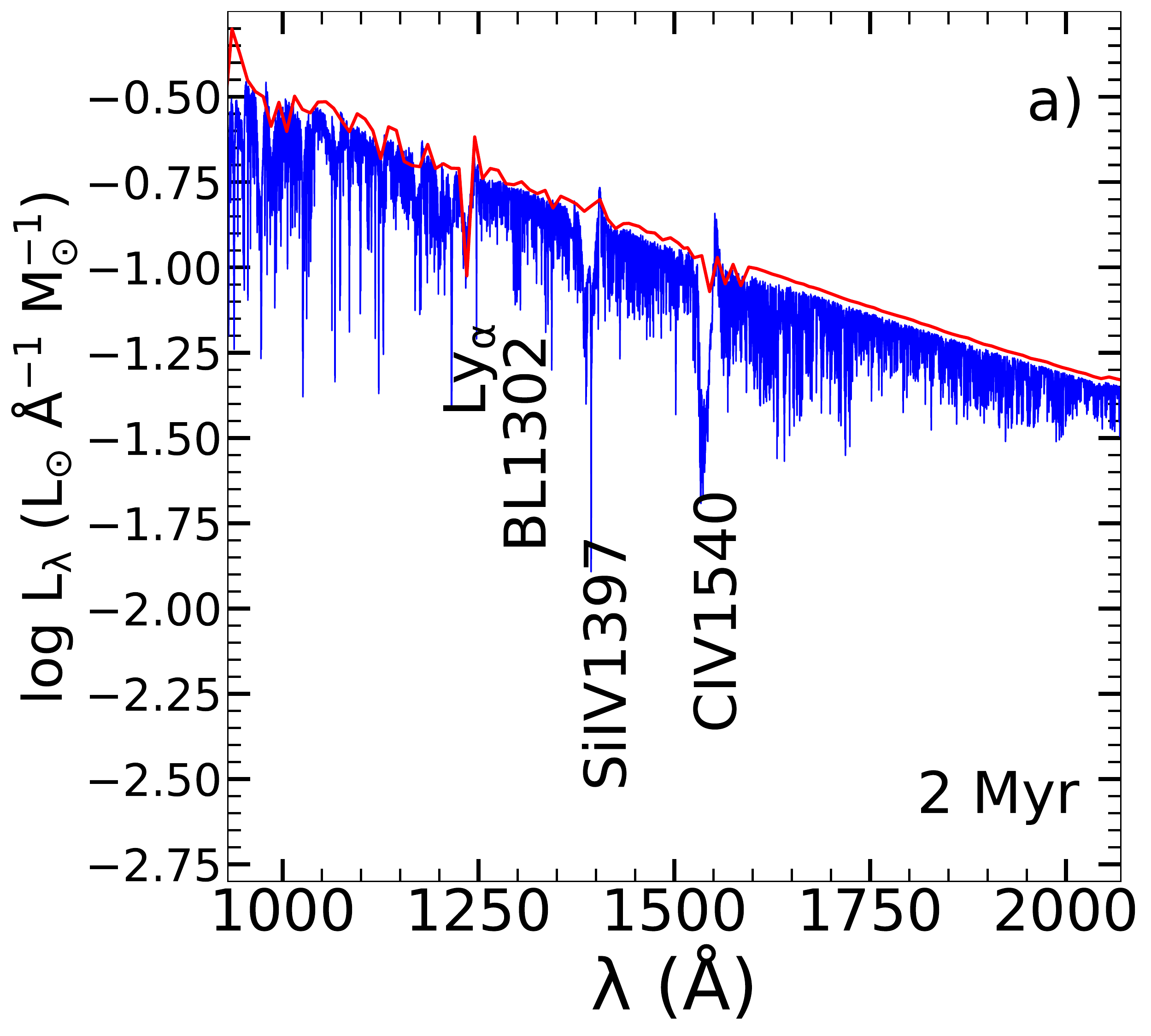}
\includegraphics[width=0.31\textwidth]{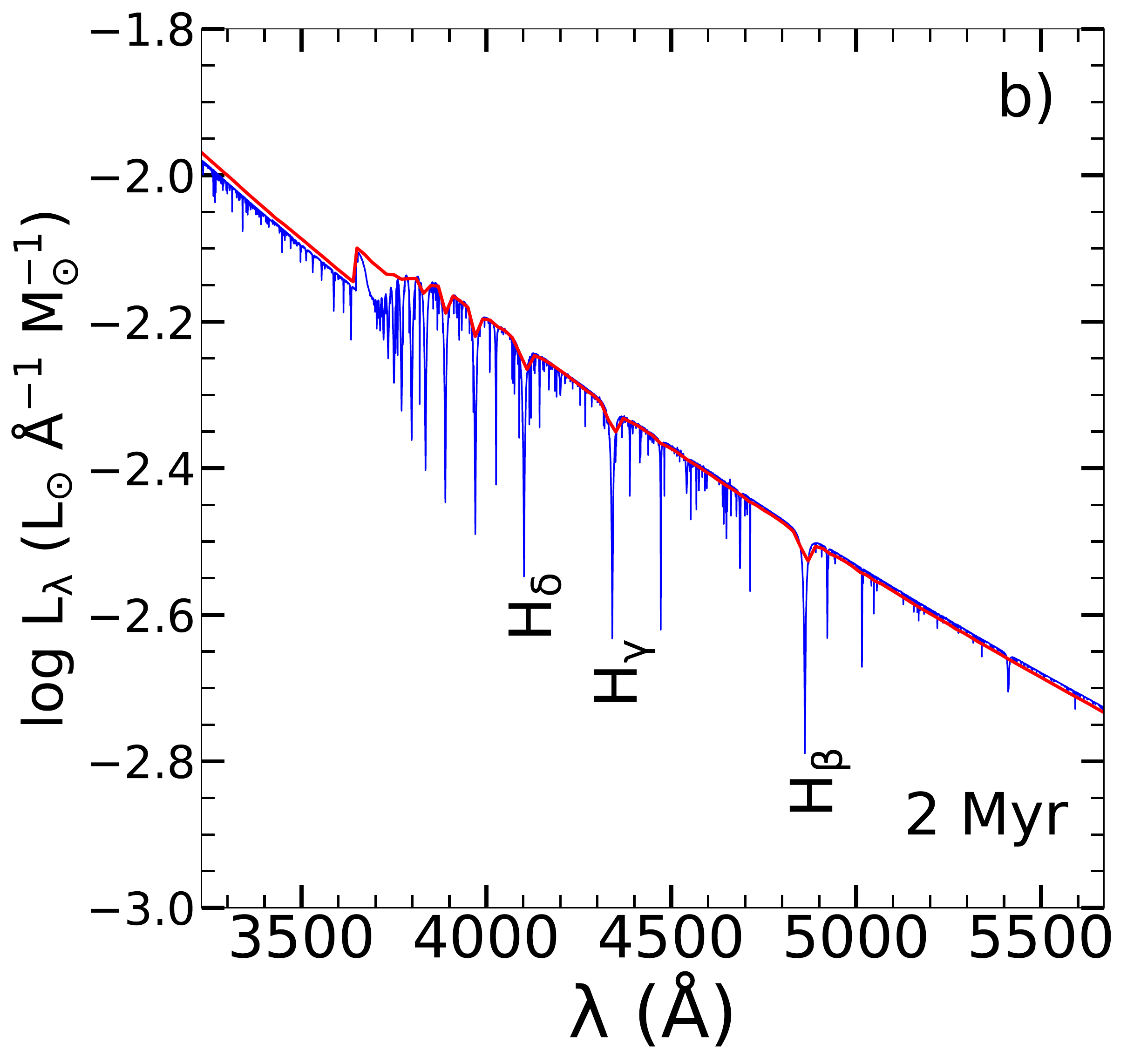}
\includegraphics[width=0.31\textwidth]{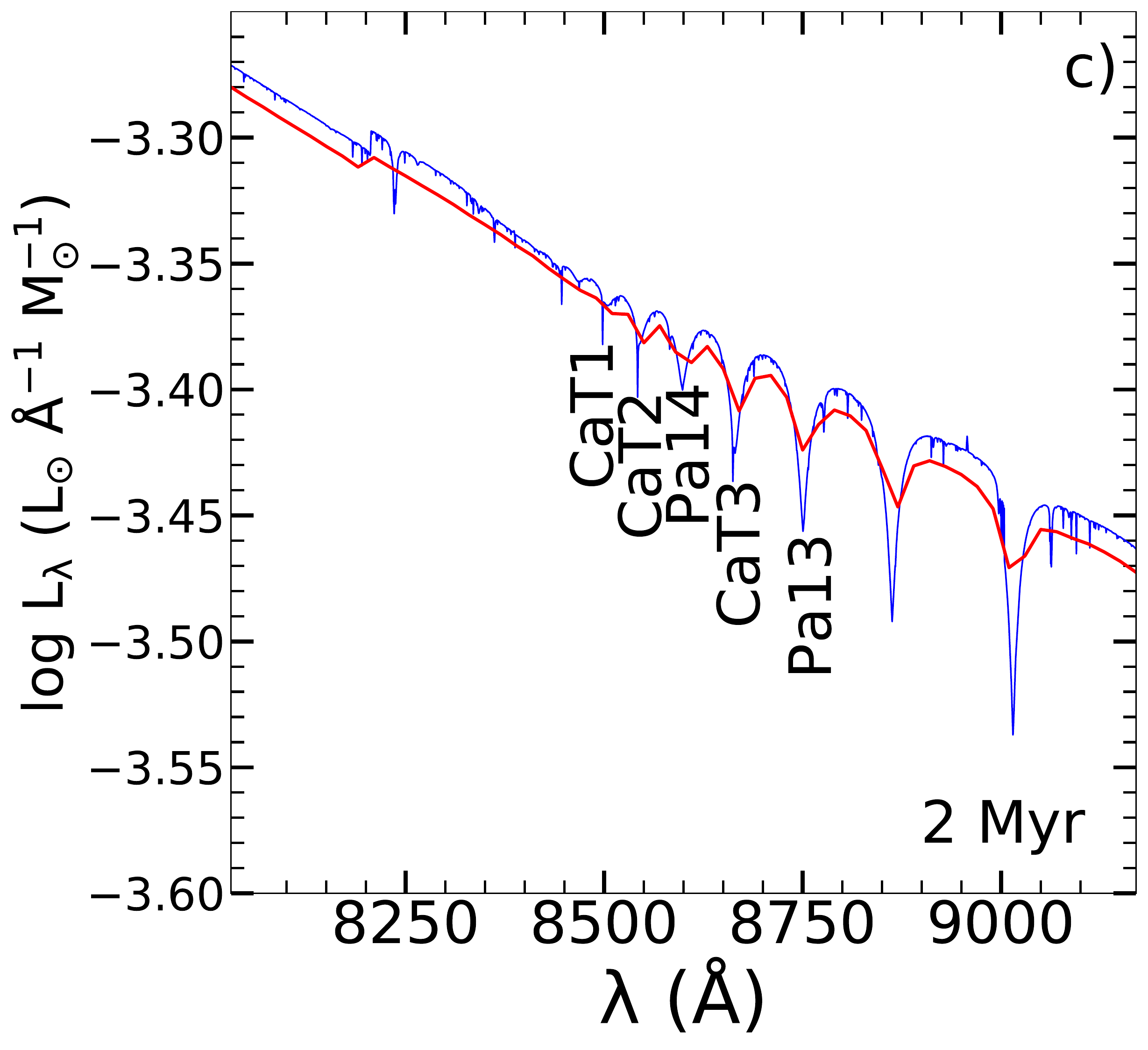}
\includegraphics[width=0.32\textwidth]{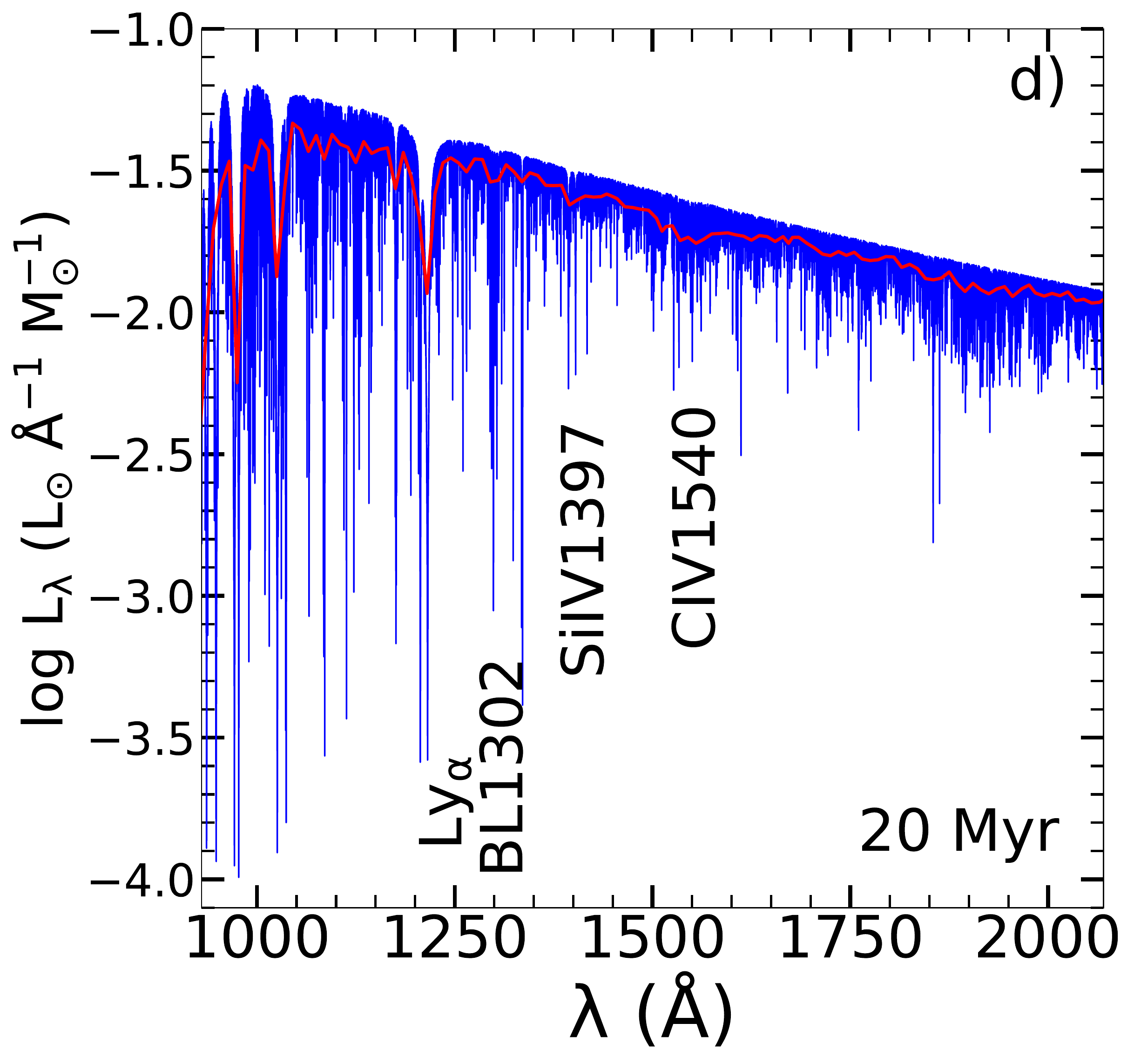}
\includegraphics[width=0.32\textwidth]{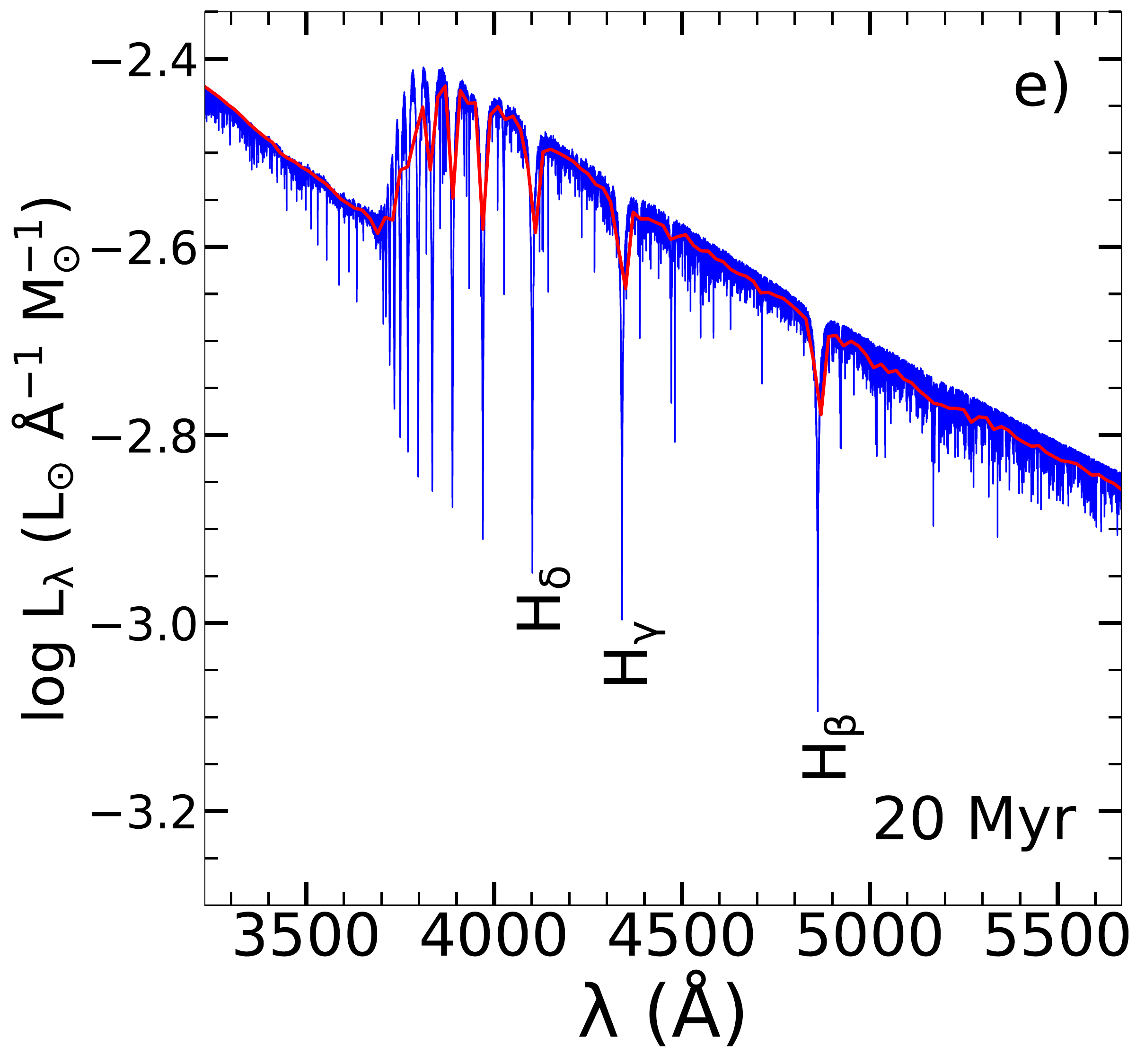}
\includegraphics[width=0.32\textwidth]{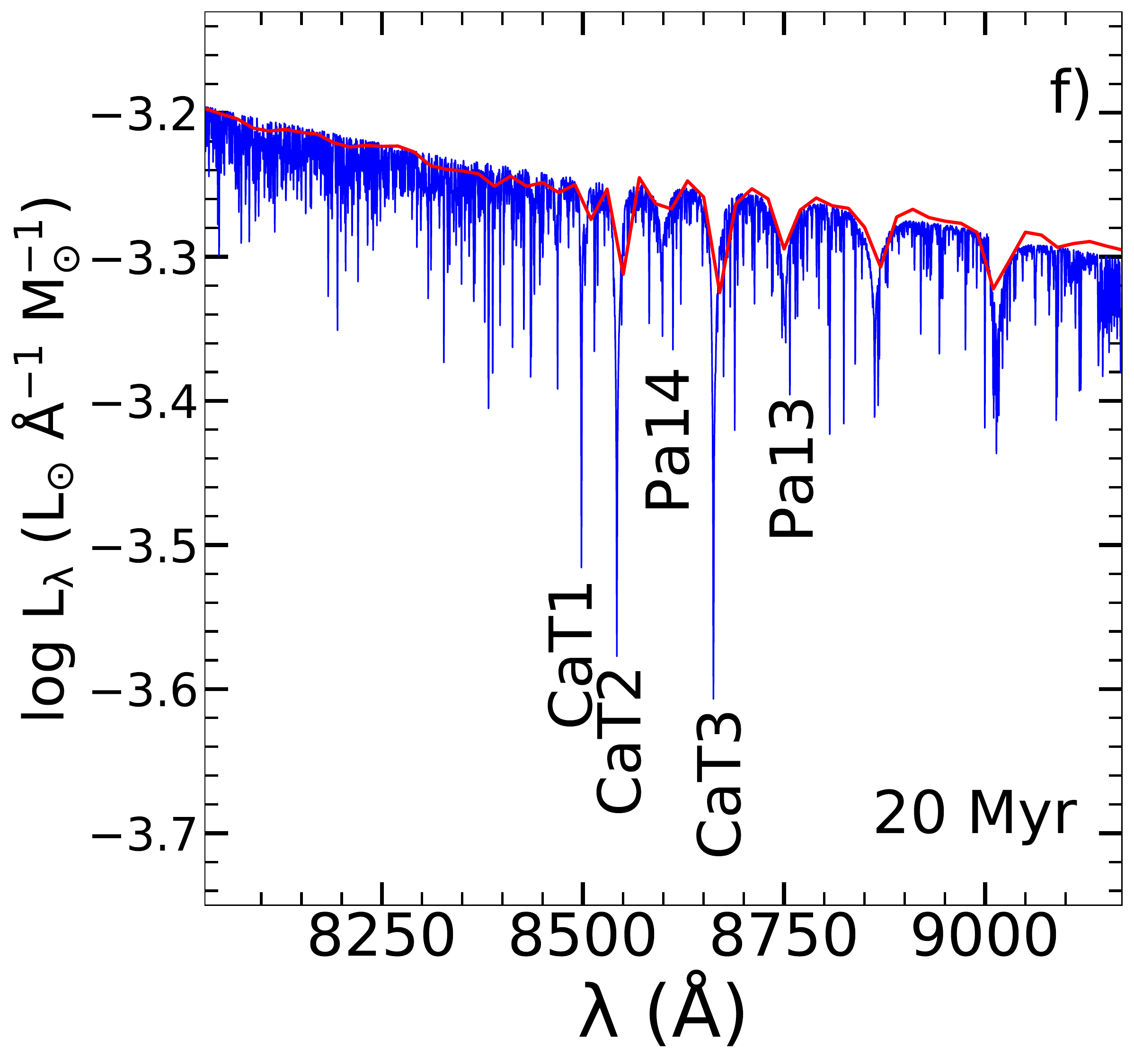}
\includegraphics[width=0.31\textwidth]{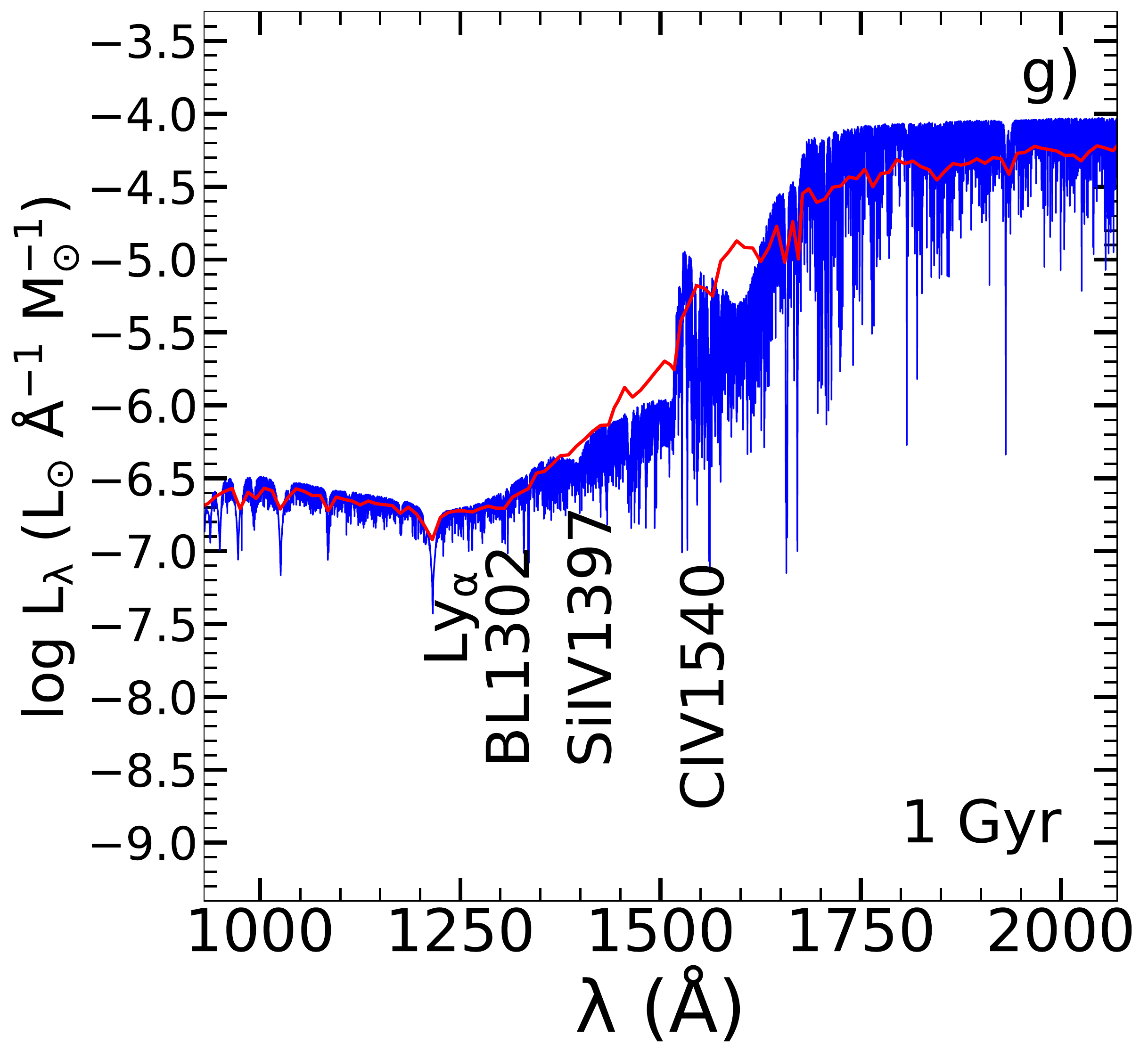}
\includegraphics[width=0.32\textwidth]{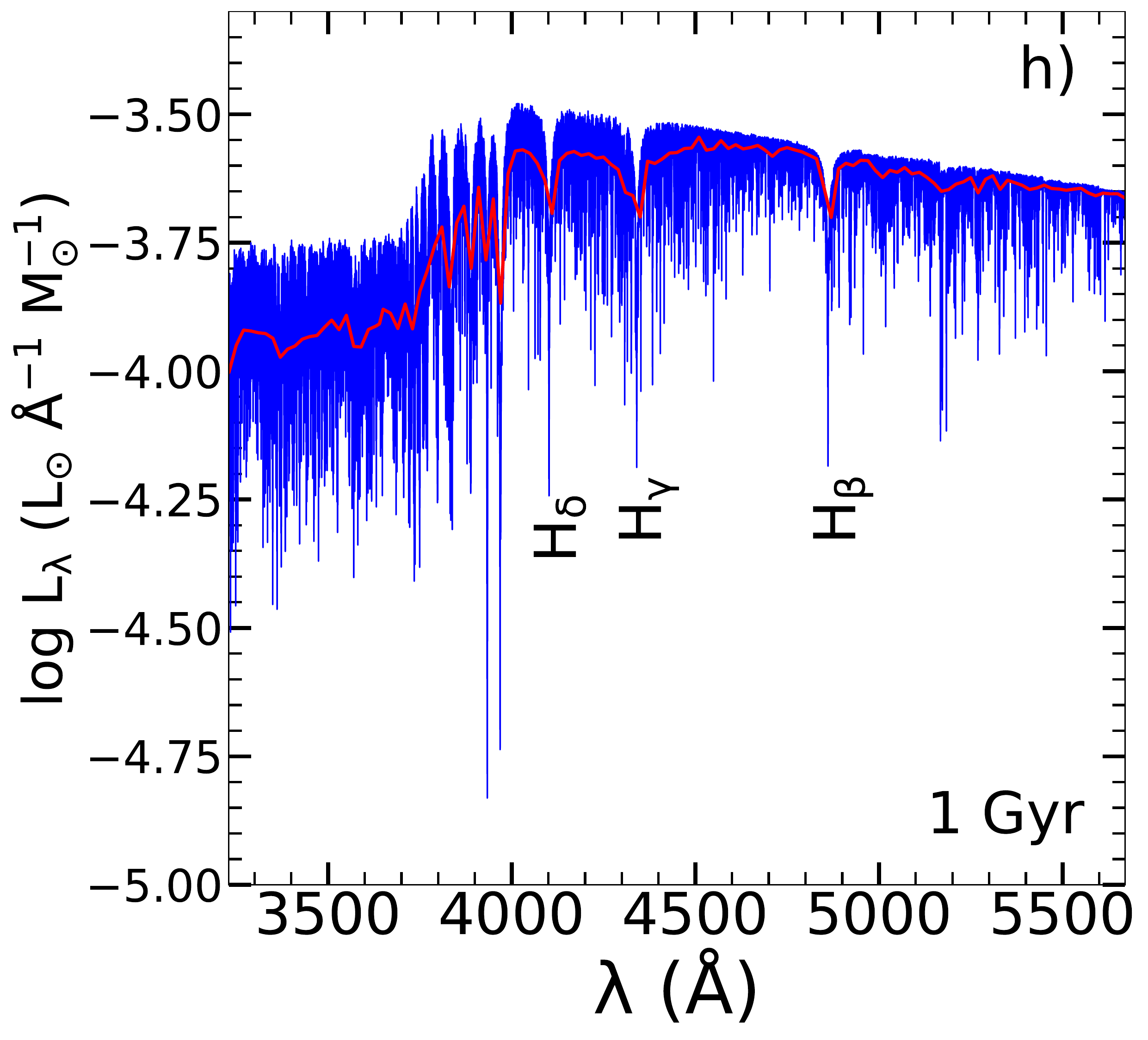}
\includegraphics[width=0.32\textwidth]{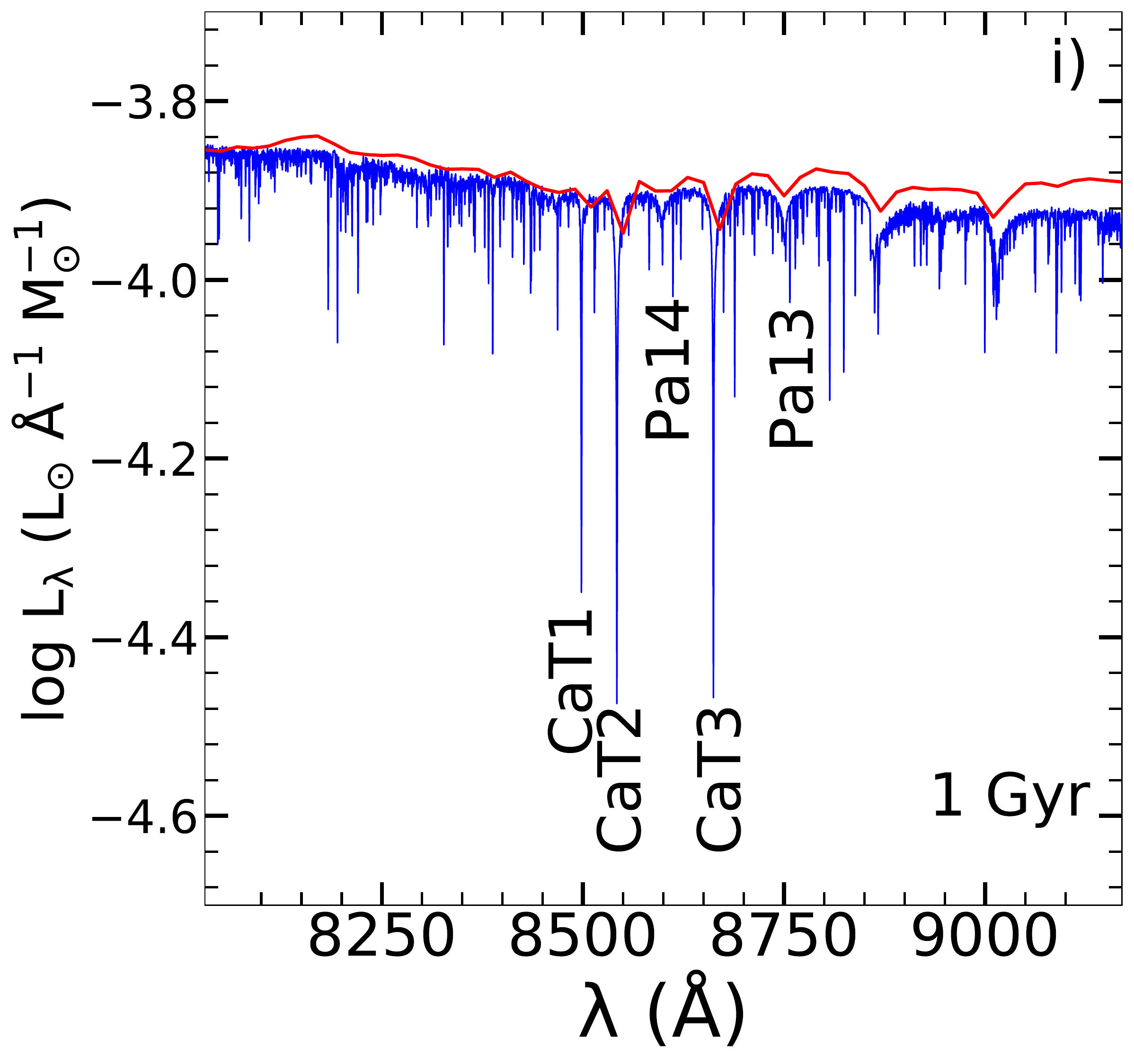}
\caption{Comparison of {\sc PopStar} (red lines) and {\sc HR-pyPopStar} (blue lines) models for $Z=0.008$ and three ages: 2\,Myr Top panels); 20\,Myr (middle panels) and 1\,Gyr (bottom panels) divided in three ranges of wavelengths.}
\label{Fig:10}
\end{figure*}
We have done a zoom of our SEDs for three ages, 2\,Myr (top panels), 20\,Myr (middle panels) and 1\,Gyr (bottom panels) comparing {\sc HR-pyPopStar} with {\sc PopStar} SEDs, represented as blue and red lines, respectively, for $Z=0.008$ in Figure~\ref{Fig:10}. In that image, we have represented three ranges of wavelength: from 1000 to 2000\,\AA\ in the left column, from 3200 to 5700\,\AA\ in the middle column, and from 8000 to 9150\,\AA\ in the right one. We show in Table~\ref{Table:8} a list of some well known spectral absorption lines that may be identified in these plots. In this table, for each line, with name in column 1, we give its central wavelength in column 2 and the corresponding band-pass wavelengths in columns 3 and 4. The references where these band-pass are defined are in column 5. Some of these lines have been indicated in the corresponding panel of Figure~\ref{Fig:11}, following the wavelength range in which they are.

This figure is a clear demonstration of the power of the new models: we can easily see the high number of absorption lines that appear in each age compared with the low --even the intermediate-- resolution spectra, and for which it is almost impossible to measure them, or, at least, it would be highly uncertain. This is particularly relevant at the oldest ages, but is is also important in young stellar populations.

In next Figure~\ref{Fig:11}, we show the profiles of the Fe\,4036, H$\alpha$ and CaT lines (one in each column) for solar metallicity and different ages, one in each row, as labelled, in order to see its variation with $\log{t}$. We mark in each panel the continuum bands and the line bandpass, as defined by the authors from Table~\ref{Table:8}. The spectra have been normalised to the continuum in each one of them. We have drawn in each panel the same age and metallicity SEDs from M\&S11 and GON05 in orange and green, respectively.  Clearly, our high wavelength-resolution allows a precise definition of the lines.

The absorption lines have been widely used for decades to analyse stellar populations, mainly for the oldest ones, to estimate metallicity ($Z$ or [M/H]), age and also the possible over-abundance of $\alpha$-elements [$\alpha$/Fe]. In order to do this, the usual technique is to compute the so-called absorption indices, defined in terms of equivalent widths or magnitudes.

It is possible to measure the equivalent width of many absorption lines over these spectra and see their dependence on age (and metallicity). Since in these spectra the lines do not overlap with the closest ones, and thus, theoretically, one may estimate the absorption indices and see the variations with $Z$ and age.
However, the measurements of the indices are dependent on the spectral resolution of instruments \citep{barbuy2003,vazdekis2010}. Obviously, their computation  over the theoretical spectra also change when they are calculated with models of different wavelength-resolution.  Therefore, the question here is if the continuum bands as defined in previous works are the most adequate, since a lot of lines might be contaminating them. On one hand,in order to do these measurements in a more realistic way than before, it would be better using the local spectra in a wider range that the own line in each case. Maybe, it could be possible to define the continuum in a more similar way to this one used by CEN01 and CEN09, (which define their "generic" index CAT* using an continuum with 10 points for all lines presented along the spectra between 8400\,\AA\ and 8800\,\AA) would be the best technique for all lines. On other hand, each line is very clear in each plot and it would be fairly simple to measure their areas. Therefore, other possible technique would be to reduce to the lines themselves to measure them, after normalizing to a common continuum for each band as we do here, and then to measure each line. Thus, for the lines Fe-I\,4026, H$\alpha$ and CaT, (defined as usual as sum of the two more intense lines CaT2 and CaT3), the measures for KRO SEDs would be those from Table~\ref{Table:9}. In a first and quick analysis, the equivalent widths, EW, measured directly in the spectra are lower than the EWs computed using the definitions of Table~\ref{Table:8} when the lines are narrow, and higher when the lines are wide. Moreover, when the lines are well defined, the equivalent widths do not include other adjacent lines. For instance, the CaT lines are now computed without Pa Lines, which decrease their indices, as it was already 
showed by \citet{vaz03} using intermediate wavelength-resolution spectra. That is, even if the evolution of these equivalent widths along the age of stellar populations seem to be similar to the one from other authors, the absolute values are not the same.  This shows that a deeper study of absorption lines, particularly weak lines, is necessary to redefine, if necessary, the band-pass of lines for high wavelength-resolution observations. We consider that a more complete and careful study about these questions of defining the continuum, the measurement of these indices and probably their definitions, too, must be performed with these new HR spectra. This task is out of the scope of this piece of work, and we will explore it in a future work.

\begin{table}
\caption{Equivalent widths, in \AA\ units, for the spectral absorption indices shown in Figure~\ref{Fig:11}.}
\label{Table:9}
    \centering
    \begin{tabular}{c|c|c|c}
Age & Fe-I\,4046 & H$_{\alpha}$ & CaT \\
(Gyr) & (\AA) & (\AA) & (\AA) \\
\hline
0.03 & 0.020 & 4.24 & 7.23\\
0.10 & 0.010 & 5.25 & 6.77 \\
1.0 & 0.53 & 3.93 & 5.59 \\
12.0 & 1.52 & 0.74 & 5.02 \\ 
\hline
    \end{tabular}
\end{table}

\subsection{The use of HR observational spectra compared with this new set of models.}

In this subsection we would like to show how our new models may be applied to have a better interpretation of high wavelength-resolution observed spectra, compared with the use of the intermediate resolution ones. Therefore, we are going to show three examples for: 1) a globular cluster, --low metallicity old stellar population; 2) a region of a starburst dwarf galaxy --intermediate metallicity young stellar population; and 3) the bulge of an early type galaxy ---old and metal-rich stellar population. 
\begin{figure}
\hspace{-0.5cm}
\includegraphics[width=0.49\textwidth]{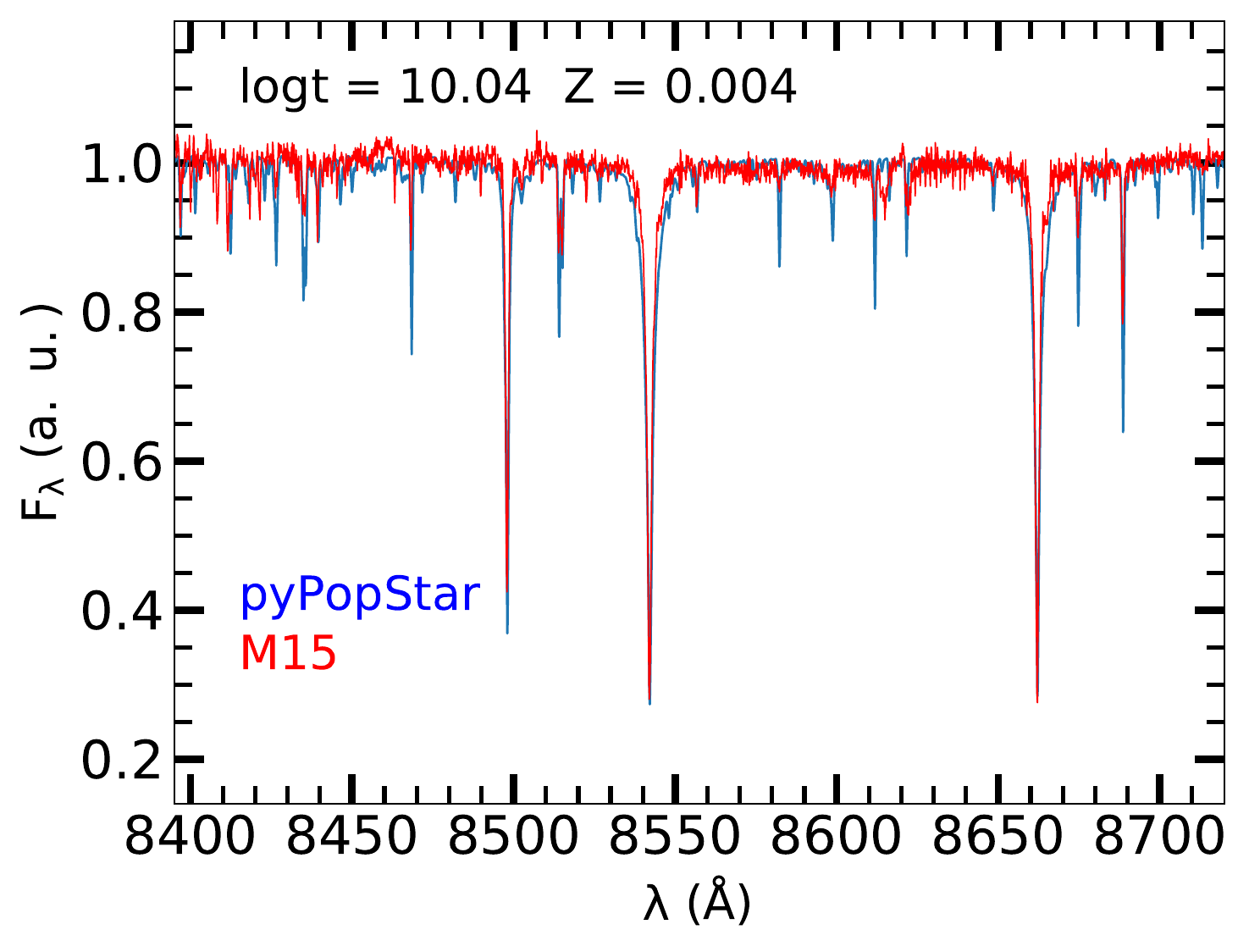}
\caption{Comparison of the spectrum for M15 obtained as sum of the stellar spectra from \citet{gv20} --red line-- with the computed {\sc HR-pyPopStar model} --blue line-- for $Z=0.004$ and age 13\,Gyr.}
\label{fig:m15}
\end{figure}

The first example corresponds to an integrated spectrum for the globular cluster M~15 (NGC ~7078). This object is an old globular cluster with age $\sim 13$\,Gyr \citep{dotter2010,vandenberg2013} of low metallicity. The accepted metallicity was $\sim$ -2.15 or -2.30 \citep{sneden2000,mcnamara2004,carreta2009}, although more recent estimates by \citet{sobeck2011} give higher values, [Fe/H]$\sim -1.66$. In \citet{gv20}, spectra for more than 50 stars in the central region of the cluster obtained from MEGARA observations with the setup HR-I were shown. Furthermore, in that work each star was fit by a stellar spectrum model from \citet{mun05}, thus obtaining the metallicity, effective temperature and gravity distributions of the 50 stars. Following the first of them, the average metallicity of the distribution is $<[\rm Fe/H]>\sim -1.72$ in good agreement with the most recent findings \citep{sobeck2011}. We have added all the stellar spectra from \citet{gv20} to obtain a total spectrum with the contribution of all these stars. We have then taken the SSPs spectrum corresponding to $Z=0.004$, the lowest metallicity we have in the models (although it is almost one order of magnitude higher than the established abundance $Z=0.0002 - 0.0003$), and an age of $t=13$\,Gyr to compare it with the obtained M15 HR-I spectrum in Figure~\ref{fig:m15}. 
\begin{figure*}
\hspace{-0.5cm}
\includegraphics[width=0.495\textwidth,angle=0]{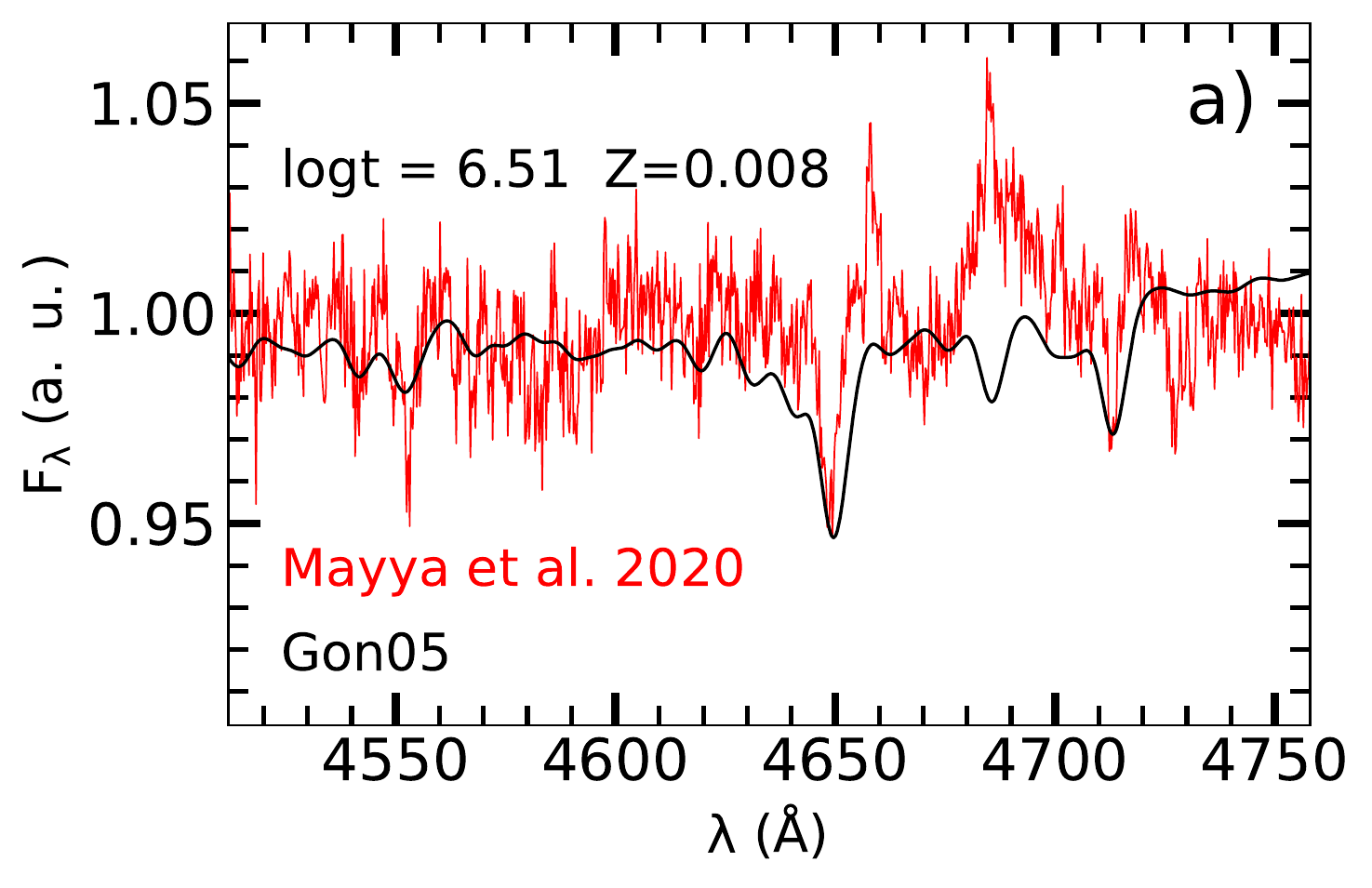}
\includegraphics[width=0.495\textwidth,angle=0]{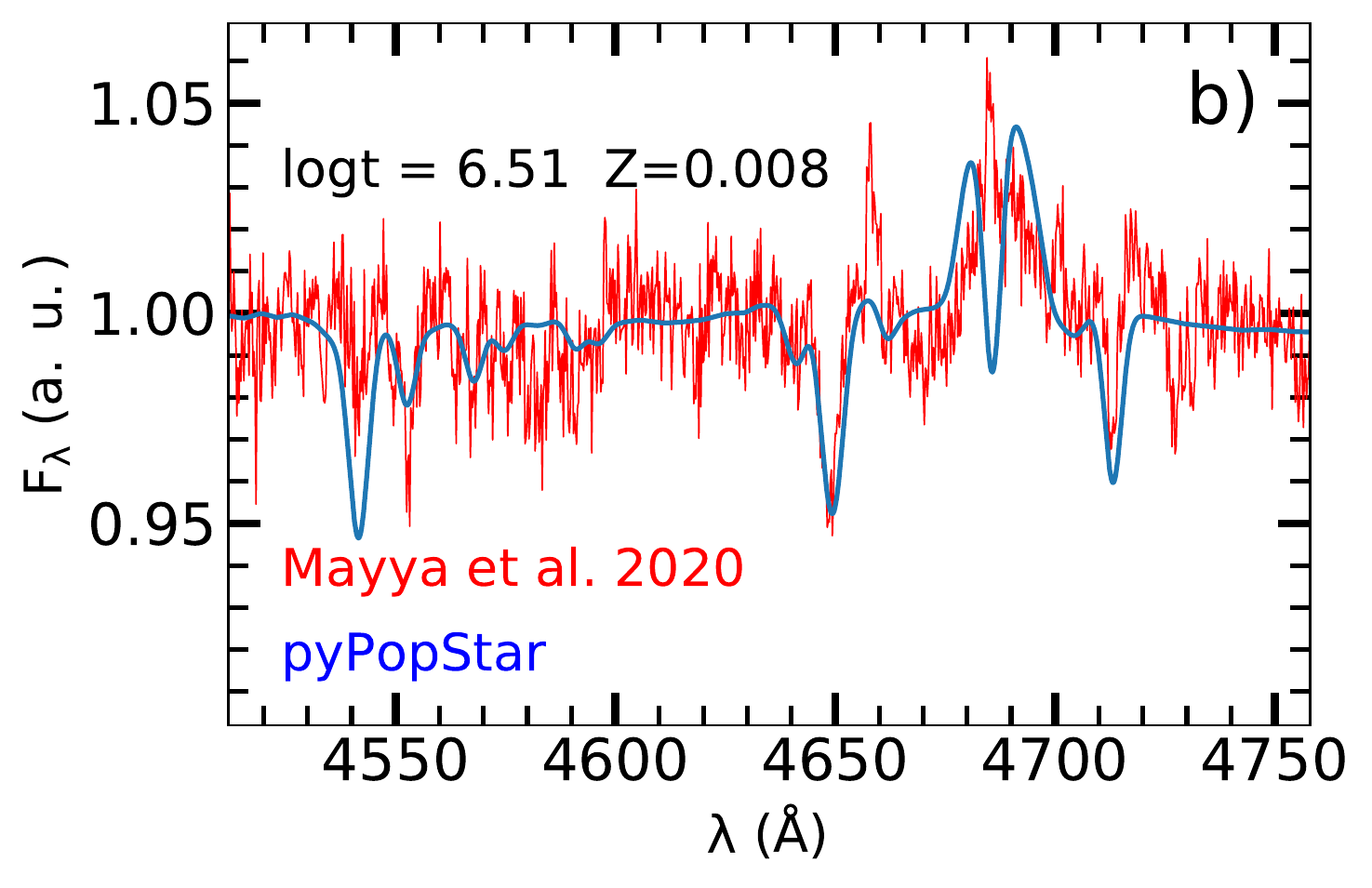}
\caption{Comparison of the spectrum for NGC ~1569 as taken by \citet{mayya2020} --red line-- compared with two models for $Z=0.008$ and age 3.5\,Myr: a) the GON05 model and b) our present {\sc HR-pyPopStar} model. Both synthetic spectra have been reddened with A$_{V}=2.3$\,mag and smoothed to the same resolution as observations.}
\label{n1569}
\end{figure*}

The velocity dispersion of this globular cluster is low enough to do not take into account its effect in the observed spectra \citep[][give values between 3 and 10\,km/s]{usher21}. On the other hand, the HR-I setup has a reciprocal linear dispersion  $\delta\lambda=0.13$\,\AA, so we have smoothed the pyPopStar models with a Gaussian of this same FWHM. As we see, the fit of the model to the observed spectrum is quite good, considering that the metallicity of our model is higher than the one of M15, as seen in the deeper stellar metallic lines of the model compared with the observed spectrum. 

Our second example is devoted to the well known starburst galaxy NGC~ 1569. This is a dwarf galaxy, located at 3.1\,Mpc, one of the nearest galaxies containing young superstellar clusters, with a gas-phase oxygen abundance close to that of the LMC, $12+ \log{(\mathrm{O/H})}=8.19$, which corresponds to $Z=0.008$. \citet{mayya2020} have carried out observations of the central region with the MEGARA instrument in GTC using in this case the setup LR-B, implying a spectral resolution of $\sim 6000$. These authors find information, such as, the number of ionising photons and H$\alpha$ luminosity from which they --as others previously-- deduce a very young age for the stellar populations, $t\sim 3.5-4.5$\,Myr, with $\sim 124$ WR stars, and an average extinction of $A_{V}=2.3$\,mag within a range between 1.6 to 4.5\,mag. The number of \ion{H}{I} ionising photons can be obtained from the observed $\rm H{\alpha}$ luminosity, $L_{\mathrm{H}{\alpha}}$ (see equation 17 of MOL09), derived from the $\rm H{\alpha}$ flux and assuming a distance to our object. This value of $Q$(\ion{H}{I}) is then compared to the number of ionising photons per solar mass from the models, strongly dependent on $Z$ and IMF (see Figure \ref{Fig:4}). This leads to an uncertainty in the ionising cluster mass that we can be reduced about one order of magnitude when choosing a metallicity similar to the abundance derived from the gas emission line spectrum. With this information, we have plotted in Figure~\ref{n1569}, panel a) the observed spectrum --red line-- compared with two models, GON05 and the ours, left and right panels, respectively for an age $t=3.23$\,Myr and $Z=0.008$. In order to do this comparison, the models have been reddened using the \citet{Fitzpatrick99} law and $A_{V}=2.3$ and all spectra --models and observed- have been normalized to the continuum. 
\begin{figure}
\hspace{-0.5cm}
\includegraphics[width=0.49\textwidth]{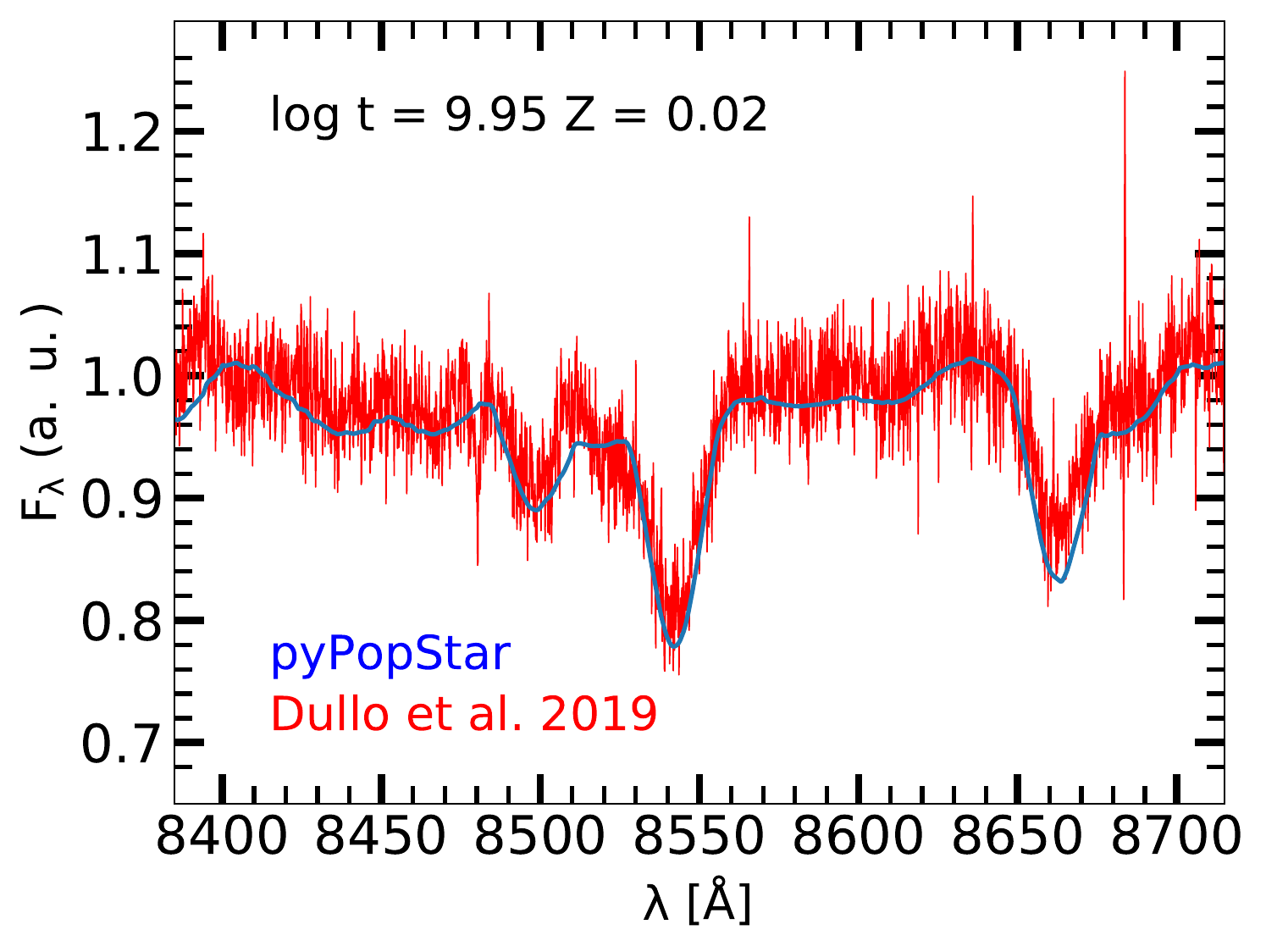}
\caption{Comparison of the spectrum for the bulge of NGC ~7025 as taken by \citet{dullo19} --red line-- with the present model --blue line--for $Z=0.02$ and an age 9\,Gyr.}
\label{n7025}
\end{figure}
Moreover, the models have been smoothed with a Gaussian of $\Delta\lambda= 0.27$\,\AA\ in order to have the same resolution as the observations. One of the most important feature clearly shown here is the WR-Bump. This feature can not be reproduced by previous HR models as seen in panel a) of Figure~\ref{n1569} where the GON05 model, the black line, (with a different colour compared with used in the previous Figure~\ref{Fig:9} in sake of clarity) has been plot. This WR bump, however, may be clearly reproduced by our {\sc HR-pyPopStar} model as seen in panel b) of the same Figure~\ref{n1569}. This way, our new {\sc HR-pyPopStar} models are the first ones with which it is possible to interpret this type of spectra, since until now the exiting HR evolutionary synthesis models either did not include young ages, such as M\&S11 or VAZ16 limited to ages older than 60-70\,Myr, or did not include WR stellar models that could reproduce the WR bump as in the case of GON05.  Moreover, the WR-bump profile produced by the WR stars has an absorption at \ion{He}{ii}~$4686\lambda$, which is filled in the observation, almost surely, with a \ion{He}{ii}~$4686\lambda$ nebular component. It is an example of how {\sc HR-pyPopStar} models allow to detect of such nebular contributions, which otherwise would be confused with the bump itself. We note that, for this and other lines, PoWR atmosphere models (both OB and WR) include emission features produced in the expanding atmosphere. Our models allow to disentangle which emission component belongs to the stars in the system and which one to the surrounding \ion{H}{ii} region, and hence would allow to improve nebular metallicity determinations.

\begin{figure}
\hspace{-0.3cm}
\includegraphics[width=0.49\textwidth]{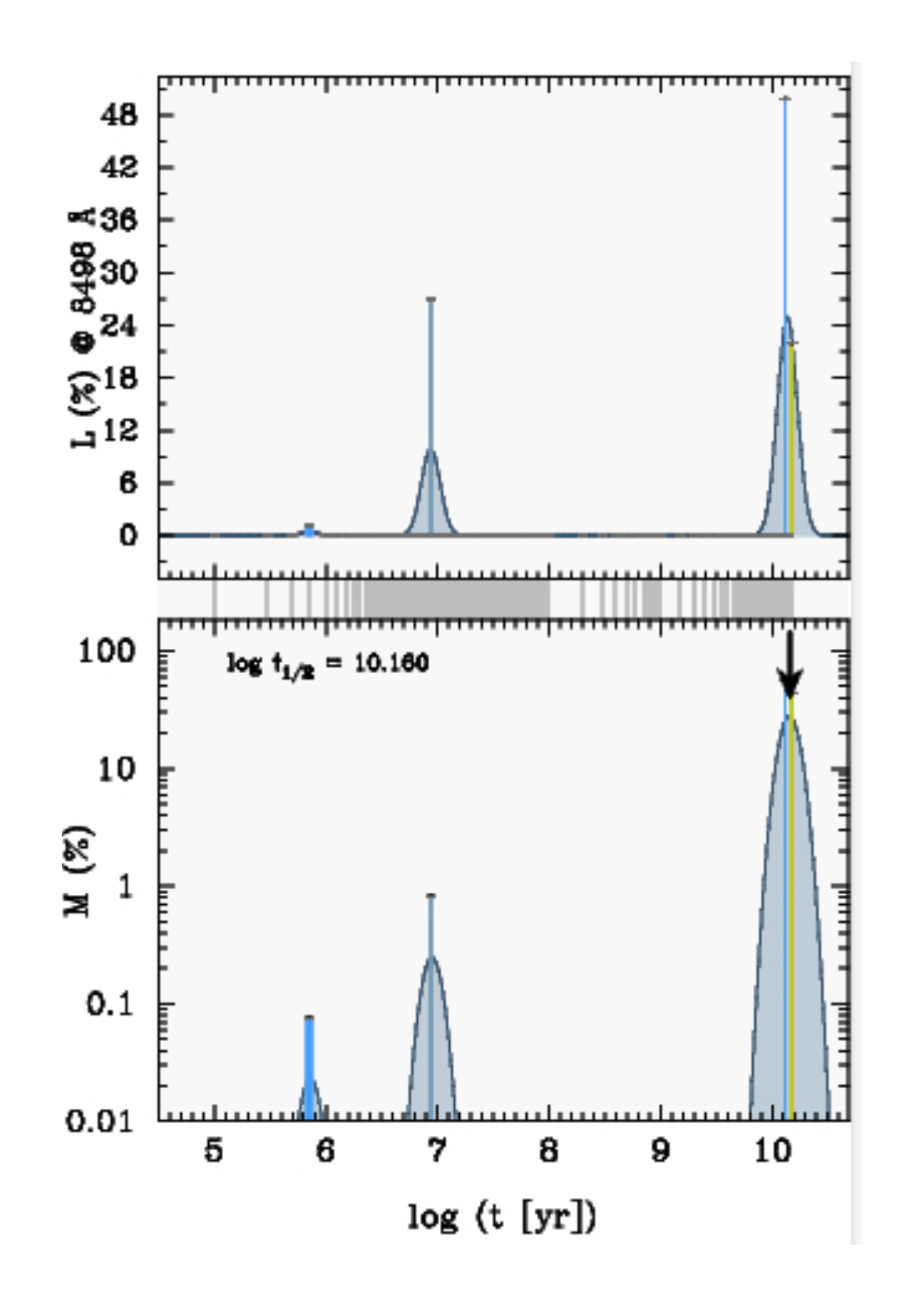}
\hspace{-0.7cm}
\includegraphics[width=0.45\textwidth]{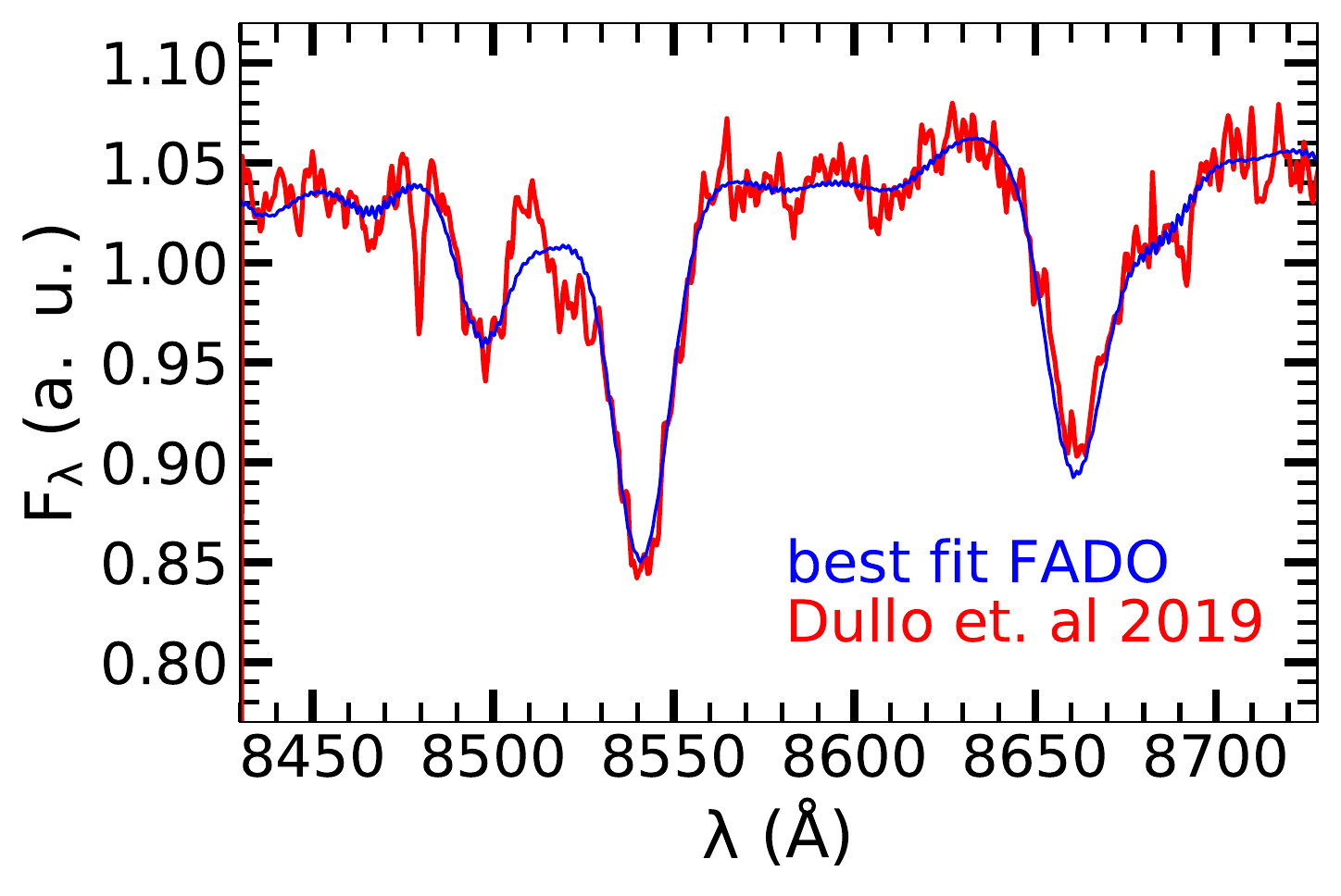}
\caption{Results of FADO for the bulge of NGC~725 after using the HR-I MEGARA spectrum from \citet{dullo19} and our HR SSP spectral models in the near-IR range. Top panels: the different components as percentage of light and mass in each age. Bottom panel: the resulting fit with these components.}
\label{n7025-FADO}
\end{figure}

It is necessary to take into account that the galaxy spectra are usually broadened by their intrinsic velocity dispersion. Therefore, these models would need to be convolved with the instrument spectral resolution to be adequately compared with data. This is indeed the case for our third case, the bulge of the galaxy NGC ~7025, studied by \citet{dullo19}. This is an isolated unbarred Sa galaxy located at a distance D=67.3\,Mpc with a bulge-to-total flux ratio (B/T) between 0.3 and 0.44, implying a stellar mass of $\sim 4.34\pm 1.70\,10^{10}\,\mathrm{M}_{\sun}$. These authors have observed its bulge with all the VPH's gratings of MEGARA in GTC. We use here the one corresponding to the HR-I, with a resolution power of $R\sim 18000$. Even using this high spectral resolution mode, the velocity dispersion of the bulge, so high as $\sigma=350$\,km\,s$^{-1}$, broadens the lines profile and smooths the spectra. We have used a old age SSP spectrum, with $t\sim 9$\,Gyr and a solar metallicity, $Z=0.02$, to compare with this bulge spectrum. We have also normalized both spectra, and we have convolved the modelled spectrum with a gaussian of $\Delta\lambda= 7.07$\,\AA\, as correspond to the given velocity dispersion of $\rm \sigma_{V} = 245$\, $\mathrm km\,s^{-1}$, before doing the comparison, shown in Figure~\ref{n7025}. This result has demonstrated the utility of using these HR models to interpret the HR spectra which are coming in the next future. 

In fact, as said in the Introduction, inverse techniques, as {\sc Starlight} or {\sc FADO}, need the SSP spectra as building blocks, with which it is possible to find the main contributions or components to reproduce a final observed spectra. We have performed a first test, analysing this bulge spectrum of NGC~ 7025 using {\sc FADO} and our new {\sc HR-pyPopStar} SSP models as bases for the code. We have fitted the wavelength range 8400 to 8800\,\AA. The result of the fit, shown in Figure~\ref{n7025-FADO}, top panels, gives three main components, one young with $\log{t}\sim 6$ and a percentage of a $\sim 0.1$\% in mass, a second intermediate age one with $\log{t}\sim 7$ and a percentage of a $\sim 1$\% in mass, and finally the old one, with an age of $\log{t}\sim 10$, being the main component with a percentage of more than a 80\% in mass. In the middle panel, the same result in terms of luminosity is shown. All components have solar metallicity and a small contribution of supersolar stellar population in the oldest one. Therefore, it is very relevant for this kind of code to have these HR spectra available as bases to run their models and to obtain the evolutionary histories of galaxies. The result of this composition is shown in the bottom panel of the same Figure where the observed spectrum is plotted smoothed, as {\sc FADO} gives, and the combination of the three contributions described above is drawn as the blue line. Since the young stellar populations participate with (very) small percentages of the total, the final fit results in a spectrum very similar to the one from Figure~\ref{n7025} only with an old stellar population. This exercise indicates to us that our models may give an important contribution to the galaxy spectra studies. 

Summarizing, with high spectral resolution instruments available, it will be possible to use these {\sc HR-pyPopStar} models in order to better interpret the spectra resulting of these observations.

\section{Conclusions}
\label{Sec:Conclusions} 

We have created {\sc HR-pyPopStar}, a new version of the {\sc PopStar} evolutionary synthesis model. Using this code, a set of high wavelength-resolution theoretical spectral energy distributions for SSPs from 91\,\AA\ to 24\,000\,\AA\ with a step in wavelength of $\delta\lambda = 0.1$\,\AA\ has been computed. We have used isochrones from  \citep{Bressan_Bertelli_Chiosi1993, Fagotto+1994a, Fagotto+1994b,Girardi+1996}, for 106 different ages in the range of $t=0.1$\,Myr to 15\,Gyr, four metallicities $Z = 0.004$, 0.008, 0.02 and 0.05,  and four IMFs from \citet{Salpeter1955},\citet{Ferrini_Penco_Palla_1990},  \citet[][with exponent $-2.7$ for the massive star range]{Kroupa2002}  and \citet{Chabrier2003}. 

All this information, the computed SEDs, as well as the tables with the associated results, will be available on the web page:
\url{http://www.pypopstar.com}. The user will be able to download the complete set of models or a required subset (with all ages) just selecting by IMF and $Z$. Moreover, since the code is totally flexible, any other model with different input, as other stellar isochrones or stellar libraries, or IMFs, might be computed upon request to the authors.

We have checked the effect of using different metallicities and IMFs on the magnitudes and ionisation photons numbers, obtaining, as expected, similar results as found with the low wavelength-resolution models: 
\begin{itemize}
\item The magnitudes have small differences for variations on $Z$, more important for the K-band where they may reach until 2 magnitudes of differences (mainly if the nebular is taken or not taken into account). The variations of the magnitudes by the effect of using different IMFs are larger than the ones with different $Z$, being larger for the youngest stellar populations and decreasing with the age of these ones.

\item The number of ionisation photons for H, \ion{He}{i}, \ion{O}{ii} and \ion{He}{ii} is very dependent on the metallicity, as known \citep{Smith_Norris_Crowther2002}.
These numbers are, however, very similar for all ages and IMF, given a metallicity, except for $Q(\mathrm H)$ at the youngest ones where these variations are evident, as due to the different number of massive stars. 

\item As the evolutionary stage of the population for a given age changes with metallicity, consequently the shape of the SED does. The variation in the four used IMF modifies mainly the total luminosity.

\end{itemize}

We have compared the results of our models with others from the literature with intermediate or similar high wavelength-resolution, finding similar shapes for the spectra, with differences explained by the choices of isochrones or stellar libraries.

The most important result is that our HR models allow to distinguish a large number of absorption lines, in a better way than existing models, which will improve very much the theoretical estimation of their equivalent widths. Even if these models would need to be broadened for using with galaxies data, we have showed the potential of high resolution spectra to discern correctly some lines or features.

In fact, we have compared our results with some particular spectra obtained with high or intermediate resolution by the MEGARA spectrograph in the GTC, this way demonstrating the ability of the models to reproduce these observations, showing clearly some features as the WR bump or other metallic lines, and helping to discriminate age or metallicity of their stellar populations. We have also tested the possibility of using these models through the code {\sc FADO}, which allows to obtain the evolutionary history of galaxies or regions of galaxies.

Although this code could be improved with new isochrones or stellar libraries when available, the present version of {\sc HR-pyPopStar} models will be useful to interpret galaxies and stellar clusters observations coming from HR instruments.

\section*{Acknowledgements}

This work has been supported by MINECO-FEDER grant AYA2016-79724-C4-3-P, AYA2016-79724-C4-1-P, AYA2017-88007-C3-1-P, PGC-2018- 0913741-B-C22, MDM-2015-0509 and MDM-2017-0737 (Unidad de Excelencia María de Maeztu  CAB). PRTC acknowledges financial support from Funda\c{c}\~{a}o de Amparo \`{a} Pesquisa do Estado de S\~{a}o Paulo (FAPESP) process number 2018/05392-8 and Conselho Nacional de Desenvolvimento Cient\'ifico e Tecnol\'ogico (CNPq) process number  310041/2018-0. 

\section{Data availability}

The data underlying this article are available in our web page: \url{http://www.pypopstar.com}, where it will be possible to download the complete set of models or only a required subset (with all ages) just selecting by IMF and $Z$.

The supplementary material as described in Appendix A and Appendix B is available in electronic format exclusively as a PDF file. Appendix A describes the ingredients of the evolutionary synthesis code inherited from {\sc PopStar}. Appendix B shows a comparison of magnitudes and numbers of ionizing photons for the new {\sc HR-pyPopStar} and the old {\sc PopStar} models.

Tables 5 and 6 are in electronic format as ASCII files. Table 5 gives the magnitudes in broadband filter of HST UV magnitudes (column 5 and 6), Johnson-Cousins-Glass system magnitudes in the Vega system (columns 7 to 15) and SDSS filters in the AB system (columns 16 to 20). Table 6 shows the number of ionizing photons for \ion{H}{i} (column 4), \ion{He}{i} (column 5), \ion{He}{ii} (column 6) and \ion{O}{ii} (column 7) for each IMF (column 1), metallicity (column 2) and age (column 3).



%\bibliographystyle{mnras}

%\clearpage

%--------------------------------------------------------------------------

\bsp	% typesetting comment
\label{lastpage}
\end{document}

\label{firstpage}
\pagerange{\pageref{firstpage}--\pageref{lastpage}}
\maketitle
\appendix
\section{Ingredients of PyPopStar}
\label{AppA}
\subsection{Isochrones}
\label{Subsec:Isochrones}

A key ingredient for stellar population synthesis models is the set of stellar isochrones. The stellar isochrones track the stellar characteristics and the actual stellar mass of the still living stars at a given age and for a given chemical composition. 

There have been a lot of studies in stellar evolution, when creating new stellar tracks and isochrones. The latest works of the Padova group have produced the codes PARSEC and COLIBRI, which take into account some of the above mentioned issues. PARSEC  code \citep{Bressan+2012} computes the stellar evolution from PMS phase to the first thermal pulse of the Asymptotic Giant Branch (AGB) phase. COLIBRI \citep{Marigo+2013} computes the rest of the AGB phase treating carefully  all the thermal pulses, taking into account the overshooting and dredge up processes mentioned above. These works, however, end the stellar track at the AGB phase without computing neither the Post-AGB nor the Planetary Nebulae (PNe) phases.

In turn, the Geneve and Bonn groups have been working in the effects of stellar rotation on the isochrones \citep{Ekstrom+2012,Georgy+2013, Yusof+2013,Groh+2019}. They have good coverage in stellar rotation and age, and the metallicity coverage is restricted to MW, LMC and SMC abundances.
However, these models do not include nor the post-AGB neither the PNe phases. The stellar rotation is important because it changes the nucleosynthesis allowing, besides other effects, the production of primary N \citep{chiappini06,chiappini11,lim18}. These last authors have followed carefully the stellar evolution of massive stars including different rotation velocity values and 4 metallicities, [Fe/H]=-3,-2,-1 and 0, allowing better studies in the chemical evolution of galaxies field \citep{prantzos+2018}, but they did not provided isochrones. 

BASTI group have created isochrone models \citep{Hidalgo+2018} for ages older than 20\,Ma, which is considered too old to reproduce observations from young stellar populations able to ionise the gas, one of the aim of our work.

All isochrones mentioned above have strengths and weaknesses, because each model approach is focused on specific aspects of stellar evolution. The plan for producing the models is a two-step process. In the first step we have updated the old {\sc PopStar} code (now written in Python) and have included the high resolution atmosphere models. We have therefore used the same Padova isochrones from \citep{Bressan_Bertelli_Chiosi1993, Fagotto+1994a, Fagotto+1994b,Girardi+1996} that we used in the previous versions of {\sc PopStar}. These isochrones have a very good coverage in age, from 0.1\,Ma to 13.8\,Ga and we use them for four metallicities, $Z$=0.004, 0.008, 0.02 and 0.05. {\bf } This first step will allow us to quantify the improvement due to the use of the high-resolution atmosphere models while keeping the same isochrones for both {\sc HR-pyPopStar} and old {\sc PopStar} models. In a second step, we will change these isochrones for more recent ones.

{\bf In order to consider WR stars, the Padova isochrones give the superficial hydrogen to distinguish a WR from a blue supergiant star. } 

\subsection{The Initial Mass Function}
\label{Subsection:IMF}

We have computed the number of stars in each mass interval for each given isochrone using 4 different Initial Mass Functions (IMF): \citet[][hereinafter SAL]{Salpeter1955}, \citet[][hereinafter FER] {Ferrini_Penco_Palla_1990}, \citet[][note that we are using the exponent $-2.7$ in the massive star range]{Kroupa2002}; and \citet[][hereinafter CHA]{Chabrier2003}:

\begin{equation}\label{IMF_SAL}
    \phi(m)_\mathrm{SAL}= 0.171 \, m^{-2.35}\,\;\; \mathrm{with}\;\;1\,\mathrm{M}_{\sun} < m \le 120\,\mathrm{M}_{\sun}
\end{equation}

\begin{eqnarray}\label{IMF_FER}
    \phi(m)_\mathrm{FER} &=& 2.055\,\frac{10^{-\sqrt{0.73+\log(m)(1.92+2.07 \ \log(m))}}}{m^{1.52}}, \nonumber \\
    && \;\;\;\; \mathrm{with}\;\;0.15\, \mathrm{M}_{\sun} < m < 100\,\mathrm{M}_{\sun}
\end{eqnarray}

\begin{equation}\label{IMF_KRO}
\phi(m)_\mathrm{KRO} = 0.318
  \begin{cases} 
%  m^{-0.35} & \text{if}\  0.08\, \mathrm{M}_{\sun} < m < 0.15 \mathrm{M}_{\sun} \\ 
   2 \,m^{-1.3} & \text{if }  0.15\,\mathrm{M_{\sun}} <  
   m < 0.50\,\mathrm{M_{\sun}}   \\ 
    m^{-2.3} & \text{if } 0.50\,\mathrm{M_{\sun}} < \mathrm{m}< 1 \ \mathrm{M}_{\sun} \\ 
    m^{-2.7} & \text{if } 1\,\mathrm{M_{\sun}} < \mathrm{m}< 100 \ \mathrm{M}_{\sun} ,
  \end{cases}
\end{equation}

\begin{equation}\label{IMF_CHA}
\phi(m)_\mathrm{CHA} = 
  \begin{cases} 
%  0.037\,m^{-1.0} \ e^{- \chi} & \text{if}\  0.15\,\mathrm{M}_{\sun} < m < 1.0\, \mathrm{M}_{\sun} \\ 
%   0.019\,m^{-2.3} & \text{if }  1.0\,\mathrm{M}_{\sun}  < 
%   m < 100\,\mathrm{M_{\sun}},
  0.068\,m^{-1.0} \ e^{- \chi} & \text{if}\  0.15\,\mathrm{M}_{\sun} < m < 1.0\, \mathrm{M}_{\sun} \\ 
   0.0193\,m^{-2.3} & \text{if }  1\,\mathrm{M}_{\sun}  < 
   m < 100\,\mathrm{M_{\sun}},
  \end{cases}
\end{equation}

\noindent where $\phi$ is the IMF i.e., the distribution that provides the probability that a star has been formed with a initial mass $m$ in a $\mathrm{d}m$ interval; and in the case of $\phi(m)_\mathrm{CHA}$, $\chi$ is defined as $ \chi = \frac{(\log{m}+\log{0.079})^2}{2\times 0.69}$. 
Each IMF is normalised in mass, which is equivalent to assume a total mass in the system of $1 \mathrm{M}_{\sun}$, it is:
\begin{equation}
    \int_{m_{low}}^{m_{up}}m\,\phi(m)\mathrm{d}m=1,
\end{equation}
where $m_{low}$ and $m_{up}$ are the limits of the used IMF, for which we use 0.15 and 100\,$\mathrm{M}_{\sun}$ except for SAL, for which we take 1 and 120 \,$\mathrm{M}_{\sun}$ in order to simulate the one from SB99.

\subsection{Assignation of an spectral model for WR stars}
\label{WR}
Wolf-Rayet stars are slightly more complex due to their optically thick winds and high mass-loss rates.
While the isochrones give the effective temperature at the radius $R_\mathrm{hydro}$, where the optical depth is $\tau=2/3$,
WR atmosphere models are computed at an optical depth $\tau=20$.
Thus, we have followed a similar method as MOL09 and assigned the WR spectral templates based on $R_{\tau=20}$. 

We begin from the optical depth equation:
\begin{equation}\label{Eq:Optical_depth}
	d\tau = -\kappa(r) \rho(r) dr,
\end{equation} 
where $\tau$ is the optical depth, $\rho$ is the density, $\kappa = 0.2 (1+ X_\mathrm{S})$ is the opacity and $X_\mathrm{S}$ is the H abundance by mass in the surface of the star, that we assume as 0.2 for WN and 0.0 for WC. 
Then, using the law of conservation of mass: 
\begin{equation}\label{Eq:Mass_loss_WR}
	\dot{M} = 4 \pi r^2 \rho(r) v(r),
\end{equation}
where $v(r)$ is the wind velocity field, and assuming that wind velocity has the following form \citep{Bertelli_Bressan_Chiosi_1984}:
\begin{equation}\label{Eq:Wind_velocity_field}
	v(r) = v_{\infty} \left(1-\frac{R_{\rm hydro}}{r}\right)^{\beta},
\end{equation}
where $v_{\infty}$ is the terminal velocity, $R_{\rm hydro}$ is the hydrostatic radius of the star and $\beta=2$,
including \eqref{Eq:Wind_velocity_field} in \eqref{Eq:Mass_loss_WR} we obtain the following formula:
\begin{equation}\label{Eq:Mass_loss_final}
    \dot{M} = 4 \pi r^2 \rho(r) v_{\infty} \left(1-\frac{R_{\rm hydro}}{r}\right)^2.
\end{equation}
Integrating \eqref{Eq:Optical_depth} and \eqref{Eq:Mass_loss_final} we obtain the relation between the hydrostatic radius and the radius $R_{\tau=20}$:
\begin{equation}
	R_{\tau=20} = R_{\rm hydro} \left[ 1+ \frac{0.01(1+X_{\rm S}) \dot{M}}{4 \pi v_{\infty} \ R_{\rm hydro}} \right]
\end{equation}

After assigning by $R_{20}$, we look for the stellar spectral model with the closest mass loss rate, estimated from the difference
\begin{equation}
    \dot{M} = \frac{M_{\rm ini}-M_{\rm current}}{\tau} ,
\end{equation}
between the initial mass $M_{\rm ini}$ of the star and the current value $M_{\rm current}$ at the age $\tau$ of the isochrone.

\subsection{Nebular emission}\label{Subsec:Nebular emission}
The computation of the nebular continuum emission is also an important part of the total spectrum of a SSP whenever the ionising spectrum exists because (a) the nebular continuum increases the total continuum (stellar + nebular) producing a larger dilution in the emission lines and (b) the contribution of the nebular continuum reddens the spectrum modifying the colours. This effect is specially relevant in young star clusters ($t<10-20$ Ma), \citep[see][]{CER1994,garcia-vargas94,GarciaVargas_Molla_MartinManjon2013}. Thus, we have computed the nebular continuum emission using the same equations than in MOL09: 

\begin{equation}
   L_{\lambda,\mathrm{neb}} = \Gamma \frac{\mathrm{c} \ Q(\ion{H}{i})}{\lambda^2 \ \alpha_\mathrm{B}(\ion{H}{i})},
   \label{eq:L_Q}
\end{equation}
where $L_{\lambda,\mathrm{neb}}$ is the luminosity per wavelength unit, $\Gamma$ 
is the emission coefficient, $Q(\rm H)$ is the number of ionising photons, $\lambda$ is the wavelength, and  $ \alpha_\mathrm{B}(\ion{H}{i})$ is the recombination coefficient for \ion{H}{i} \citep[see][]{Osterbrock_Ferland2006}.

Therefore, it is necessary to compute the $Q(\ion{H}{i})$ of each stellar SED computed as explained in the above sections. In addition to the number of ionising photons of \ion{H}{i}, we have also computed $Q(\ion{He}{i})$, $Q(\ion{He}{ii})$ and $Q(\ion{O}{ii})$. In order to compute each $Q$, we have to integrate the stellar luminosity: 
\begin{equation}
Q(i)(Z,t) = \int_{0}^{\lambda_i}\frac{L_{\lambda}(Z,t)}{\mathrm{h} \, \mathrm{c}} \, \lambda \, \mathrm{d}\lambda
\end{equation} 
where $\lambda_i$ is the ionisation edge wavelength for an element in a given ionisation state $i$, the wavelengths for the elements that we are considering  
being in Table~\ref{Table:A1}, 
$L_{\lambda}$ is the stellar luminosity of each SSP per wavelength. 

The emission coefficient, $\Gamma$ has three main components: the emission from \ion{H}{i}, the emission from \ion{He}{i} and the two photon continuum emission. Thus, the emission coefficient is computed by summing up the three components: 
\begin{equation}
    \Gamma = \Gamma_{\ion{H}{i}} + \Gamma_{\ion{He}{i}} \frac{N(\ion{He}{i} )}{N(\mathrm{H})}+ \Gamma_{2q}
\end{equation}
The two-photon continuum emission is an important source of nebular continuum emission. The emissivity of this source is the following:
\begin{equation}
    \rm \Gamma (2q) = \frac{\alpha_{\rm eff} g_{\nu}}{1 + q_2/A_{2q}},
\end{equation}
where $\alpha_{\rm eff}$ is the effective recombination coefficient,  $A_{2q}=8.2249$ is the two-photon continuum transition probability and $q_2$ is the collisional transitional rate. All these data have been taken from \citet{Osterbrock_Ferland2006}, except $\rm g_{\nu}$, which comes from \citet{Nussbaumer_Schmutz1984}. 

The emissivity of \ion{H}{i}, \ion{He}{i} and \ion{He}{ii} is divided in two parts \citep[see][for further details]{molla09}: free-free emission due to collisions between electrons and protons in the ionised gas surrounding the stellar population is computed according to \citet{Osterbrock_Ferland2006}, whereas free-bound emission due to recombination is estimated from the emissivity tables given by \citet{Ercolano_Storey2006}.

\begin{table}
	\begin{center}
\caption{Ionisation edge wavelength for species whose $Q$ is evaluated.}
\label{Table:A1}
			\begin{tabular}{|c|c|}
				\hline
				Element & Wavelength \\
				\hline
				$\rm H \ I$ &  $ 911.8$\, \AA \\
				$\rm He \ I$ & $504.0$\, \AA \\
				$\rm O \ II$ & $ 353.3$\, \AA \\
				$\rm He \ II$ & $ 228.0$\, \AA \\
				\hline
			\end{tabular}
	\end{center}
\end{table}

\begin{figure*}
\includegraphics[width=0.4\textwidth]{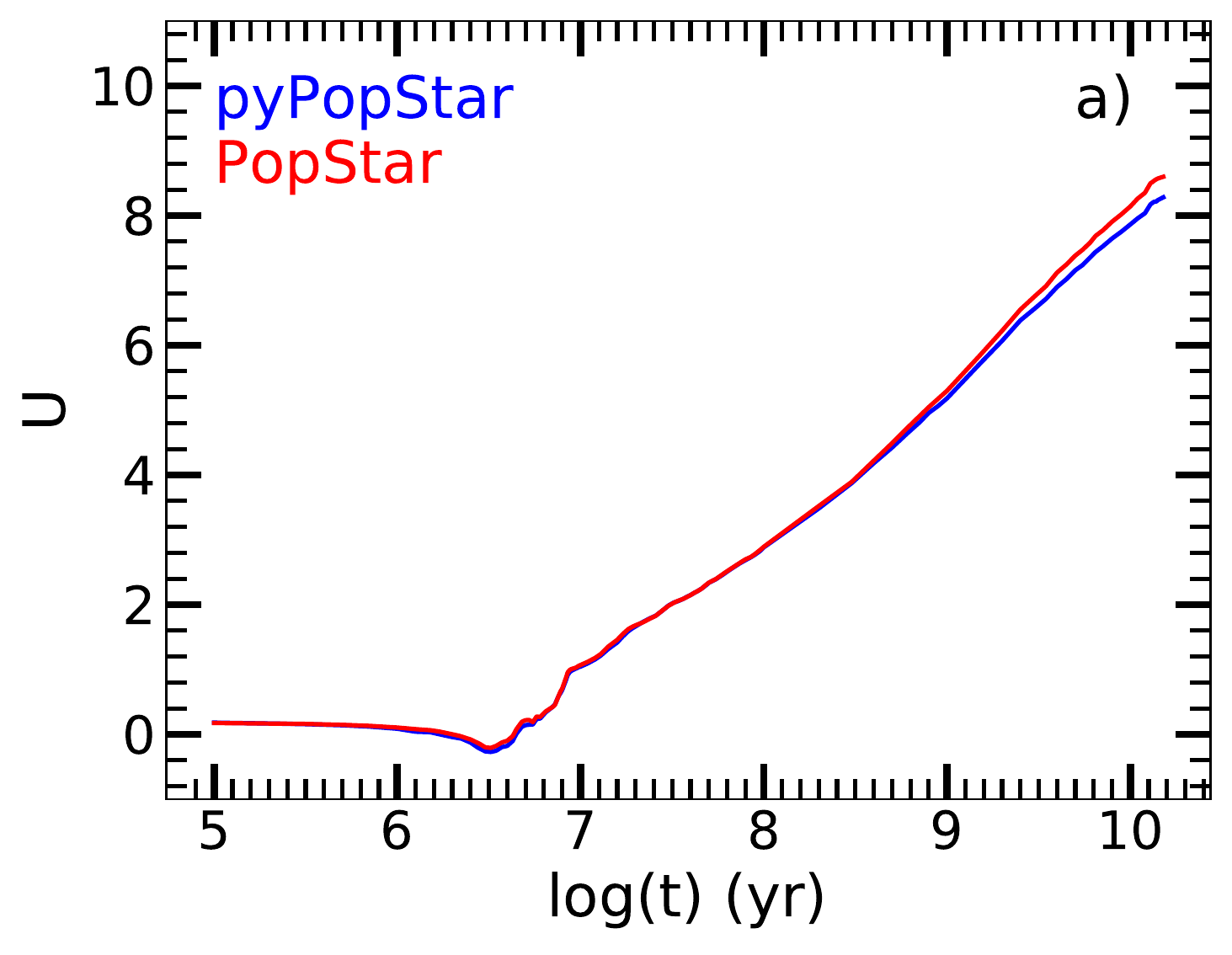}
\includegraphics[width=0.4\textwidth]{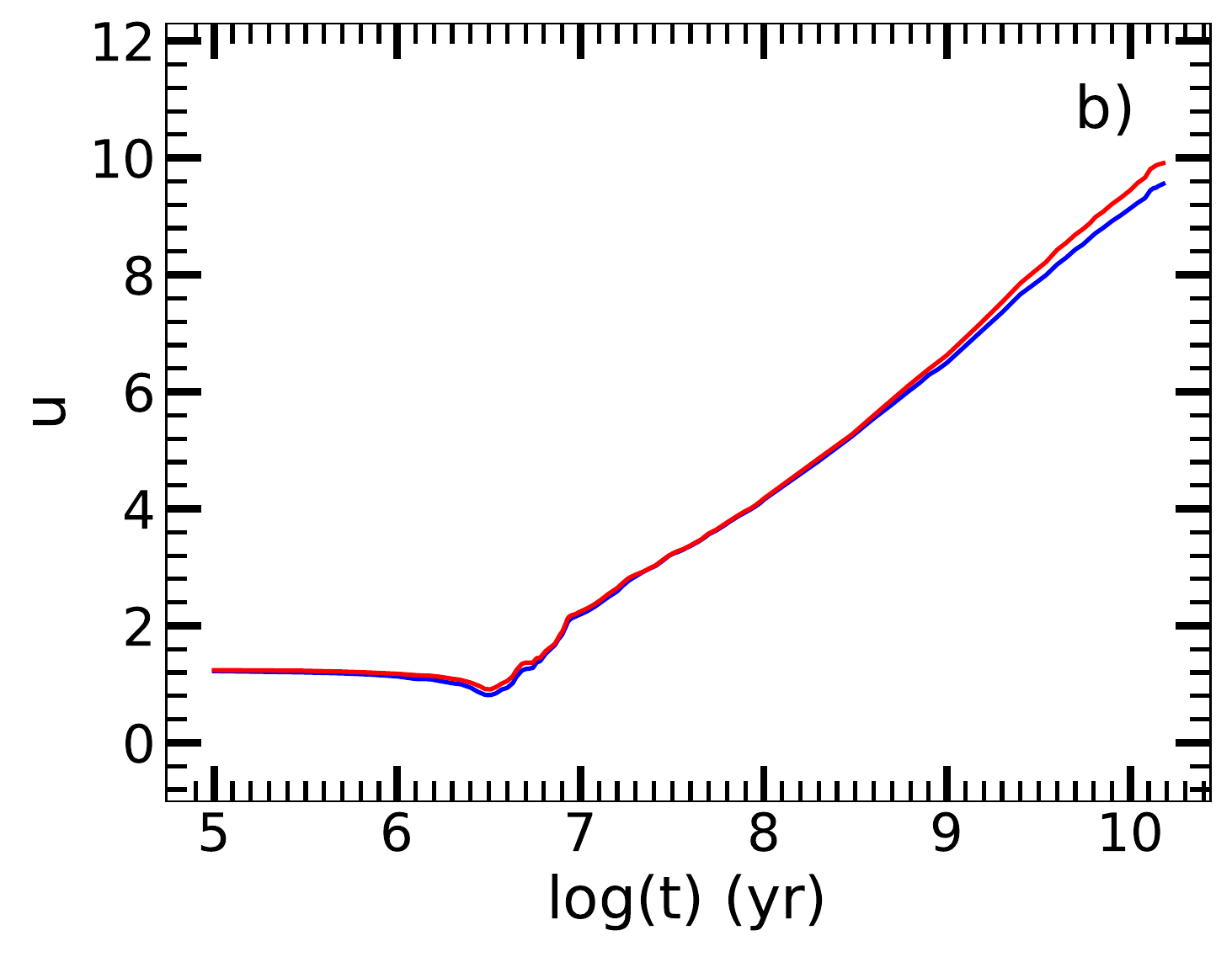}
\includegraphics[width=0.4\textwidth]{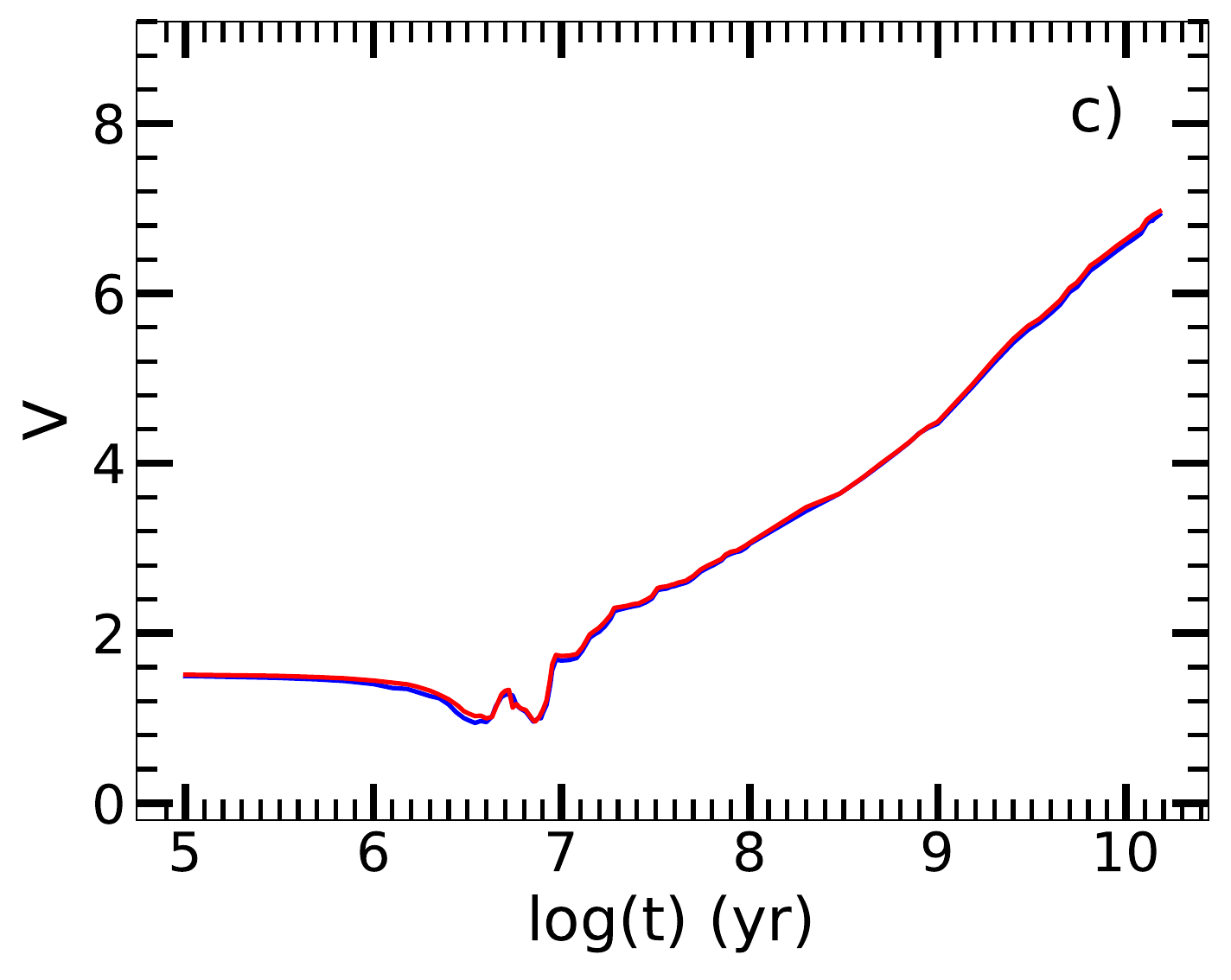}
\includegraphics[width=0.4\textwidth]{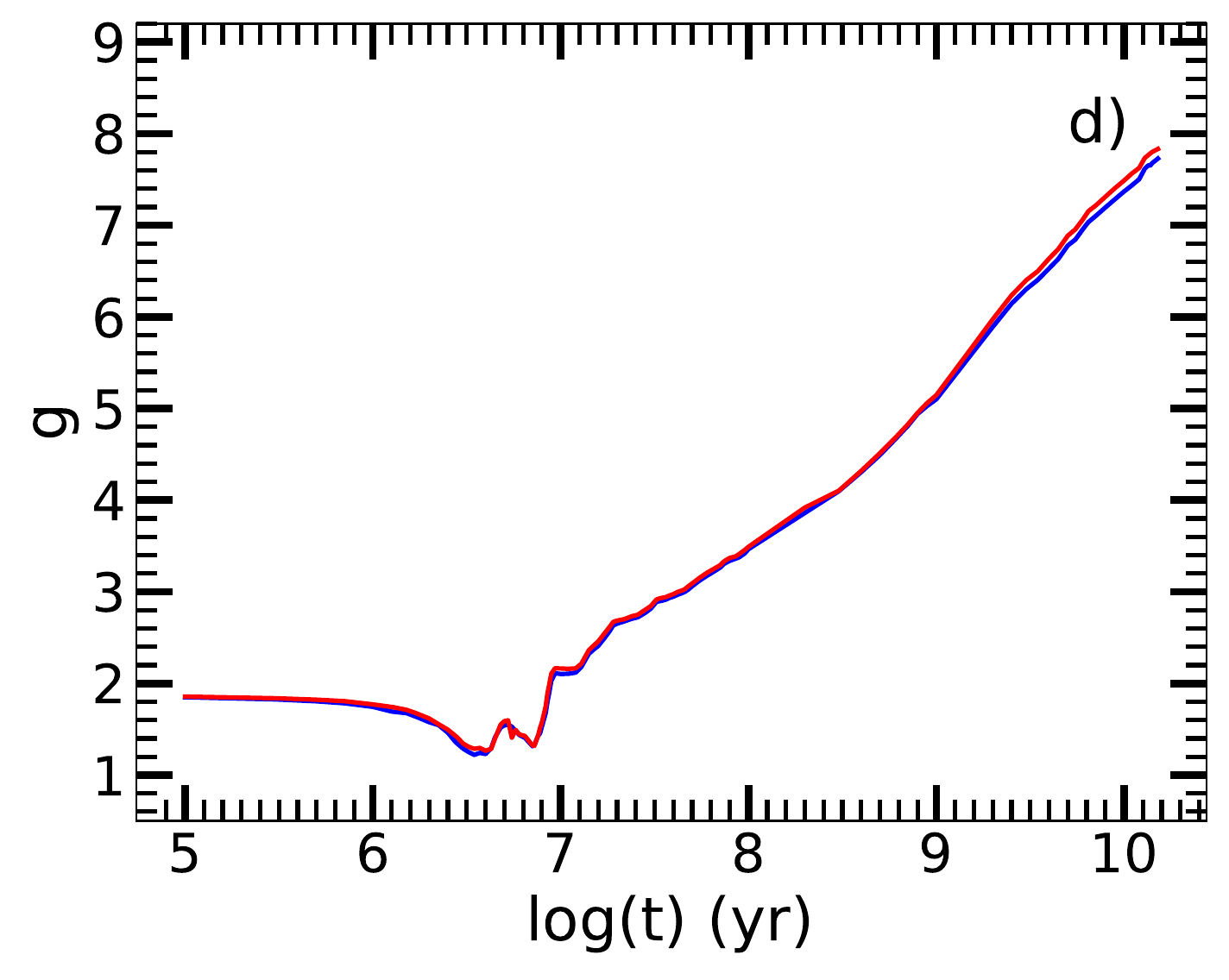}
\includegraphics[width=0.4\textwidth]{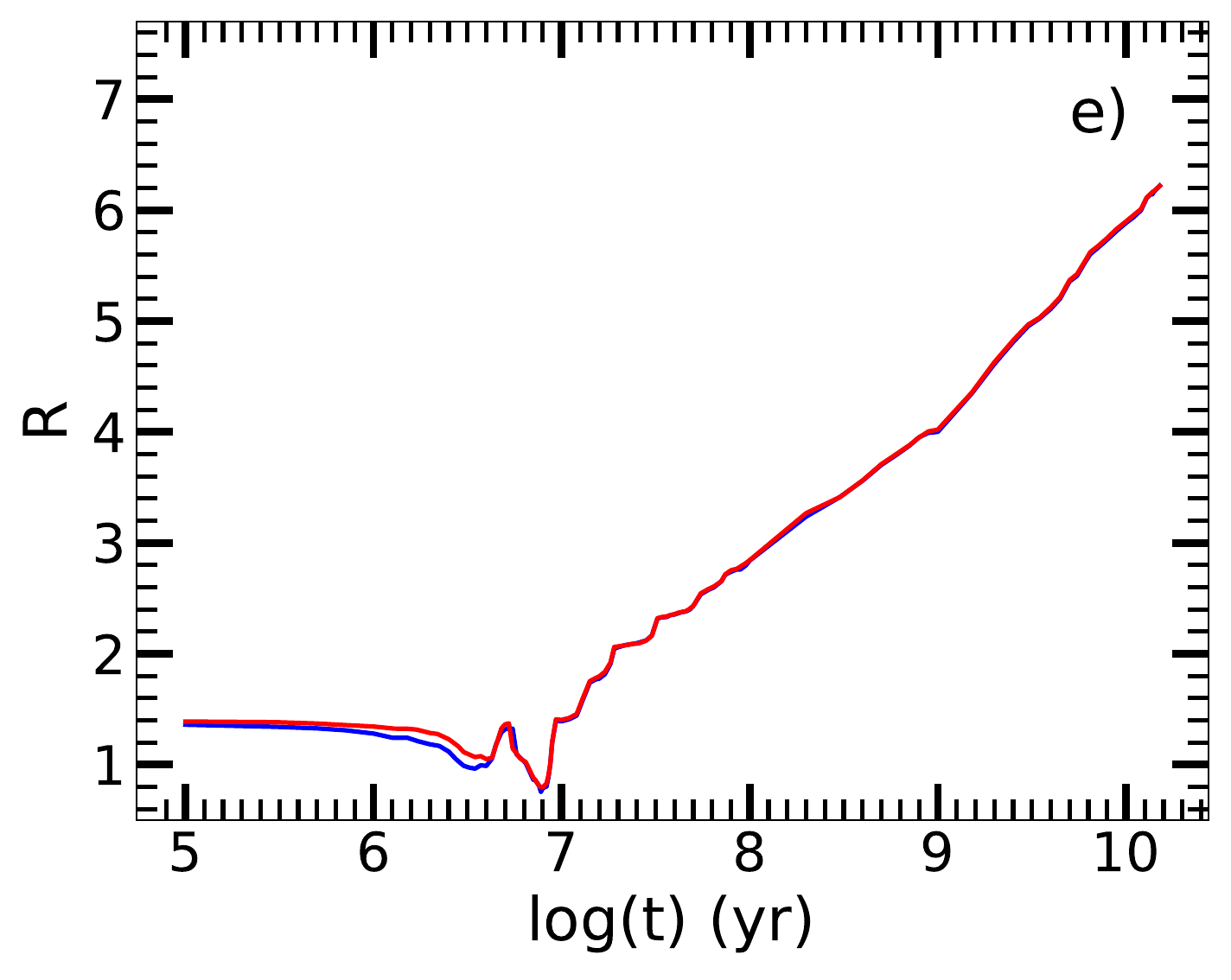}
\includegraphics[width=0.4\textwidth]{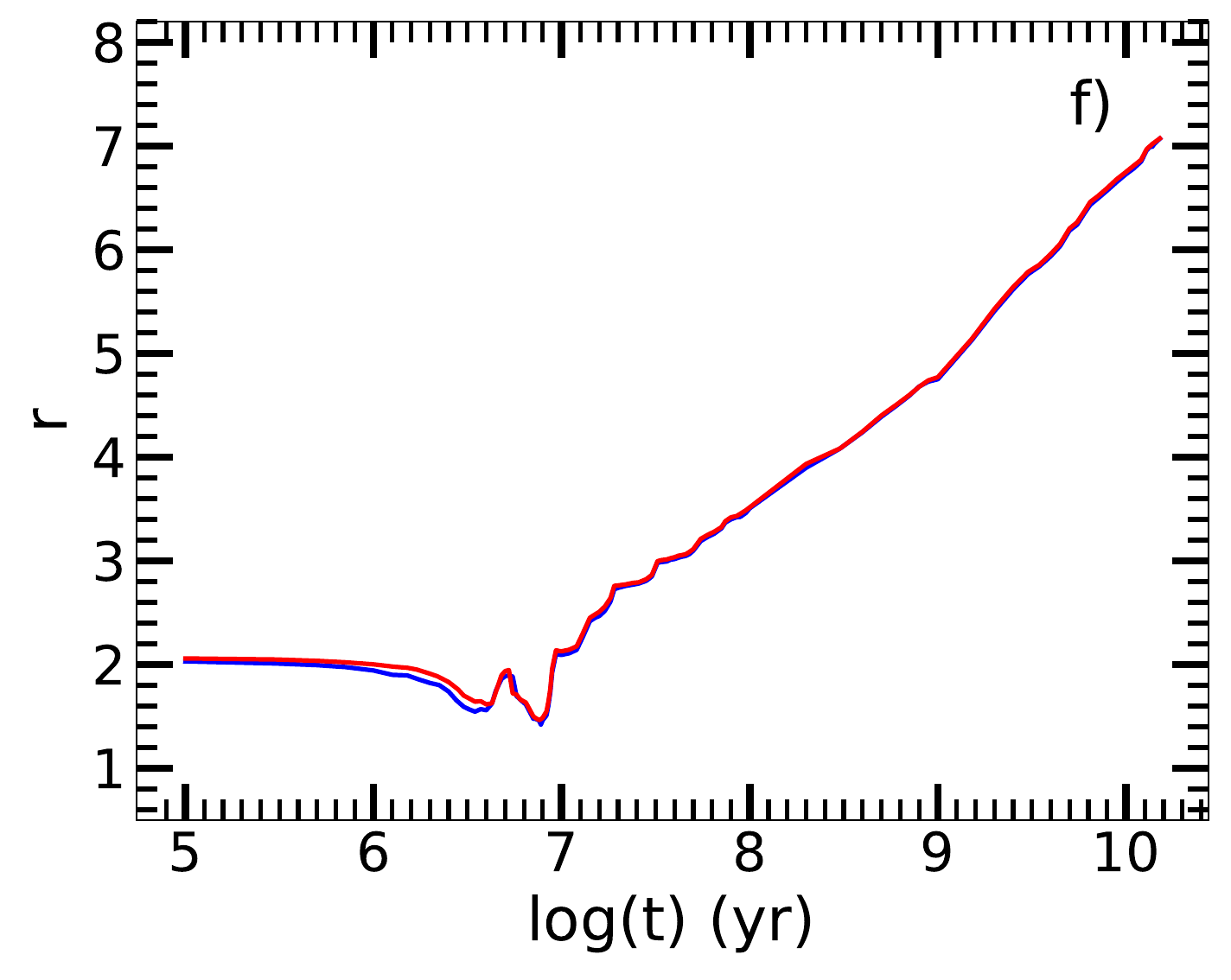}
\includegraphics[width=0.4\textwidth]{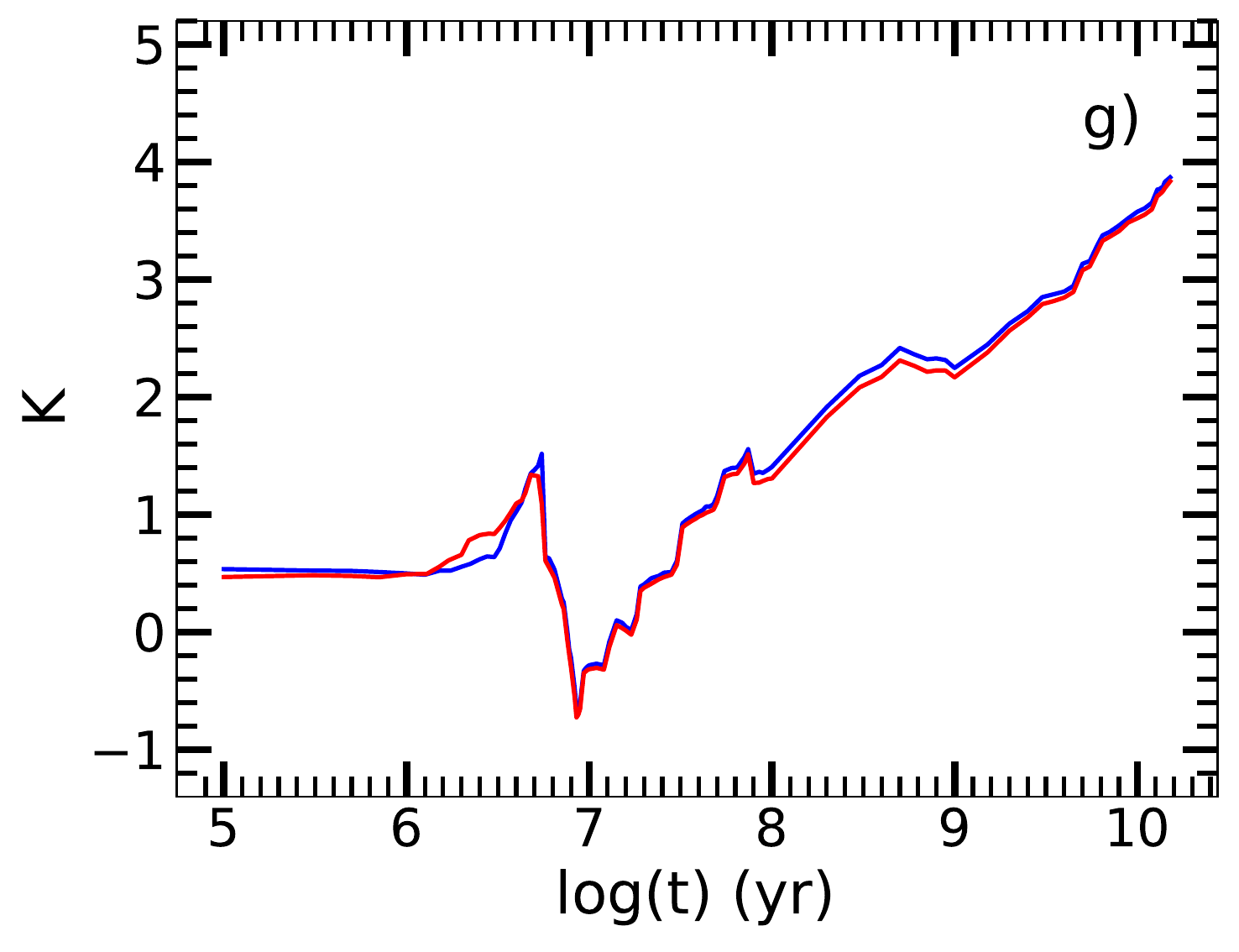}
\includegraphics[width=0.4\textwidth]{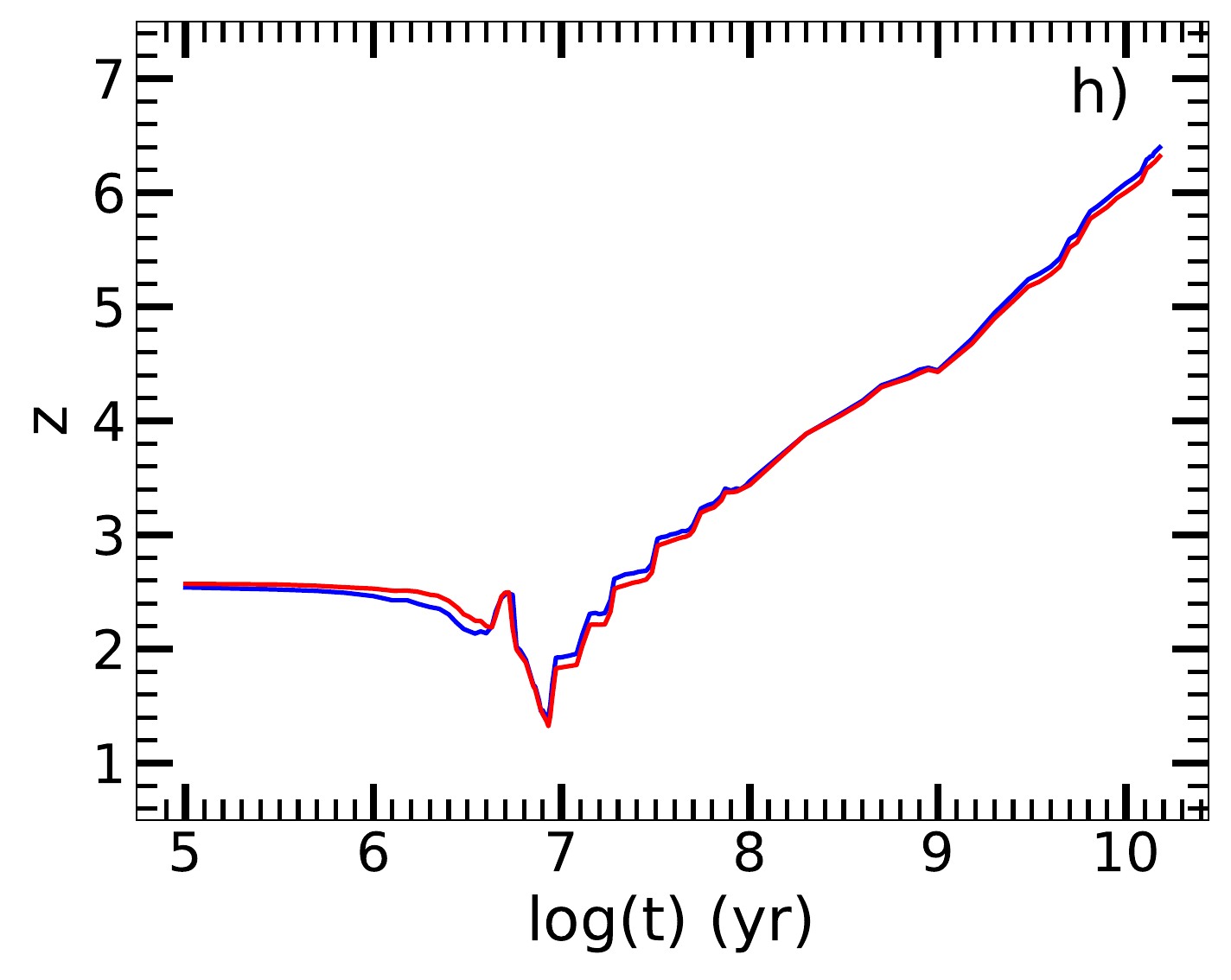}
\caption{Comparison of U,B,V $\&$ K (left panels);
and of $u$, $g$, $r$, and $z$ (right panels) time evolution for solar metallicity of {\sc PopStar} (red lines) and {\sc HR-pyPopStar} (blue lines).}
\label{Fig:B1}
\end{figure*}
\section{Comparison HR-pyPopStar with PopStar}
\label{AppB}

In subsection 3.4 we have compared our new {\sc HR-pyPopStar} SED's with the results of the old ones {\sc Popstar} MOL09 in sake of consistency and to check the new results are equally good.

As we want to ensure the consistency between the {\sc Popstar} and {\sc HR-pyPopStar} models, we are also going to check that the magnitudes are equivalent for the only metallicity in common, the solar one.  
The comparison of the time evolution of magnitudes U, V, R and K for both models is shown in Fig.~\ref{Fig:B1}, left panels, while the corresponding comparison with the SDSS magnitudes is in the right panels for $u$, $g$, $r$ and $z$. All of them computed with the total (stellar $+$ nebular spectra). 
As expected, there are no significant difference between the old and the new models. 

\begin{figure}
\centering
\includegraphics[width=0.45\textwidth]{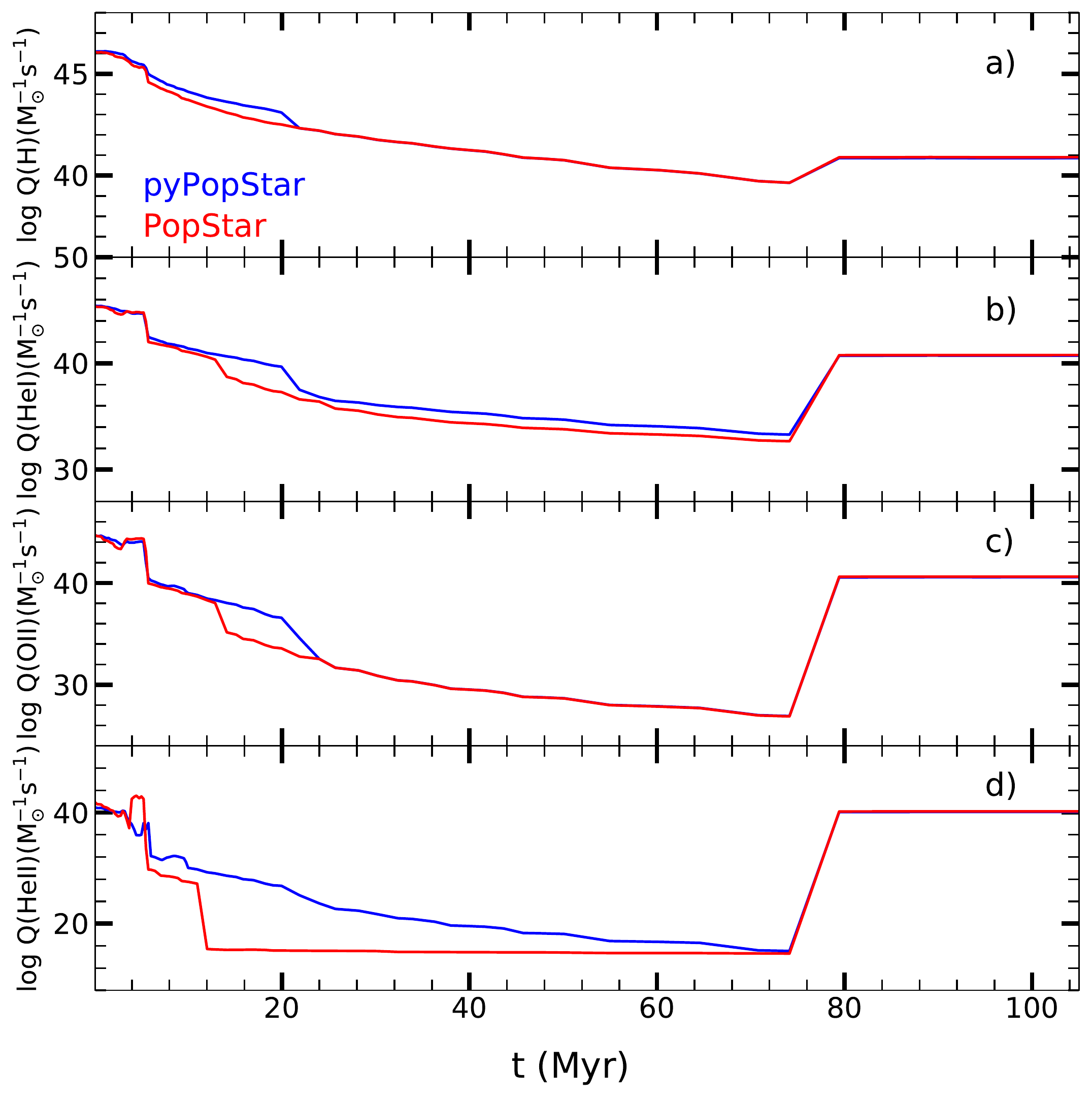}
\caption{Time evolution of $Q$(H), $Q$(HeI), $Q$(OI) and $Q$(HeII) for solar metallicity compared with the corresponding evolution from {\sc PopStar}.}
\label{Fig:B2}
\end{figure}

We have compared the evolution of all $Q$'s with the old {\sc PopStar} ones, in order to quantify the difference arisen from the change of the stellar libraries.   In Fig.\ref{Fig:B2} we represent  the evolution of $Q$(\ion{H}{i}), $Q$(\ion{He}{i}), $Q$(\ion{O}{ii}) and $Q$(\ion{He}{ii}) of {\sc PopStar} (red lines) and {\sc HR-pyPopStar} (blue lines) for the solar abundance.  
Since the stellar libraries for the PN are the same as used in {\sc PopStar}, we only represent the evolution for ages younger than 100\,Ma, ($\log{\tau} \le 8$). For the \ion{H}{i}, there are slight differences in the young populations due to the changes in the O and B libraries that are harder than the previously used in the old {\sc PopStar}. In the case of \ion{He}{i}, the difference is noticeable until 70 Ma. The evolution of the \ion{O}{ii} is very similar to the evolution of \ion{H}{i}. 
Finally, the \ion{He}{ii} have larger differences in the youngest ages, because the new O \& B and WR star models are very different to the old ones in the range of 91-250 \AA.

\bsp	% typesetting comment
\label{lastpage}